\documentclass[floats,floatfix,amssymb,prd,
  superscriptaddress,nofootinbib,preprintnumbers,
  onecolumn,notitlepage
  ]{revtex4-1}
	
\usepackage{amsmath}	
\usepackage{graphicx,xcolor}

\begin{document}

{

\graphicspath{{Figures/}}

\title{The History of Primordial Black Holes}
\author{Bernard J. Carr~\footnote{School of Physics and Astronomy, Queen Mary University of London, UK,  \email{b.j.carr@qmul.ac.uk}} and Anne M. Green~\footnote{School of Physics and Astronomy, University of Nottingham, UK,  \email{anne.green@nottingham.ac.uk} }}

\begin{abstract}
We overview the history of primordial black hole (PBH) research from the first papers around 50 years ago to the present epoch.
The history may be divided into four periods, the dividing lines being marked by three key developments: inflation on the theoretical front and the detection of microlensing events by the MACHO project and gravitational waves by the LIGO/Virgo/KAGRA project on the observation front. However, they are also characterised by somewhat different focuses of research.  The period 1967-1980 covered the groundbreaking work on PBH formation and evaporation. The period 1980-1996 mainly focussed on their formation, while the period 1996-2016 consolidated the work on formation but also collated the constraints on the PBH abundance.  In the period  2016-2024 there was a shift of emphasis to the search for evidence for PBHs and -- while opinions about the strength of the purported evidence vary --  this has motivated more careful studies of some aspects of the subject. Certainly the soaring number of papers on PBHs in this last period indicates a growing interest in the topic.
\end{abstract}

\maketitle\label{chapter:history}

\section{Introduction}
\label{sec:intro}

General relativity predicts that a region of mass $M$ forms a black hole (i.e.~a region where the gravitational field is so strong that not even light can escape) if it falls within its Schwarzschild radius $R_{\rm S} \equiv 2G M / c^{2}$.  Black holes could exist over a wide range of mass scales, although astrophysical processes could only produce them above a solar mass.  Those larger than several solar masses would form at the endpoint of evolution of ordinary stars and there should be billions of these even in the disc of our own Galaxy. ``Intermediate Mass Black Holes'' (IMBHs) would derive from stars bigger than $100 M_\odot$, which are radiation-dominated and collapse due to an instability during oxygen-burning, and the first stars may have been in this range. ``Supermassive Black Holes'' (SMBHs), with masses from $10^{6} M_\odot$ to $10^{10}M_\odot$, are thought to reside in galactic nuclei, with our own Galaxy harbouring one of mass $4 \times 10^{6} M_{\odot}$ and quasars being powered by ones with mass of around $10^{8} M_\odot$. There is now overwhelming evidence for these astrophysical types of black holes, but they can only provide a small fraction of the dark matter density. 

Black holes could also have formed in the early Universe and these are termed ``primordial''.  Comparing the cosmological density at a time $t$ after the Big Bang
and the density required for a region of mass $M$ to fall within its Schwarzschild radius,
implies that primordial black holes~(PBHs) would initially have around the cosmological horizon mass $M \sim c^{3} t/G$ at formation.  So they would have the Planck mass $( M_{\rm Pl} \sim 10^{-5} \rm g )$ if they formed at the Planck time ($10^{-43}$s), $1 M_\odot$ if they formed at the quantum chromodynamics (QCD) epoch ($10^{-5}$s) and $10^{5} M_\odot$ if they formed at $t \sim 1 $s.  Therefore PBHs could span an enormous mass range and are the only black holes which could be 
smaller than a solar mass. In particular, only PBHs could be
light enough for Hawking radiation to be important,  those lighter than the Earth being hotter than the cosmic microwave background (CMB) and those lighter than $10^{15}$g evaporating within the current age of the Universe.

The wide range of masses of black holes and their crucial r{\^o}le in linking macrophysics and microphysics is summarised in Figure \ref{fig:urob}. 
The edge of the orange circle can be regarded as a sort of ``clock'' in which the scale changes by a factor of $10$ for each minute, from the Planck scale at the top left to the scale of the observable Universe at the top right. 
The top itself
corresponds to the Big Bang because at the horizon distance one is peering back to an epoch when the Universe was very small, so the very large meets the very small there. 
The various types of black holes 
are labelled by their mass and positioned according to their Schwarzschild radius.
On the right are the astrophysical black holes, with the well-established stellar and supermassive ones corresponding to the segments between  $5$ and $50 \, M_\odot$ and between $10^6$ and $10^{10}M_\odot$, respectively.  On the left{\,---\,}and possibly extending somewhat to the right{\,---\,}are the more speculative PBHs. 

The vertical line between the bottom (planetary-mass black holes) and the top (Planck-mass black holes) provides a convenient division between the microphysical and macrophysical domains.
Quantum emission is suppressed by accretion of the CMB to the right of the bottom point, so this might be regarded as the transition between quantum and classical black holes.  The effects of extra dimensions could be important at the top, especially if they are compactified on a scale much larger than the Planck length.  In this context, there is a sense in which the whole Universe might be regarded as a PBH; this is because in brane cosmology (in which one extra dimension is extended) the Universe can be regarded as emerging from a five-dimensional black hole.

\begin{figure}[t]
	\centering
	\includegraphics[scale = 0.25]{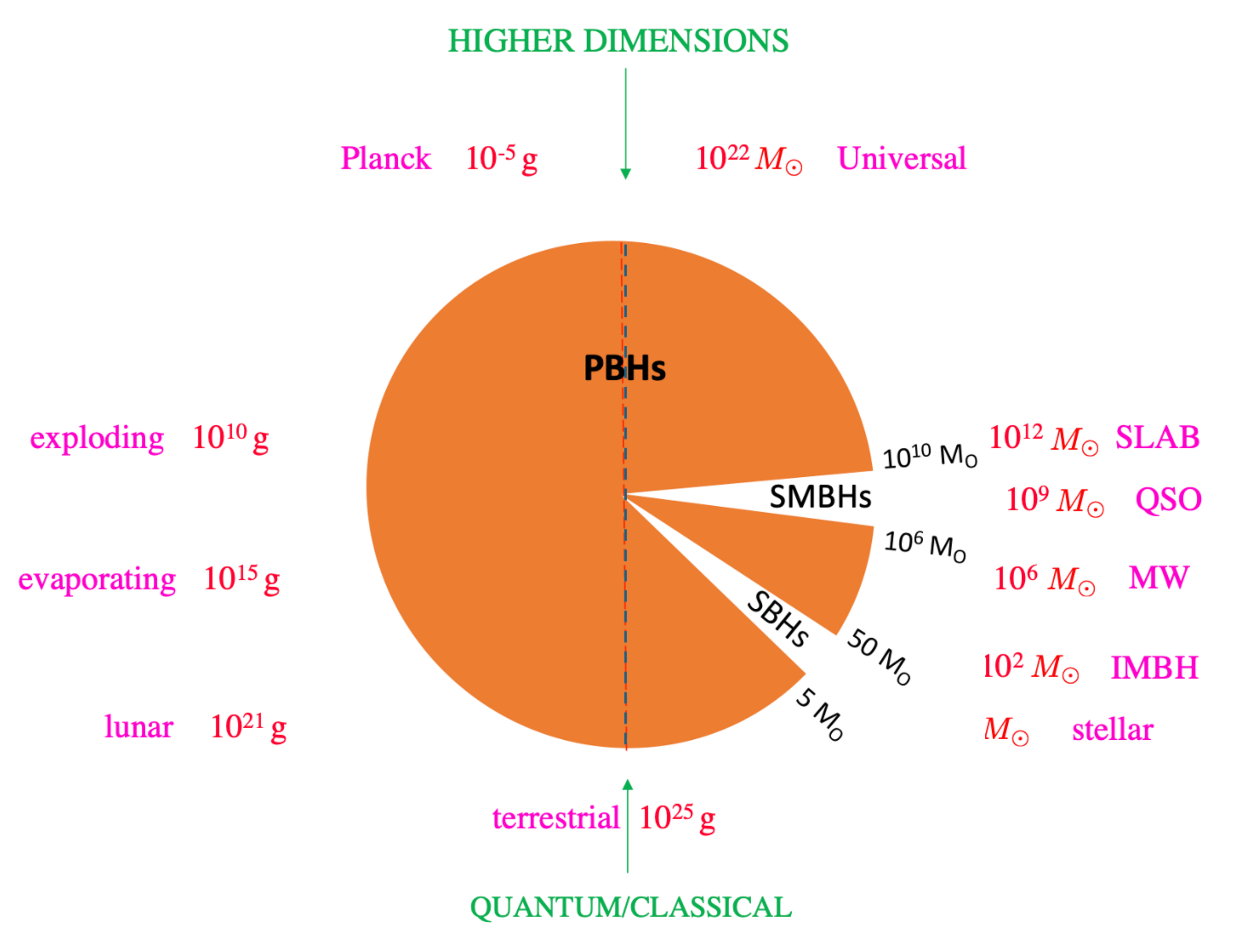}
	\caption{%
		This diagram indicates 
		the mass and size of various types of black holes, with
		the division between the micro and macro domains 
		being indicated by the vertical line. 
		QSO stands for ``Quasi-Stellar Object'', 
		MW for ``Milky Way'', 
		IMBH for ``Intermediate Mass Black Hole'', 
		SLAB for ``Stupendously Large Black Hole".
Stellar black holes (SBHs) 
			and supermassive black holes (SMBHs) occupy 
			only small slivers, 
			whereas PBHs occupy a much wider range (shown in orange). } 
\label{fig:urob}
\end{figure}

The study of PBHs provides a unique probe of four areas of physics:
	(1) the early Universe ($M < 10^{15}$g);
	(2) gravitational collapse ($M > 10^{15}$g);
	(3) high energy physics ($M \sim 10^{15}$g);
		and
	(4) quantum gravity ($M \sim 10^{-5}$g).  Although we still cannot be certain that PBHs formed in {\it any} of these mass ranges, these numerous interesting applications suggest that nature would be remiss if their existence were precluded.

From a historical perspective, it should be stressed that PBHs have attracted increasing attention in recent years. Following the founding papers in the 1970s, there were only a dozen or so publications per year for the next two decades, although Hawking radiation obviously attracted attention. The rate rose to around a hundred per year after the MACHO microlensing results in 1996 but the most dramatic rise occurred after the first LIGO-Virgo gravitational wave events in 2016. This is illustrated in Figure~\ref{fig:PBH-Citations}.

The structure of this chapter is primarily chronological, with successive sections covering the periods 1967-1980, 1980-1996, 1996-2016 and 2016-2024.  
The first period covered the initial groundbreaking work on PBH formation and evaporation. The second period focussed mainly on their formation, in particular from the collapse of large inflationary density perturbations, so we take this period to start with the inflationary proposal in 1980. The third period continued the studies of formation but also saw much work on PBH constraints, this being  intensified after the detection of
microlensing events by the MACHO project in 1996.  The fourth period was initiated by the LIGO/Virgo detection in 2016 of gravitational waves from the mergers of multi-solar mass black holes, some of which could be PBHs, but it also saw claims of positive evidence for PBHs from several other observations. The authors of this chapter have somewhat different opinions about the strength of this purported evidence 
but this merely reflects the different opinions of the PBH community as a whole. So this chapter may be more balanced than it might have been otherwise.

\begin{figure}[t]
	\centering
	\includegraphics[width = 0.4\linewidth]{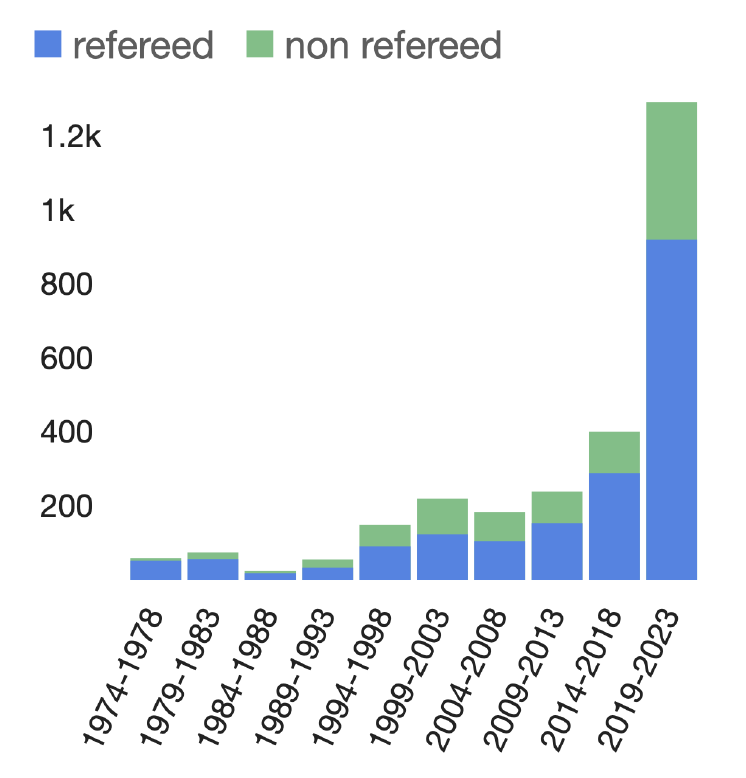}
	\caption{
Number of articles  in SAO/NASA Astrophysics Data System  with ``Primordial Black Hole'' 
in title in five-year bins.}
	\label{fig:PBH-Citations}
\end{figure}

\section{Early history (1967-1980)}
\label{sec:early}

 In this section, we describe the early history of PBHs.  The first paper on the topic was by Zeldovich and Novikov~\cite{1967SvA....10..602Z} in 1967 but they concluded (incorrectly) that PBHs would have accreted catastrophically and therefore could not have formed.  This did not deter Hawking~\cite{Hawking:1971ei} from proposing a specific scenario in 1971. This was also partly flawed but it ultimately led to his discovery of Hawking radiation in 1974~\cite{Hawking:1974rv}. This prompted more general interest in PBHs but it also implied observational constraints which suggested their existence was less likely.

\subsection{Formation and accretion}
\label{sec:earlyformationaccretion}

Since the cosmological density at a time $t$ after the Big Bang is $\rho \sim 1 / ( G t^{2} )$ and the density required for a region of mass $M$ to fall within its Schwarzschild radius is $\rho \sim c^{6} / ( G^{3} M^{2} )$,  PBHs would initially have a mass
\begin{equation}
	\label{eq:Moft}
	M
		\sim
					\frac{ c^{3}t }{ G }
		\sim
					10^{15}\mspace{-1mu}
					\left(
						\frac{ t }{ 10^{-23} \rm s }
					\right)
					\rm g
					\, .
\end{equation}
This is just the cosmological horizon mass at their formation epoch.
The first proposal for the existence of PBHs was in a paper by Hawking just over $50$ years ago~\cite{Hawking:1971ei}. He argued that PBHs of the Planck mass would be electrically charged and thereby capture electrons or protons to form ``atoms''. These could then leave tracks in bubble chambers and collections of them might accumulate in the centres of stars. This might explain the low flux of neutrinos coming from the Sun (which was then unexplained). Later it was realised that such small black holes would lose their charge through quantum effects.

In fact, the first discussion of PBHs, including Eq.~\eqref{eq:Moft} for the mass, was in a paper by Zeldovich and Novikov~\cite{1967SvA....10..602Z} several years before Hawking's paper. However, they concluded that the existence of PBHs was unlikely on the basis of a Bondi accretion analysis. This suggested that the PBH mass would increase according to 
\begin{equation}
	\label{eq:ZN}
	M
		=
					\frac{ \eta c^{3} t / G }
					{ 1 + ( t / t_{\rm f} )( \eta c^{3} t_{\rm f} / G M_{\rm f} - 1) }
		\approx
					\begin{cases}
						M_{\rm f}
						\quad
						( M_{\rm f} \ll \eta c^{3} t_{\rm f} / G )
						\\[0.2mm]
						\eta c^3 t/G
						\quad
						( M_{\rm f} \sim \eta c^{3} t_{\rm f} / G )
						\, ,
					\end{cases}
\end{equation}
where $M_{\rm f}$ is the mass at its formation time $t_{\rm f}$ and $\eta$ is a constant of order unity. Thus PBHs with initial size comparable to the horizon (as expected) should grow as fast as the horizon and reach a mass of $10^{17} M_\odot$ by the end of the radiation-dominated era. Since the existence of such huge black holes is precluded, this might suggest that PBHs never formed. However, this argument neglects the cosmic expansion, which is important for PBHs with the horizon size and would inhibit accretion, and 
in 1974 Carr and Hawking showed that there is no self-similar solution in general relativity in which a back hole formed from local collapse can grow as fast as the horizon~\cite{Carr:1974nx}. Furthermore, the black hole would soon become much smaller than the horizon, at which point Eq.~\eqref{eq:ZN} should apply, so one would not expect much growth at all. This removed the concerns raised by Zeldovich-Novikov and reinvigorated PBH research.

At this time 
more detailed studies of PBH formation were initiated.  Only overdensities larger than the Jeans length at maximum expansion can collapse against the pressure and this is around $\sqrt{w}$ times the horizon size for $p = w \rho$.   If PBHs formed from Gaussian density perturbations of root-mean-square amplitude $\epsilon(M)$ at the horizon epoch, then the fraction of the Universe collapsing into PBHs of mass $M$ should be~\cite{Carr:1975qj}  
\begin{equation}
\beta (M) \sim \epsilon (M) \exp \left[-\frac{w^2}{2 \epsilon(M)^2} \right] \, .
\label{beta}
\end{equation}
At that time it was expected that $\epsilon(M)$ should be scale-invariant, corresponding to a Harrison-Zeldovich spectrum~\cite{1970PhRvD...1.2726H,1972MNRAS.160P...1Z}.   In this case, $\beta$ is also scale-invariant and the PBH mass function should have the form~\cite{Carr:1975qj}
\begin{equation}
dN/dM \propto  M^{-\left( \frac{1+3w}{1+w}\right) -1} \, ,
\end{equation}
which just falls as $M^{-5/2}$ for a radiation-dominated universe ($w=1/3$), as expected.  However, this does not apply for the fluctuations expected in the inflationary scenario since
(as discussed later) these are not exactly scale-invariant.
The first numerical studies of PBH formation were carried out by Nadezhin et al.  \cite{1978SvA....22..129N}, modelling overdense regions as $k=+1$ FRW models matched to a $k=0$ background by a vacuum region.  These roughly confirmed the analytic prediction but included the effect of pressure gradients,  resulting in PBHs somewhat smaller than the horizon. 

On cosmological scales the amplitude of the fluctuations at horizon crossing is only around $10^{-5}$, so the exponential dependence in Eq.~\eqref{beta} implies that the collapse fraction should be tiny if $\epsilon$ is scale-independent.
Observations also require this since,  if
the current density parameter
of PBHs which form at redshift $z$ is $\Omega_{\rm PBH}$, in units of the critical density,  then the initial collapse fraction is~\cite{Carr:1975qj}
\begin{equation}
	\beta 
		=
					\frac{ \Omega_{\rm PBH}}{
					\Omega_{\rm R}}
					( 1 + z )^{-1}
		\approx
					10^{-6}
					\Omega_{\rm PBH}
					\left(
						\frac{ t }{ \rm s }
					\right)^{1/2}
		\approx
					10^{-18}
					\Omega_{\rm PBH}
					\left(
						\frac{ M }{ 10^{15}\rm g }
					\right)^{1/2}
					,
\end{equation}
where $\Omega_{\rm R} \approx 10^{-4}$ is the current density parameter of radiation and we have used Eq.~\eqref{eq:Moft} at the last step.  The $( 1 + z )$ factor arises because the radiation density scales as $( 1 + z )^{4}$, whereas the PBH density scales as $( 1 + z )^{3}$. 
So $\beta$ must be tiny even if PBHs provide all of the dark matter.  
This is a potential criticism of the PBH dark matter proposal, since it requires fine-tuning of the collapse fraction and even greater fine-tuning of the density fluctuations. There is also the puzzling feature that the PBH and baryon densities are so close if PBHs provide the dark matter.  However, as discussed later, there is one scenario in which this arises naturally.

\subsection{Evaporation and constraints}

The realisation that PBHs might be small prompted Hawking to study their quantum properties. This led to his famous discovery~\cite{Hawking:1974rv} that black holes radiate thermally with a temperature
\begin{equation}
	T_{\rm BH}
		=
					\frac{ \hbar c^{3} }{ 8\pi G M k_{\rm B} }
		\approx
					10^{-7}
					\left(
						\frac{ M }{M_\odot}
					\right)^{-1}
					{\rm K}
					\, 
\end{equation}
and ones which are sufficiently massive that they emit only massless particles evaporate on a timescale 
\begin{equation}
 \tau( M )
 	\approx
					8000 \, \frac{ G^{2} M^{3} } { \hbar c^{4} }
		\approx
					10^{67}
					\left(
						\frac{ M }{M_\odot}
					\right)^{3}
					{\rm yr}
					\, .
\end{equation}
The numerical factor in the second expression would be $5120 \pi \approx 16000$ if the radiation were exactly black-body but this does not apply because the spectrum also depends upon the spin of the emitted particles.  The numerical factor is reduced for PBHs light enough to emit massive particles.  Those with initial mass $M_{*} \sim 10^{15}$g, which formed at $10^{-23}$s and had the size of a proton, would be evaporating now and lighter ones would  have evaporated at an earlier epoch.  However, 
evaporation would be suppressed for PBHs heavier than the Earth, $10^{24}$g, since they would be cooler than the cosmic microwave background (CMB) and so would accrete rather than evaporate.

Hawking's discovery has not yet been confirmed experimentally and there remain major conceptual puzzles associated with the process. Nevertheless, it is generally recognised as one of the key developments in 20th-century physics because it beautifully unifies general relativity, quantum mechanics and thermodynamics. The fact that Hawking was only led to this discovery through contemplating the properties of PBHs illustrates that it has been useful to study them even if they do not exist. However, at first sight it was bad news for PBH enthusiasts. Since PBHs with a mass of $10^{15}$g would be producing photons with energy of order $100$MeV at the present epoch, the observational limit on the $\gamma$-ray background intensity at $100$MeV immediately implied that their density could not exceed $10^{-8}$ times the critical density~\cite{Page:1976wx}. This implied that there was little chance of detecting black hole explosions at the present epoch, which would have confirmed the existence of both PBHs and Hawking radiation. 

Nevertheless, the evaporation of PBHs smaller than $10^{15}$g could still have many interesting cosmological consequences~\cite{1976ApJ...206....8C}, each associated with the different types of particle emitted, and it also implied interesting constraints on the collapse fraction $ \beta(M)$.
These constraints were first brought together by Novikov \textit{et al.} \cite{1979A&A....80..104N} in 1979. The strongest one is the $\gamma$-ray limit associated with the $10^{15}$g PBHs evaporating at the present epoch. Others are associated with the generation of entropy and modifications to the cosmological production of light elements,
PBHs with $ M \sim 10^{10}\,\mathrm g $
 having a lifetime $ \tau \sim 10^3\,\mathrm s $ and therefore evaporating at the big bang nucleosynthesis (BBN) epoch.
Injection of high-energy neutrinos and antineutrinos \cite{1978AZh....55..231V} changes the weak interaction freeze-out time and hence the neutron-to-proton ratio at the onset of BBN, which changes $ {}^4\mathrm{He} $ production.
PBHs with $ M = 10^{10}\text{--}10^{13}\,\mathrm g $ evaporated after BBN,  increasing the baryon-to-entropy ratio at nucleosynthesis and resulting in overproduction of $ {}^4\mathrm{He} $ and underproduction of $ \mathrm D $ \cite{Miyama:1978mp}.
Emission of high-energy nucleons and antinucleons \cite{1977PAZh....3..208Z} increases the primordial deuterium abundance due to the capture of free neutrons by protons and spallation of $ {}^4\mathrm{He} $.
The emission of photons by PBHs with $ M > 10^{10}\,\mathrm g $ dissociates the deuterons produced in nucleosynthesis \cite{1980MNRAS.193..593L}.
The limits associated with these effects are shown in the left panel of Fig.~\ref{fig:bc1}.
The equivalent figure 20 years later is shown on the right and is not very different.

PBHs were also invoked to explain certain observations.  For example, evaporating PBHs of around $10^{15}$g might explain the 511 keV annihilation line radiation from the Galactic centre \cite{1980A&A....81..263O} or antiprotons in cosmic rays~\cite{1976ApJ...206....8C}.  PBHs more massive than $10^{15}$g might provide the dark matter since these are unaffected by Hawking radiation, with Chapline~\cite{1975Natur.253..251C} suggesting this in 1975.   In this case,  as pointed out by
M{\'e}sz{\'a}ros~\cite{Meszaros:1975ef} in the same year,
sufficiently large ones could generate cosmic structures through the $\sqrt{N}$ Poisson effect.

\begin{figure}[t]
	\centering
	\includegraphics[scale=0.25]{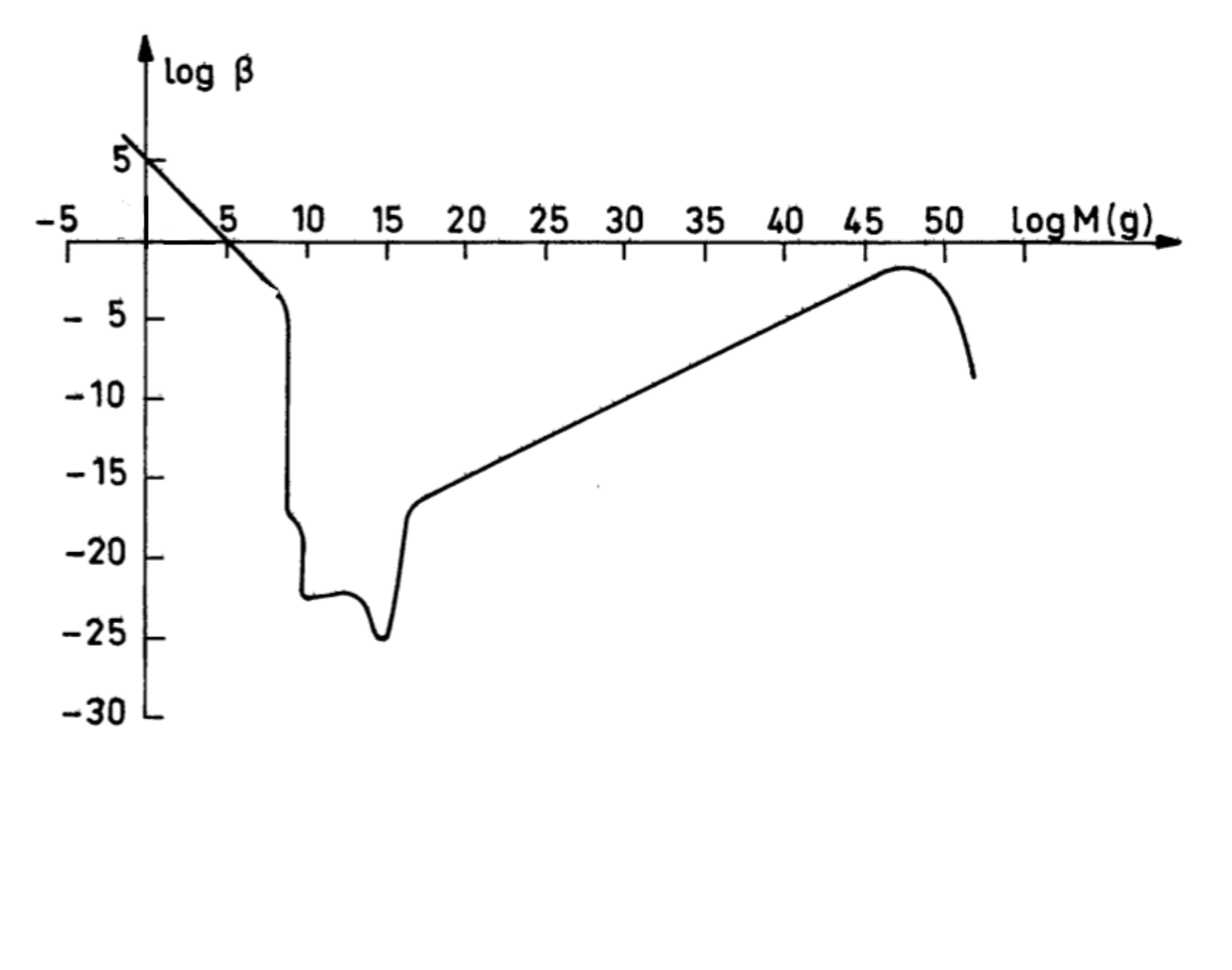}
\includegraphics[scale=0.5]{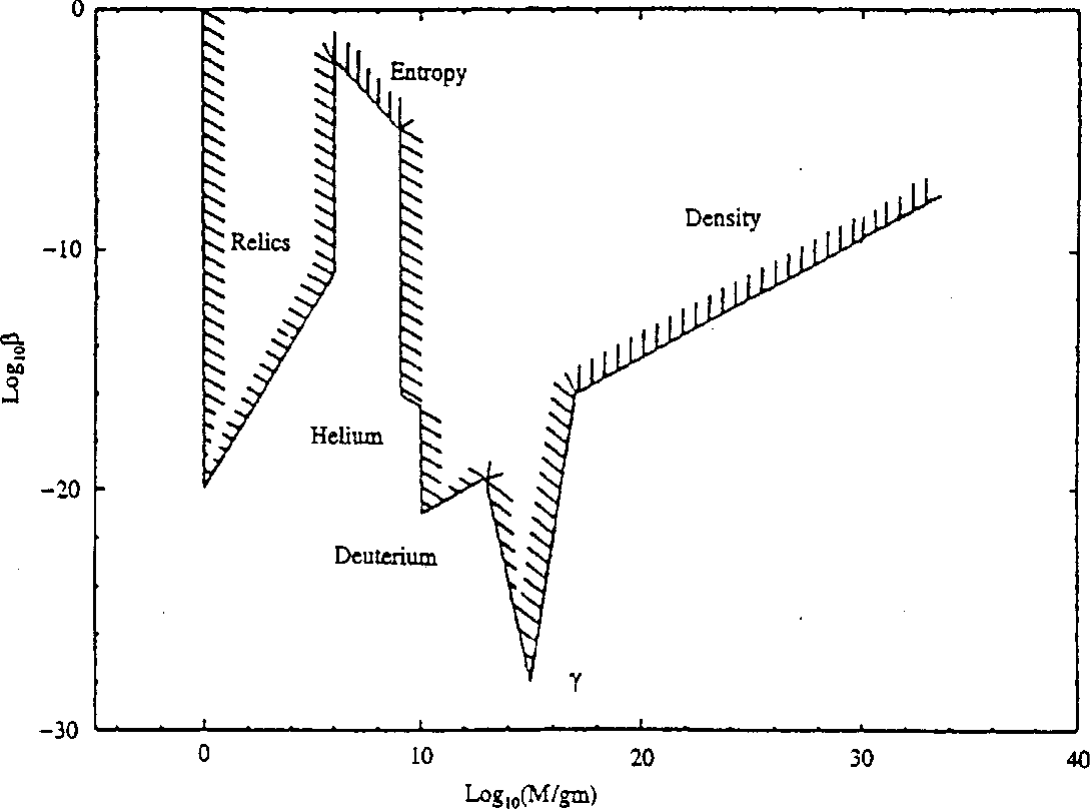}
	\caption{%
		Evaporation constraints on $\beta( M )$, 
		the fraction of Universe collapsing into PBHs of mass $M$, 
		from Ref.~\cite{1979A&A....80..104N} in 1979 (left) and Ref.~\cite{Carr:1994ar} in 1994 (right).
}
	\label{fig:bc1}
\end{figure}

\section{
Studies of PBH formation (1980-1996)}
\label{sec:80s}

This section overviews developments in the 1980s and early 1990s, in particular studies of  PBH formation. 
This period includes the development of inflationary theory and, since inflation generates density fluctuations, there was interest in whether these could produce PBHs.  Indeed, constraints can be imposed on inflationary models by the requirement that they do not overproduce them.  
Other PBH formation mechanisms were also proposed, and improved calculations of the evaporation of PBHs led to refinements in abundance constraints.

\subsection{Formation from inflationary density perturbations}
\label{subsec:inflation}

Inflation, a period of accelerated expansion in the early Universe, was proposed in the early 1980s~\cite{Starobinsky:1980te,Guth:1980zm,Sato:1980yn,Albrecht:1982wi}.  
It was quickly realised that quantum fluctuations in the scalar field driving inflation would lead to density perturbations from which structures can form~\cite{Guth:1982ec,Hawking:1982cz,Linde:1982uu,Starobinsky:1982ee}, so there was naturally interest in whether these fluctuations could be  large enough to generate an interesting abundance of PBHs. Also, since inflation exponentially reduces the number density of any PBHs formed earlier, the horizon mass at the end of inflation sets a lower limit on the mass of PBHs which are subsequently observable. 
The tensor contribution to the CMB temperature anisotropies on large angular scales limits the reheat temperature to be less than $10^{16}$GeV, which corresponds to a horizon mass of $\sim 1 \, {\rm g}$ \cite{Carr:1993aq}. 

The amplitude of the density perturbations on CMB scales is of order $\sim 10^{-5}$~\cite{COBE:1992syq,Wright:1992tf}.
From Eq.~(\ref{beta}) this amplitude is too small to generate even one PBH per current horizon volume. 
To produce a non-negligible abundance, the amplitude
must be far larger on some small scale.  In single-field slow-roll inflation models the amplitude of the density perturbation is $\epsilon \propto V^{3/2} / V^{\prime}$ where $V$ is the potential and ${V}^{\prime}$ is its slope.  One requires either a blue spectrum, where
 the primordial perturbations increase monotonically on smaller scales, or a spike on some scale.  Both of these are possible with single-field inflation but one needs to go beyond the slow-roll approximation.   
In 1993 Carr and Lidsey \cite{Carr:1993aq}
studied PBH formation in models with blue spectra.
In this case, PBHs could form abundantly and avoiding their overproduction placed a tighter constraint on the spectral index of the primordial perturbations than CMB observations did at that time~\cite{Carr:1994ar}. In 1994 Ivanov {\it et al.} pointed out that a plateau in the potential, with $V^{\prime} \rightarrow 0$, could generate large PBH-forming perturbations~\cite{Ivanov:1994pa}.  
However, more complicated models were also proposed,  with Garc{\'\i}a-Bellido {\it et al.}~\cite{Garcia-Bellido:1996mdl} invoking hybrid inflation; this has two fields, with one having large quantum fluctuations as it undergoes a phase transition which ends inflation. 
It was pointed out by Dolgov and Silk \cite{PhysRevD.47.4244} that PBHs with a lognormal mass function could be formed from baryon isocurvature fluctuations generated by various inflation models. 

\subsection{
PBH formation during matter-domination}

It is usually assumed that the Universe is radiation-dominated from matter-radiation equality ($t_{\rm eq} = 1.7 \times 10^{12} \, {\rm s}$) back to very early times.
There could, however, be an early 
period of matter domination before BBN due, for instance, to long-lived non-relativistic particles dominating the Universe and then decaying.
In the 1980s Khlopov and Polnarev~\cite{Khlopov:1980mg,1981AZh....58..706P} studied PBH formation during such a period of early matter domination.  A matter-like phase might also arise at the end of inflation if reheating is slow~\cite{Carr:1994ar}.  In these cases, the criteria for PBH formation from the collapse of density perturbations are somewhat different from those during radiation-domination (as covered in Sec. ~\ref{sec:earlyformationaccretion}). 
A perturbation must be close to spherically symmetric to form a PBH, rather than collapsing to form a pancake-like or cigar-like configuration~\cite{Khlopov:1980mg}. It must also collapse within its Schwarzschild radius before a caustic can form at its centre~\cite{1981AZh....58..706P}.  

\subsection{PBH formation from phase transitions}
\label{subsec:bubblecollisions}

In the early 1980s, various authors showed that PBHs, with mass of order the horizon mass, could form from the collisions of bubbles formed during a first-order phase transition~\cite{Crawford:1982yz,Hawking:1982ga,Kodama:1982sf}. However, to form a non-negligible PBH abundance the bubble nucleation rate had to be fine-tuned, so that bubble collisions occur but without the phase transition completing too quickly. 

Phase transitions in the early Universe can also lead to the formation of a network of one-dimensional topological defects, known as cosmic strings~\cite{Kibble:1976sj}. These include oscillating loops that are smaller than the Hubble radius 
and in the late 1980s it was shown a PBH can form from the collapse of a loop 
if it becomes smaller than its Schwarzschild radius~\cite{Hawking:1987bn,Polnarev:1988dh}. The PBH mass is proportional to the loop mass, which is proportional to the horizon mass, and the rate at which loops collapse to form PBHs is independent of time. Therefore, as 
for PBHs formed from scale-invariant density perturbations, the number density of PBHs formed is ${\rm d} n/ {\rm d} M \propto M^{-5/2}$~\cite{MacGibbon:1997pu}.  
As indicated by numerical simulations in the early 1990s,  the fraction of 
loops collapsing to PBHs  is very sensitive to the string tension $(G \mu)$~\cite{PhysRevD.40.973,PhysRevLett.64.119,Caldwell:1991jj}
and there were already strong constraints on this parameter from other observations.

\subsection{Observational probes}

In the early 1990s MacGibbon and Webber~\cite{MacGibbon:1990zk,MacGibbon:1991tj} studied the evaporation of PBHs in more detail.
Since black holes emit all elementary particles with rest mass less than the black hole temperature, $T_{\rm BH}$,  they argued that   PBHs with mass $M \lesssim 2 \times 10^{14} \, {\rm g}$, which are hotter than than the QCD confinement scale,  should emit quark-gluon jets which then fragment. The total emission is thus the combination of the primary and secondary emission.  MacGibbon and Carr~\cite{MacGibbon:1991vc} studied the observational consequences of this, in particular their possible contribution to cosmic rays and the extragalactic gamma-ray background. 
They placed constraints on the contribution from PBHs, and hence their abundance,  but they did not claim positive evidence for them.

In 1985 Lacey and Ostriker argued that the observed puffing of the Galactic disc could be due to halo black holes with mass around $10^{6}\,M_\odot$~\cite{1985ApJ...299..633L}, older stars being heated more than younger ones.  Although dynamical constraints now exclude such large PBHs providing the dark matter, clusters of smaller PBHs could also explain disc heating \cite{1987ApJ...316...23C}. However, heating by a combination of spiral density waves and giant molecular clouds may also explain the data~\cite{1991dodg.conf..257L}. 
The suggestion that sufficiently massive PBHs could generate cosmic structures through the `seed' or `Poisson' effect, 
as first pointed out by Meszaros~\cite{Meszaros:1975ef}, 
was explored in detail in several papers during the 1980s \cite{1983ApJ...268....1C,1983ApJ...275..405F,1984MNRAS.206..801C}.  
 Bond and Carr ~\cite{1984MNRAS.207..585B} also considered the possibility that LIGO could detect the bursts and gravitational wave background generated by binary black holes, although they did not require them to be primordial. 
In the early 1990s Hawkins~\cite{Hawkins:1993yud} argued that observed variations in the brightness of quasars indicate 
a large abundance of compact objects. 

\section{Studies of PBH formation and constraints (1996-2016)}
\label{subsec:developments}

Until the mid-1990s there was no evidence for PBHs and no compelling reason to 
associate them with the dark matter problem.  However, there was a flurry of excitement in 1996, when microlensing searches for massive compact halo objects (MACHOs) suggested that the dark matter could be black holes of mass $0.5M_\odot$~\cite{MACHO:1996qam}. Alternative microlensing candidates could be excluded and it was realised that PBHs of this mass might naturally form at the quark-hadron phase transition~\cite{Jedamzik:1998hc}.  With further data, the MACHO collaboration found that solar-mass compact objects could comprise only $20\%$ of the dark matter~\cite{Alcock:2000ph} and other microlensing observations during this period subsequently excluded compact objects in the mass range $10^{-7}M_\odot$ to $10M_\odot$ from providing all of it~\cite{Tisserand:2006zx}. 
 Putative evidence for PBHs thus became a constraint on their abundance.  Indeed,  much of the focus of research in this period was on the collation of constraints  over a huge range of mass scales from $10^{-5}$g to $10^{12}M_{\odot}$.  These constraints were associated with numerous astrophysical and cosmological effects -- related to evaporation,  microlensing, dynamics, accretion and gravitational waves -- and  expressed as upper limits on $f_{\rm PBH}(M)$, the fraction of the dark matter in PBHs with mass  $M$. They are summarised briefly below and discussed in more detail in Part V.  See Refs.~\cite{Carr:2009jm} and \cite{Carr:2016drx} for the status of constraints in 2010 and 2016, respectively.  However,  much of the literature in this period 
 continued the study of PBH formation, so we will start by reviewing this.

\subsection{Further studies of formation from inflation}

It was pointed out in the late 1990s that since PBHs form from rare large perturbations, their abundance 
 depends sensitively on the shape of the large-amplitude tail of the probability distribution of the perturbations~\cite{Bullock:1996at,Ivanov:1997ia}. For the flat inflaton potential required to generate large perturbations in single-field models (see Sec.~\ref{subsec:inflation}), 
 it was  also argued that quantum fluctuations play a significant role in the dynamics of the field and can generate a non-Gaussian probability distribution for the fluctuations~\cite{Young:2015kda}. 

Various inflation models that can produce large, PBH-forming perturbations were explored, such as
hilltop inflation (where the field evolves away from a local maximum)~\cite{Kohri:2007qn}
and double inflation (with two periods of inflation)~\cite{Kawasaki:1997ju}.
It was also realised that the reheating era, at the end of inflation, could produce large perturbations~\cite{Green:2000he,Bassett:2000ha}.  Several authors pointed out that the amplitude of large density perturbations could be indirectly constrained via limits on the gravitational waves they induce ~\cite{Ananda:2006af,Baumann:2007zm,Saito:2008jc} (see Chapter 18). 
In the mid-2000s, Green {\it et al.}
\cite{Green:2004wb} calculated the abundance of PBHs using the classic Bardeen, Bond, Kaiser and Szalay (BBKS)~\cite{Bardeen:1985tr} peaks theory rather than the traditional Press-Schechter approach which leads to Eq.~(\ref{beta}). This is discussed in more detail in Chapter 7 and is not only relevant to inflation.

\subsection{Formation from other processes}

It was realised in the 1990s that due to critical phenomena~\cite{Choptuik:1992jv}, the mass of a black hole depends on the amplitude of the perturbation from which it forms: $M \propto M_{\rm H} (\delta -\delta_{\rm c})^{\gamma}$, where the scaling exponent $\gamma$ is constant for a given equation of state. Niemeyer and Jedamzik~\cite{Niemeyer:1997mt} pointed out that this applies to PBH formation and calculated the resulting PBH mass function (assuming all PBHs form at the same time). They also carried out simulations of PBH formation 
for different shaped perturbations~\cite{Niemeyer:1999ak}. Subsequent work by Musco and his collaborators~\cite{Musco:2004ak,Musco:2008hv,Musco:2012au} in this period explored critical collapse further, in particular verifying the mass scaling for small ($\delta-\delta_{\rm c}$) and studying a range of equation of state parameters ($0 < w < 0.6$). See Chapter 5 for further details.

Jedamzik pointed out that the reduction in pressure at the QCD phase transition leads to enhanced PBH formation  at this time, with the horizon mass (and hence PBH mass) being around $1 \, M_\odot$~\cite{Jedamzik:1996mr}. There were also improvements in specifying the criteria for PBH formation, with Ref.~\cite{Nakama:2013ica} showing that the threshold depends on the shape of the perturbation. Numerical studies of PBH formation during matter-domination found that the initial mass fraction of PBHs in this case is given by $\beta \approx 0.21 \epsilon^{13/2}$ for $\epsilon \ll 1$~\cite{Harada:2016mhb}. Since angular momentum is significant for PBHs formed 
during matter-domination, they have larger spins
than those formed during radiation-domination~\cite{Harada:2017fjm}.

\subsection{
Evaporation constraints}

In a series of papers starting in 1997 Cline and colleagues argued that exploding PBHs could explain some short-duration gamma-ray bursts \cite{Cline:1996zg}.  Unlike cosmological gamma-ray bursts, these would be located within the Galactic halo and therefore anisotropically distributed.  If true,  this would require some new effect when the black hole temperature reaches the QCD scale,  possibly associated with the formation of a photosphere due to electron-positron or QCD interactions~\cite{Heckler:1995qq}.  However, in 2008 MacGibbon, Carr and Page~\cite{MacGibbon:2007yq} showed that
evaporating black holes do not form photospheres.
They also found that a more precise value for the initial mass of a PBH evaporating today is $M_* \approx 5 \times 10^{14} \, {\rm g}$. 
The constraints on $\beta(M)$ for PBHs with initial mass $M \lesssim 10^{17} \, {\rm g}$ were extended and improved
 by numerous studies of the effects of PBH evaporations on the early Universe.  These constraints were summarised in Ref.~\cite{Carr:2009jm}, which also updated the  cosmological nucleosynthesis bound,  and are shown in Fig.~\ref{fig:bc2}.  

\begin{figure}[t]
	\centering
\includegraphics[scale=0.9]{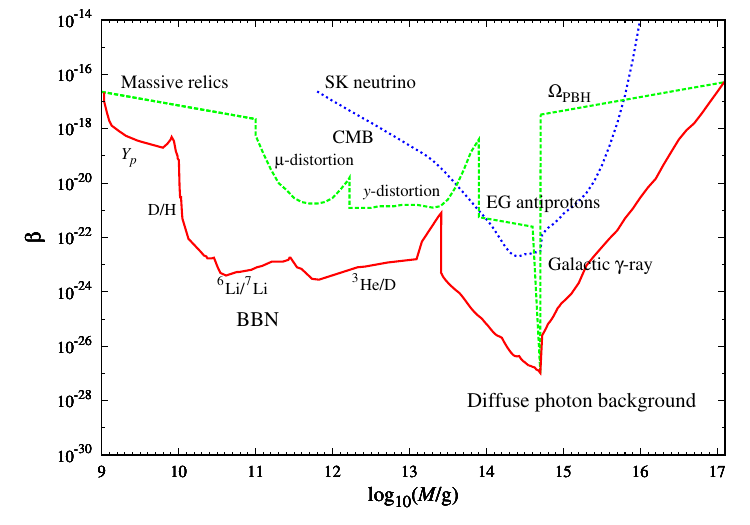}
	\caption{%
		Evaporation constraints on $\beta( M )$, 
		the fraction of Universe collapsing into PBHs of mass $M$, 
		from Ref.~\cite{Carr:2009jm} in 2010. }
	\label{fig:bc2}
\end{figure}

 If PBHs of mass $M_*$
are clustered inside the Galactic halo, as expected, then their quantum evaporation should generate a Galactic $ \gamma $-ray background. Since this would be anisotropic, it should be separable from the extragalactic $ \gamma $-ray background. 
While the dominant contribution to the latter comes from the time-integrated emission of PBHs with initial mass $M_*$, the Galactic background is dominated by the instantaneous emission of those with  initial mass slightly larger than $M_*$ and current mass below $M_*$. 
In 1996 Wright \cite{1996ApJ...459..487W} claimed that a Galactic background had been detected in EGRET observations between $ 30\,\mathrm{MeV} $ and $ 120\,\mathrm{MeV} $ and attributed this to PBHs.
A later analysis of EGRET data in 2009,
assuming a variety of PBH distributions, 
was given by Lehoucq \textit{et al.} \cite{Lehoucq:2009ge}.
In 2016 Ref.~\cite{Carr:2016hva} reassessed this limit by including a realistic model for the PBH mass spectrum and a more precise relationship between the initial and current PBH mass. Specifically, the PBHs which generate the Galactic background  have an initial mass $M_*(1+ \mu)$  and a current mass $(3 \mu)^{1/3} M_* $ with $\mu \ll 1$.

\subsection{Lensing and gravitational-wave constraints}
\label{subsec:lensing}

Stellar microlensing, the temporary achromatic amplification of a star which occurs when a compact object crosses the line of sight to a star~\cite{Paczynski:1985jf}, is a long-standing probe of the abundance of planetary and solar-mass compact objects.
In the 1990s the EROS, MACHO and OGLE projects started 
microlensing surveys of the Milky Way.  In particular, they monitored millions of stars in the Magellanic Clouds to probe compact objects in our
halo. In their 2-year results~\cite{MACHO:1996qam} the MACHO collaboration found significantly more events than expected from known stellar populations. Their number 
and durations were consistent with roughly half of our
 halo being in compact objects with mass of around
$0.5 M_{\odot}$.  

Baryon budget arguments excluded such a large population of astrophysical objects (e.g.~white dwarfs)~\cite{Fields:1999ar} and PBHs were a plausible explanation for these events.  However, in the MACHO project's subsequent (5.7 year) results, the best-fit halo fraction dropped by a factor of roughly two,
with the absence of long-duration events excluding compact objects with mass $(1-30) M_{\odot}$ from comprising all of the dark matter~\cite{MACHO:2000qbb}.  Subsequent results from EROS and OGLE,  under the assumption that the halo is an isothermal sphere with a flat rotation curve,  excluded planetary and stellar mass compact objects making up more than about $10 \%$ of the
halo~\cite{EROS-2:2006ryy,Wyrzykowski:2011tr}.

Nakamura et al.~\cite{Nakamura:1997sm,Ioka:1998nz} pointed out that if PBHs make up a significant fraction of the dark matter, then PBH binaries could form at early times. Pairs of PBHs that happen to be close together would decouple from the expansion of the Universe before matter-radiation equality, with the torque from nearby PBHs leading to the formation of binaries with high eccentricity. The coalescence times of these binaries could be comparable to the age of the Universe, so
gravitational waves from their mergers would be detectable by interferometers such as LIGO providing the binaries survive within halos to late times.

\subsection{Dynamical and accretion constraints}
\label{subsec:dynamical}

Carr and Sakellariadou~\cite{Carr:1997cn} provided a comprehensive collection of  dynamical constraints on dark compact objects. Encounters of compact objects with mass $M \gtrsim 10 M_{\odot}$ with wide binary stars will increase the energy of those binaries, increasing their separation and disrupting the widest ones~\cite{1985ApJ...290...15B}. Limits on the fraction of the MW halo in PBHs can therefore be obtained by comparing observed and simulated wide binary separation distributions~\cite{Yoo:2003fr,2014ApJ...790..159M,2009MNRAS.396L..11Q}. Reliable constraints require a large sample of confirmed halo binaries and also accurate modelling of their initial properties, 
see e.g.~Ref.~\cite{2009MNRAS.396L..11Q}.  
Afshordi et al.~\cite{Afshordi:2003zb} showed that the Poisson fluctuations in the PBH distribution would lead to the formation of PBH clusters
shortly after matter-radiation equality and 
the enhanced clustering on subgalactic scales
would have important implications for observations of Lyman-alpha clouds. 
It was also suggested that asteroid-mass PBHs ($10^{17} \, {\rm g} \lesssim M \lesssim 10^{22} \, {\rm g}$) could be probed via the consequences of their  interactions with stars~\cite{Capela:2013yf,Pani:2014rca,Graham:2015apa}. 

Gas accretion onto PBHs with $M \gtrsim 10 M_{\odot}$ has potentially observable consequences. The resulting emission can modify the recombination history of the Universe~\cite{1981MNRAS.194..639C} and Ricotti and collaborators calculated the constraints from the anisotropies and spectral features 
of the CMB~\cite{Ricotti:2007au}.  For more details see Ref.~\cite{Montero-Camacho:2019jte} and Chapter 22.
Mack et al.~\cite{Mack:2006gz} showed that if dark matter is composed of a mixture of PBHs and particles,
then PBHs will accrete halos of particle dark matter which have a steep density profile. Lacki and Beacom~\cite{Lacki:2010zf} pointed out that PBHs and WIMPs cannot 
both make up a significant fraction of the dark matter
 as WIMP annihilation in these halos would produce a higher than observed flux of $\gamma-$rays.

\section{
 Improved calculations and search for evidence (2016-2024)}
\label{sec:evidence}

The detection of gravitational waves from the mergers of black holes with masses of $10 - 100 \, M_{\odot}$
by LIGO/Virgo has generated a huge wave of interest in PBHs as the origin of some of these events and as dark matter.  Indeed,  perhaps the most significant change of emphasis in this fourth period has been the search for {\it evidence} for PBHs.  Partly prompted by this development,  there have also been significant improvements in calculations of the abundance and mass function of PBHs formed from
large density perturbations, and other formation mechanisms have been advocated.  New probes of the PBH abundance 
have also been proposed and improvements made to 
existing constraints and signatures (see Chapters 20 and 21).  The status of the constraints in 2021 was reviewed in Refs.~\cite{Carr:2020gox,Green:2020jor} and a recent form of the $f_{\rm PBH}(M)$ diagram is shown in Fig.~\ref{fig:constraints}.  The shift towards the search for positive evidence for PBHs was signalled by the 2018 paper of Clesse and Garc{\'\i}a-Bellido~\cite{Clesse:2017bsw}. This was followed by their 2021 paper with Carr and Kuhnel, in which much of the claimed evidence was shown to be compatible with a `natural' thermal history model in which the sound speed dips at various times and most strikingly at the QCD transition at $10^{-5}$s~\cite{Carr:2019kxo}.  More recently the claimed evidence has been reviewed in Ref.~\cite{Carr:2023tpt}, which promotes what has been termed a `positivist' approach. Some of the arguments are mentioned below but they are discussed  in more detail in Chapter  21.

\begin{figure}[t]
	\centering
	\includegraphics[width=0.65\textwidth]{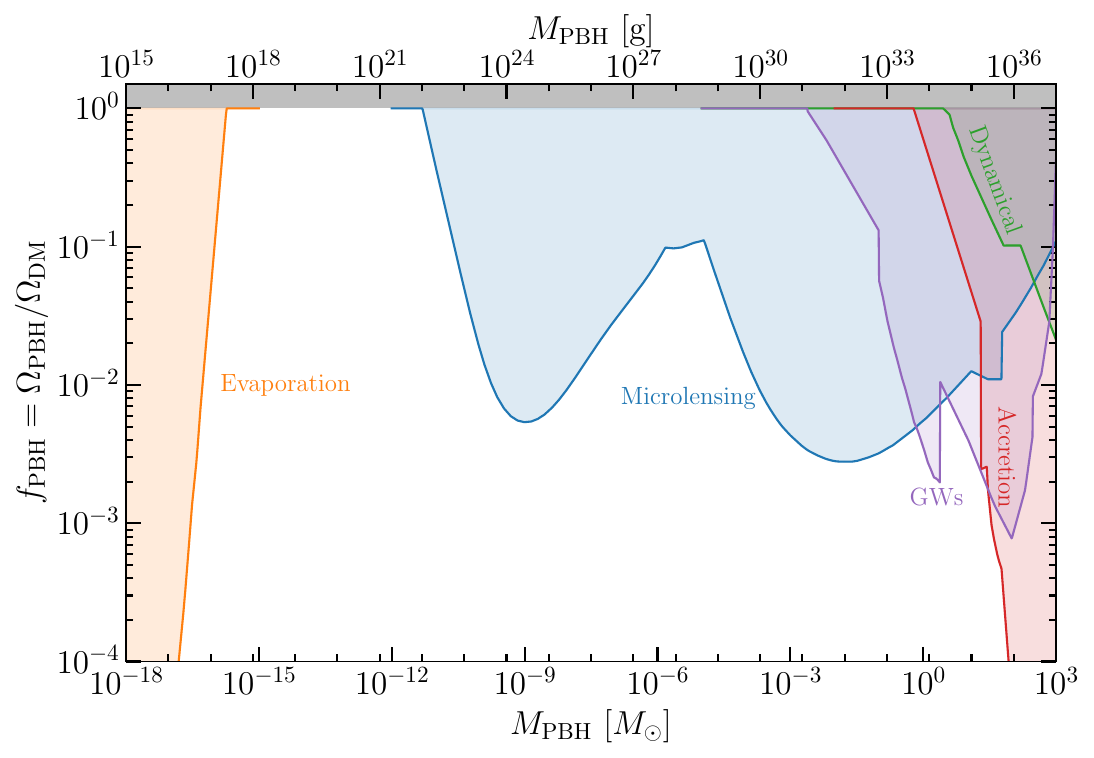}
	\caption{Constraints on the fraction of DM in the form of PBHs, $f_{\rm PBHs}$, as a function of mass, $M_{\rm PBH}$, assuming all PBHs have the same mass. 
   The bounds shown are (left to right) from evaporation (orange), microlensing (blue), gravitational waves (purple), accretion (red) and dynamical effects (green).
   For each type of bound the tightest constraint at each mass is shown and the shaded regions are excluded under standard assumptions. Figure created using 
Kavanagh's PBHbounds code~\cite{PBHbounds},  which is regularly updated to include the latest constraints.
				}
	\label{fig:constraints}
\end{figure}

\subsection{Further work on PBH formation}

Byrnes et al.~\cite{Byrnes:2018clq} calculated the mass function for PBHs formed at the QCD epoch, using an accurate form for the softened equation of state then. In general the threshold for PBH formation, $\delta_{\rm c}$, will 
decrease at any phase transition where there is a reduction in pressure,  so the PBH mass function will have peaks at the corresponding horizon mass (see Chapter 6). Carr et al.~\cite{Carr:2019kxo} extended this idea by pointing out that the thermal history of the Universe produces four dips in the sound speed in the period $(10^{-8}-10)$~s, leading to corresponding bumps in the PBH mass function if the amplitude of the primordial perturbations is large across the corresponding range of scales.  

There has been further extensive work exploring which inflation models can generate large PBH-forming density perturbations, while also satisfying the constraints on the amplitude and scale dependence of perturbations on cosmological scales~\cite{Planck:2018jri}. This is not possible in single-field slow-roll models (e.g. Refs.~\cite{Ballesteros:2017fsr,Hertzberg:2017dkh}) but can be achieved with a feature in the potential, such as an inflection point (e.g.~Ref.~\cite{Garcia-Bellido:2017mdw,Ballesteros:2017fsr}) or a small dip/peak (e.g.~Ref.~\cite{Hertzberg:2017dkh,Mishra:2019pzq}), which leads to a period of ultra-slow-roll inflation. Other methods for generating large inflationary density perturbations have been proposed, such as
multi-field models with rapid turns in field space~\cite{Palma:2020ejf,Fumagalli:2020adf}); for an overview see Ref.~\cite{Ozsoy:2023ryl}.  New mechanisms for the production of PBH-forming fluctuations have also been proposed, for instance, Q-balls produced by the fragmentation of scalar field condensate~\cite{Cotner:2016cvr,2022PhRvL.129f1302D}, long-range forces mediated by scalar fields~\cite{Amendola:2017xhl,Flores:2020drq} and the collapse of cosmic string cusps~\cite{Jenkins:2020ctp}. For further details see Chapter 10.

It was pointed out in Ref.~\cite{Germani:2018jgr} that the shape of density perturbations, and hence the threshold for collapse, depends on the form of the primordial power spectrum. Subsequently, Escriv\'a {\it et al.}
~\cite{Escriva:2019phb} showed that the threshold is universal if specified in terms of the average of the compaction function, which is similar to the Schwarzschild gravitational potential. See Chapters 3 and 7 for further details. It was realised that even if the curvature perturbations are Gaussian, the non-linear relationship between density and curvature perturbations means that the distribution of large density perturbations is inevitably non-Gaussian~\cite{Kawasaki:2019mbl,DeLuca:2019qsy,Young:2019yug}.  Chapter 8 provides a detailed discussion of non-Gaussianity and an approach to calculating the probability distribution of the primordial fluctuations for stochastic inflation is described in Chapter 9. 

The issue of the fine-tuning required to generate an interesting PBH abundance has also been addressed.  Fine-tuning is required in three different senses: the collapse fraction $\beta$ must be tiny even if they constitute all of the dark matter; if PBHs form from the collapse of density perturbations, their amplitude must be even more fine-tuned; if the perturbations are generated by inflation, the parameters of the inflation model must also be fine-tuned. The first problem has been addressed by Garc{\'\i}a-Bellido {\it et al.}~\cite{Garcia-Bellido:2019vlf}, who point out that PBHs which form at the QCD epoch would necessarily have $\beta \sim \eta$, where $\eta$ is the cosmological baryon-to-photon ratio, if they provide the dark matter.  They propose a baryogenesis scenario in which the PBH formation at the QCD epoch naturally generates this relation and this also explains why the PBH and baryon densities are comparable.  The second problem is then viewed from an anthropic perspective. The third problem has been discussed by Cole {\it et al.}~\cite{Cole:2023wyx}.

\subsection{Extended mass functions and clustering}
\label{subsec:emf}

Constraints on the PBH abundance are usually calculated assuming that PBHs have a smooth density distribution and a single mass (i.e.~a delta-function or monochromatic mass function). 
However, most formation mechanisms produce PBHs that have enhanced clustering on small scales and an extended mass function. The latter problem was first addressed in Refs.~\cite{Carr:2016drx} and \cite{Green:2016xgy} and a general method
for applying constraints for a monochromatic mass function to an extended mass function was presented in Ref.~\cite{Carr:2017jsz}. 
The fraction of dark matter in PBHs is constrained to be less than $1$ over a wider range of masses than for a monochromatic mass function~\cite{Green:2016xgy} but the tightest value of each constraint is weakened~\cite{Carr:2017jsz}.  

The clustering of PBHs as a result of their Poisson fluctuations has featured in all the historical periods  but several groups studied this affect more extensively in this fourth one.  In particular,  the fact that the  first baryonic clouds could form earlier than in the standard scenario would have interesting observational consequences, as stressed by Kashlinsky~\cite{2016ApJ...823L..25K}.  Later
Inman and Ali-Ha\"\i{}moud carried out numerical simulations of the clusters that form shortly after matter-radiation equality if PBHs make up a fraction of the dark matter~\cite{Inman:2019wvr}. 
This clustering modifies some of the PBH constraints but in different ways. For the diffuse clusters formed from Gaussian perturbations the change in the microlensing constraints is small~\cite{Gorton:2022fyb,Petac:2022rio}. However for compact clusters (which could be formed from non-Gaussian perturbations) 
the microlensing constraints would be weakened~\cite{Calcino:2018mwh}, while other constraints
would be tightened~\cite{Young:2019gfc,DeLuca:2022uvz}.

Adamek {\it et al.}~\cite{Adamek:2019gns} studied the ultracompact minihalos that form around PBHs when dark matter is a mixture of PBHs and WIMPs. The limits for such mixed dark matter models
were updated in Ref.~\cite{Carr:2020mqm}, using the most recent and detailed simulations of the particle dark matter halos around PBHs~\cite{Boudaud:2021irr}. 
As also emphasised in Ref.~\cite{Bertone:2019vsk}, definitive evidence for the existence of PBHs would rule out WIMPs making up a significant fraction of the dark matter.  See Chapters 15 and 21 for more details.

\subsection{
Evidence and constraints from gravitational waves and microlensing}

In 2016 the LIGO-Virgo collaboration announced the discovery of gravitational waves from mergers of black holes
~\cite{LIGOScientific:2016aoc}.
Soon afterwards several papers suggested that these 
could be PBHs and also make up some of the dark matter~\cite{Bird:2016dcv,Clesse:2016vqa,Sasaki:2016jop}.
 Whereas Ref.~\cite{Bird:2016dcv} focussed on PBH binaries 
formed within dark matter halos at late times,  Ref.~\cite{Sasaki:2016jop} focussed on binaries  forming at early times, but the latter process is probably dominant.
If all of the dark matter is in such PBHs, then (provided the orbits of the binaries are not changed significantly during structure formation) the expected present-day merger rate is 
several orders of magnitude larger than that measured by LIGO-Virgo~\cite{Sasaki:2016jop}.  However,  
a subdominant PBH component of the dark matter could still be responsible for some of the observed events~\cite{Garcia-Bellido:2017fdg,2018CQGra..35f3001S,Hall:2020daa,DeLuca:2020qqa,Hutsi:2020sol,Franciolini:2021tla} and this is discussed further in Chapter 25. Although the mainstream view is that all the LIGO-Virgo-KAGRA detections derive from astrophysical black holes, the masses are larger than initially expected~\cite{Kovetz:2017rvv} and some lie in mass gaps where stellar remnants should not be found~\cite{Carr:2023tpt}.  Also, the observations suggest smaller spins than would be expected for astrophysical black holes~\cite{Garcia-Bellido:2017fdg}. The PBH origin proposal is still controversial but future observations of the mass, spin and redshift distribution of the objects should clarify the issue soon. Further details and more recent developments are covered in Part IV.

In 2017
Niikura et al.~\cite{Niikura:2017zjd} carried out a microlensing survey of the Andromeda galaxy (M31) which was sensitive to short-duration events and hence lighter compact objects. They found only one candidate event and extended the range of masses constrained by microlensing down to $\sim 10^{22} \, {\rm g}$~\cite{Niikura:2017zjd, Smyth:2019whb,Croon:2020ouk}. Lighter PBHs cannot be probed by optical observations of main-sequence stars; both the finite size of the source stars and wave-optics effects (from the PBH Schwarzschild radius being comparable with the wavelength of light)  reduce the microlensing event rate (see Refs.~\cite{Montero-Camacho:2019jte,Sugiyama:2019dgt} and references therein).  

The Milky Way microlensing constraints have now been extended to larger masses, $M \sim 10^{3} M_{\odot}$, by combining data from different surveys to obtain sensitivity to longer duration events~\cite{Blaineau:2022nhy,Mroz:2024mse}. 
However, Ref.~\cite{Garcia-Bellido:2024yaz} argues that these limits can be weakened when one allows for the extended PBH mass function expected in the thermal history model and the falling rotation curve indicated by recent observations of the Milky Way. 
 
 Various other microlensing constraints on stellar mass PBHs have been studied, including the magnification distribution of type 1a supernovae (SNe)~\cite{Zumalacarregui:2017qqd}, supermagnified stars such as Icarus~\cite{Oguri:2017ock,Muller:2024pwn} (where a background star at a relatively high redshift passes close to a caustic in a galaxy cluster and is magnified by a huge factor~\cite{Venumadhav:2017pps}) and quasar microlensing~\cite{Esteban-Gutierrez:2023qcz}.  See Chapter 24 for details.

\subsection{Evidence and constraints from evaporation, dynamical, accretion and nucleosynthetic effects}

 In 2019 Arbey and Auffinger publicly released the \texttt{BlackHawk} code which calculates the evaporation spectra produced by any population of PBHs ~\cite{Arbey:2019mbc,Arbey:2021mbl}. Various observations (e.g. of MeV gamma-rays) were used to tightly constrain the abundance of PBHs lighter than 
$10^{17} \, {\rm g}$ and limits on evaporating PBHs were overviewed  in Ref.~\cite{Carr:2020gox} in 2021 and Ref.~\cite{Auffinger:2022khh} in 2023.  These results are discussed further in Chapter 20. 

Also in 2019 Tian et al. ~\cite{Tian_2019} found that the separation distribution of halo-like wide binaries from GAIA-DR2 steepens at large radii, but is not sharply truncated, as would be expected if they are perturbed by compact objects. They argued that this, combined with the uncertainty in the initial separation distribution, makes it hard to probe PBHs using wide binaries.  However, constraints were placed on heavier
PBHs from their dynamical effects on stars in dwarf galaxies.  Two-body interactions equalize the kinetic energies of populations, so if a significant fraction of the dark matter in dwarf galaxies is composed of PBHs with $M  \gtrsim 10 M_{\odot}$, the stars there
will gain energy and their distribution will expand~\cite{Brandt:2016aco,Zhu:2017plg,Stegmann:2019wyz}.  
In 2017 the effects of PBH accretion 
on the recombination history of the Universe were refined~\cite{Poulin:2017bwe} and other present-day accretion probes were proposed~\cite{Gaggero:2016dpq}.  In particular, } accretion of interstellar gas onto PBHs in the Milky Way would lead to observable X-ray and radio emission~\cite{Manshanden:2018tze,Inoue:2017csr} and heat the gas in dwarf galaxies~\cite{Lu:2020bmd}. See Chapters 14 and 26 for more details.  

Fuller {\it et al.}~\cite{Fuller:2017uyd} showed that some $r$-process elements (i.e.~those generated by fast nuclear reactions) can be produced by the interaction of PBHs with neutron stars if they have $f > 0.01$ in the mass range $10^{-14}$ -- $10^{-8}\,M_\odot$. Abramowicz and Bejger~\cite{Abramowicz:2017zbp} argued that collisions of neutron stars with PBHs of mass $10^{23}\,$g may explain the millisecond durations and large luminosities of fast radio bursts. 

\subsection{Evidence and constraints from cosmic structure,  dwarf galaxies and supermassive black holes}

An updated discussion of the effects of PBHs on the generation of cosmic structure,  through the seed and Poisson effects was provided in Ref.~\cite{Carr:2018rid}. The mass binding at redshift $z_{\rm B}$ is $4000\,M\,z_{\rm B}^{-1}M_\odot$ for the seed effect and $10^{7}\,f\,M\,z_{\rm B}^{-2}M_\odot$ for the Poisson effect.  
Having $f = 1$ requires $M < 10^{3}\,M_\odot$ and so the Poisson effect could only bind a scale below $10^{10}\,z_{\rm B}^{-2}\,M_\odot$, which is necessarily subgalactic. However, 
providing the mass and dark matter fraction of the PBHs are large enough, this effect could generate the first bound objects at a much earlier epoch than in the standard cosmological scenario
with interesting observational consequences.  
This may explain the cosmic-infrared-background fluctuation excess first detected in Spitzer data~\cite{2005Natur.438...45K}, with a consistency check being provided by correlations of the source-subtracted infrared and X-ray cosmic backgrounds. 
 This line of evidence has been studied extensively by Kashlinsky and colleagues~\cite{2013ApJ...769...68C,2018RvMP...90b5006K,2019ApJ...871L...6K} and more recently by Hasinger {\it et al.}~\cite{Hasinger:2020ptw} and Cappelutti {\it et al.} \cite{Cappelluti:2021usg}.

There are problems with the standard CDM scenario which Silk claims can be resolved with PBHs in the intermediate mass range
~\cite{Silk:2017yai}. 
In particular, he has argued that
such PBHs could be ubiquitous in early dwarf galaxies, being mostly passive today but active in their gas-rich past. This is suggested by observations of AGN~\cite{2013ARAA..51..511K, 
2016ApJ...831..203P, Baldassare:2016cox} and early feedback from these objects could provide a unified explanation for many dwarf galaxy anomalies. Besides providing a phase of early galaxy formation and seeds for SMBHs at high $z$, they could:
	(1) suppress the number of luminous dwarfs; 
	(2) generate cores in dwarfs by dynamical heating; 
	(3) resolve the ``too big to fail'' problem; 
	(4) create bulgeless disks; 
	(5) form ultra-faint dwarfs and ultra-diffuse galaxies; 
	(6) reduce the baryon fraction in Milky-Way-type galaxies; 
	(7) explain ultra-luminous X-ray sources in the outskirts of galaxies;
	(8) trigger star formation in dwarfs via AGN.

The mainstream view is that the SMBHs in galactic nuclei form from dynamical processes {\it after} galaxies, but this proposal is becoming increasingly challenged by the high mass and redshift of some SMBHs.  
For example, there is a population of red galaxies 
at $7.4 < z < 9.1$ with central SMBHs~\cite{Labb__2023} and
ALMA has observed an AGN at $z \sim 7$ whose large luminosity suggests a $10^{9} M_{\odot}$ black hole~\cite{Endsley_2023}.  It is unclear that such objects can form quickly enough in the standard model, so this 
suggests that the SMBHs{\,---\,}or at least their seeds{\,---\,}could form {\it before} galaxies~\cite{Dolgov:2023ijt}.  
Thus Liu and Bromm~\cite{Liu:2022bvr} have argued that unusually massive galaxies at $z > 10$ could be evidence that structure formation is accelerated by $10^{9} M_{\odot}$ PBHs making up $10^{-6}\,\text{--}\,10^{-3}$ of the dark matter (see Chapter 12). Ref.~\cite{Carr:2019kxo} claims that the $10^6M_{\odot}$ PBHs formed at the electron-positron annihilation epoch could provide the required seeds but Byrnes {\it et al.}~\cite{Byrnes:2018txb} have argued that is precluded by limits on the  $\mu$-distortion in the CMB~\cite{Chluba:2012we}.

\section{Conclusion}

We are at a tantalizing point in the history of PBHs since their existence is likely to be either confirmed or excluded within the next decade. The community has a range of views as to which outcome is most likely,  and even the authors of this chapter have different expectations.  However, there is no doubt that there has been a huge surge of interest in PBHs in recent years and this is important because it has prompted more sophisticated studies of their expected characteristics and consequences.  For researchers who advocate PBHs as a solution of the dark matter problem,  it should be stressed that there is a division between those who favour PBHs in the asteroidal and solar mass ranges.  The former view has the advantage that it is not excluded by current observations. 
The latter view is driven by the claimed evidence but is more controversial due to the constraints in this mass region.


\bibliography{main}

\begin{thebibliography}{205}%
\makeatletter
\providecommand \@ifxundefined [1]{%
 \@ifx{#1\undefined}
}%
\providecommand \@ifnum [1]{%
 \ifnum #1\expandafter \@firstoftwo
 \else \expandafter \@secondoftwo
 \fi
}%
\providecommand \@ifx [1]{%
 \ifx #1\expandafter \@firstoftwo
 \else \expandafter \@secondoftwo
 \fi
}%
\providecommand \natexlab [1]{#1}%
\providecommand \enquote  [1]{``#1''}%
\providecommand \bibnamefont  [1]{#1}%
\providecommand \bibfnamefont [1]{#1}%
\providecommand \citenamefont [1]{#1}%
\providecommand \href@noop [0]{\@secondoftwo}%
\providecommand \href [0]{\begingroup \@sanitize@url \@href}%
\providecommand \@href[1]{\@@startlink{#1}\@@href}%
\providecommand \@@href[1]{\endgroup#1\@@endlink}%
\providecommand \@sanitize@url [0]{\catcode `\\12\catcode `\$12\catcode
  `\&12\catcode `\#12\catcode `\^12\catcode `\_12\catcode `\%12\relax}%
\providecommand \@@startlink[1]{}%
\providecommand \@@endlink[0]{}%
\providecommand \url  [0]{\begingroup\@sanitize@url \@url }%
\providecommand \@url [1]{\endgroup\@href {#1}{\urlprefix }}%
\providecommand \urlprefix  [0]{URL }%
\providecommand \Eprint [0]{\href }%
\providecommand \doibase [0]{http://dx.doi.org/}%
\providecommand \selectlanguage [0]{\@gobble}%
\providecommand \bibinfo  [0]{\@secondoftwo}%
\providecommand \bibfield  [0]{\@secondoftwo}%
\providecommand \translation [1]{[#1]}%
\providecommand \BibitemOpen [0]{}%
\providecommand \bibitemStop [0]{}%
\providecommand \bibitemNoStop [0]{.\EOS\space}%
\providecommand \EOS [0]{\spacefactor3000\relax}%
\providecommand \BibitemShut  [1]{\csname bibitem#1\endcsname}%
\let\auto@bib@innerbib\@empty
\bibitem [{\citenamefont {{Zel'dovich}}\ and\ \citenamefont
  {{Novikov}}(1967)}]{1967SvA....10..602Z}%
  \BibitemOpen
  \bibfield  {author} {\bibinfo {author} {\bibfnamefont {Y.}~\bibnamefont
  {{Zel'dovich}}}\ and\ \bibinfo {author} {\bibfnamefont {I.}~\bibnamefont
  {{Novikov}}},\ }\href@noop {} {\bibfield  {journal} {\bibinfo  {journal}
  {Sov. Astron.}\ }\textbf {\bibinfo {volume} {10}},\ \bibinfo {pages} {602}
  (\bibinfo {year} {1967})}\BibitemShut {NoStop}%
\bibitem [{\citenamefont {Hawking}(1971)}]{Hawking:1971ei}%
  \BibitemOpen
  \bibfield  {author} {\bibinfo {author} {\bibfnamefont {S.}~\bibnamefont
  {Hawking}},\ }\href@noop {} {\bibfield  {journal} {\bibinfo  {journal} {Mon.
  Not. Roy. Astron. Soc.}\ }\textbf {\bibinfo {volume} {152}},\ \bibinfo
  {pages} {75} (\bibinfo {year} {1971})}\BibitemShut {NoStop}%
\bibitem [{\citenamefont {Hawking}(1974)}]{Hawking:1974rv}%
  \BibitemOpen
  \bibfield  {author} {\bibinfo {author} {\bibfnamefont {S.~W.}\ \bibnamefont
  {Hawking}},\ }\href {\doibase 10.1038/248030a0} {\bibfield  {journal}
  {\bibinfo  {journal} {Nature}\ }\textbf {\bibinfo {volume} {248}},\ \bibinfo
  {pages} {30} (\bibinfo {year} {1974})}\BibitemShut {NoStop}%
\bibitem [{\citenamefont {Carr}\ and\ \citenamefont
  {Hawking}(1974)}]{Carr:1974nx}%
  \BibitemOpen
  \bibfield  {author} {\bibinfo {author} {\bibfnamefont {B.~J.}\ \bibnamefont
  {Carr}}\ and\ \bibinfo {author} {\bibfnamefont {S.~W.}\ \bibnamefont
  {Hawking}},\ }\href {\doibase 10.1093/mnras/168.2.399} {\bibfield  {journal}
  {\bibinfo  {journal} {Mon. Not. Roy. Astron. Soc.}\ }\textbf {\bibinfo
  {volume} {168}},\ \bibinfo {pages} {399} (\bibinfo {year}
  {1974})}\BibitemShut {NoStop}%
\bibitem [{\citenamefont {Carr}(1975)}]{Carr:1975qj}%
  \BibitemOpen
  \bibfield  {author} {\bibinfo {author} {\bibfnamefont {B.}~\bibnamefont
  {Carr}},\ }\href {\doibase 10.1086/153853} {\bibfield  {journal} {\bibinfo
  {journal} {Astrophys.~J.}\ }\textbf {\bibinfo {volume} {201}},\ \bibinfo
  {pages} {1} (\bibinfo {year} {1975})}\BibitemShut {NoStop}%
\bibitem [{\citenamefont {{Harrison, Phys.~Rev.~{\bf D1}, 2726-2730
  (1970)}}(1970)}]{1970PhRvD...1.2726H}%
  \BibitemOpen
  \bibfield  {author} {\bibinfo {author} {\bibfnamefont {E.}~\bibnamefont
  {{Harrison, Phys.~Rev.~{\bf D1}, 2726-2730 (1970)}}},\ }\href {\doibase
  10.1103/PhysRevD.1.2726} {\enquote {\bibinfo {title} {{Fluctuations at the
  threshold of classical cosmology}},}\ } (\bibinfo {year} {1970})\BibitemShut
  {NoStop}%
\bibitem [{\citenamefont {{Zeldovich}}(1972)}]{1972MNRAS.160P...1Z}%
  \BibitemOpen
  \bibfield  {author} {\bibinfo {author} {\bibfnamefont {Y.}~\bibnamefont
  {{Zeldovich}}},\ }\href {\doibase 10.1093/mnras/160.1.1P} {\bibfield
  {journal} {\bibinfo  {journal} {Mon.~Not.~R. astron.~Soc.}\ }\textbf
  {\bibinfo {volume} {160}},\ \bibinfo {pages} {1P} (\bibinfo {year}
  {1972})}\BibitemShut {NoStop}%
\bibitem [{\citenamefont {{Nadezhin}}\ \emph {et~al.}(1978)\citenamefont
  {{Nadezhin}}, \citenamefont {{Novikov}},\ and\ \citenamefont
  {{Polnarev}}}]{1978SvA....22..129N}%
  \BibitemOpen
  \bibfield  {author} {\bibinfo {author} {\bibfnamefont {D.~K.}\ \bibnamefont
  {{Nadezhin}}}, \bibinfo {author} {\bibfnamefont {I.~D.}\ \bibnamefont
  {{Novikov}}}, \ and\ \bibinfo {author} {\bibfnamefont {A.~G.}\ \bibnamefont
  {{Polnarev}}},\ }\href@noop {} {\bibfield  {journal} {\bibinfo  {journal}
  {Sov. Astron.}\ }\textbf {\bibinfo {volume} {22}},\ \bibinfo {pages} {129}
  (\bibinfo {year} {1978})}\BibitemShut {NoStop}%
\bibitem [{\citenamefont {Page}\ and\ \citenamefont
  {Hawking}(1976)}]{Page:1976wx}%
  \BibitemOpen
  \bibfield  {author} {\bibinfo {author} {\bibfnamefont {D.}~\bibnamefont
  {Page}}\ and\ \bibinfo {author} {\bibfnamefont {S.}~\bibnamefont {Hawking}},\
  }\href {\doibase 10.1086/154350} {\bibfield  {journal} {\bibinfo  {journal}
  {Astrophys.~J.}\ }\textbf {\bibinfo {volume} {206}},\ \bibinfo {pages} {1}
  (\bibinfo {year} {1976})}\BibitemShut {NoStop}%
\bibitem [{\citenamefont {{Carr}}(1976)}]{1976ApJ...206....8C}%
  \BibitemOpen
  \bibfield  {author} {\bibinfo {author} {\bibfnamefont {B.~J.}\ \bibnamefont
  {{Carr}}},\ }\href {\doibase 10.1086/154351} {\bibfield  {journal} {\bibinfo
  {journal} {Astrophys. J.}\ }\textbf {\bibinfo {volume} {206}},\ \bibinfo
  {pages} {8} (\bibinfo {year} {1976})}\BibitemShut {NoStop}%
\bibitem [{\citenamefont {{Novikov}}\ \emph {et~al.}(1979)\citenamefont
  {{Novikov}}, \citenamefont {{Polnarev}}, \citenamefont {{Starobinsky}},\ and\
  \citenamefont {{Zeldovich}}}]{1979A&A....80..104N}%
  \BibitemOpen
  \bibfield  {author} {\bibinfo {author} {\bibfnamefont {I.~D.}\ \bibnamefont
  {{Novikov}}}, \bibinfo {author} {\bibfnamefont {A.~G.}\ \bibnamefont
  {{Polnarev}}}, \bibinfo {author} {\bibfnamefont {A.~A.}\ \bibnamefont
  {{Starobinsky}}}, \ and\ \bibinfo {author} {\bibfnamefont {Y.~B.}\
  \bibnamefont {{Zeldovich}}},\ }\href@noop {} {\bibfield  {journal} {\bibinfo
  {journal} {Astron. Astrophys.}\ }\textbf {\bibinfo {volume} {80}},\ \bibinfo
  {pages} {104} (\bibinfo {year} {1979})}\BibitemShut {NoStop}%
\bibitem [{\citenamefont {{Vainer}}\ and\ \citenamefont
  {{Naselskii}}(1978)}]{1978AZh....55..231V}%
  \BibitemOpen
  \bibfield  {author} {\bibinfo {author} {\bibfnamefont {B.~V.}\ \bibnamefont
  {{Vainer}}}\ and\ \bibinfo {author} {\bibfnamefont {P.~D.}\ \bibnamefont
  {{Naselskii}}},\ }\href@noop {} {\bibfield  {journal} {\bibinfo  {journal}
  {Astron. Zh.}\ }\textbf {\bibinfo {volume} {55}},\ \bibinfo {pages} {231}
  (\bibinfo {year} {1978})},\ \bibinfo {note} {[Sov.\ Astron. \textbf{22}, 138
  (1978).]}\BibitemShut {NoStop}%
\bibitem [{\citenamefont {Miyama}\ and\ \citenamefont
  {Sato}(1978)}]{Miyama:1978mp}%
  \BibitemOpen
  \bibfield  {author} {\bibinfo {author} {\bibfnamefont {S.}~\bibnamefont
  {Miyama}}\ and\ \bibinfo {author} {\bibfnamefont {K.}~\bibnamefont {Sato}},\
  }\href {\doibase 10.1143/PTP.59.1012} {\bibfield  {journal} {\bibinfo
  {journal} {Prog. Theor. Phys.}\ }\textbf {\bibinfo {volume} {59}},\ \bibinfo
  {pages} {1012} (\bibinfo {year} {1978})}\BibitemShut {NoStop}%
\bibitem [{\citenamefont {{Zel'dovich}}\ \emph {et~al.}(1977)\citenamefont
  {{Zel'dovich}}, \citenamefont {{Starobinskii}}, \citenamefont {{Khlopov}},\
  and\ \citenamefont {{Chechetkin}}}]{1977PAZh....3..208Z}%
  \BibitemOpen
  \bibfield  {author} {\bibinfo {author} {\bibfnamefont {Y.~B.}\ \bibnamefont
  {{Zel'dovich}}}, \bibinfo {author} {\bibfnamefont {A.~A.}\ \bibnamefont
  {{Starobinskii}}}, \bibinfo {author} {\bibfnamefont {M.~I.}\ \bibnamefont
  {{Khlopov}}}, \ and\ \bibinfo {author} {\bibfnamefont {V.~M.}\ \bibnamefont
  {{Chechetkin}}},\ }\href@noop {} {\bibfield  {journal} {\bibinfo  {journal}
  {Pisma Astron.\ Zh.}\ }\textbf {\bibinfo {volume} {3}},\ \bibinfo {pages}
  {208} (\bibinfo {year} {1977})},\ \bibinfo {note} {[Sov.\ Astron.\ Lett.
  \textbf{3}, 110 (1977).]}\BibitemShut {NoStop}%
\bibitem [{\citenamefont {{Lindley}}(1980)}]{1980MNRAS.193..593L}%
  \BibitemOpen
  \bibfield  {author} {\bibinfo {author} {\bibfnamefont {D.}~\bibnamefont
  {{Lindley}}},\ }\href {\doibase 10.1093/mnras/193.3.593} {\bibfield
  {journal} {\bibinfo  {journal} {Mon. Not. Roy. Astron. Soc.}\ }\textbf
  {\bibinfo {volume} {193}},\ \bibinfo {pages} {593} (\bibinfo {year}
  {1980})}\BibitemShut {NoStop}%
\bibitem [{\citenamefont {{Okele}}\ and\ \citenamefont
  {{Rees}}(1980)}]{1980A&A....81..263O}%
  \BibitemOpen
  \bibfield  {author} {\bibinfo {author} {\bibfnamefont {P.~N.}\ \bibnamefont
  {{Okele}}}\ and\ \bibinfo {author} {\bibfnamefont {M.~J.}\ \bibnamefont
  {{Rees}}},\ }\href@noop {} {\bibfield  {journal} {\bibinfo  {journal}
  {Astron. Astrophys.}\ }\textbf {\bibinfo {volume} {81}},\ \bibinfo {pages}
  {263} (\bibinfo {year} {1980})}\BibitemShut {NoStop}%
\bibitem [{\citenamefont {{Chapline}}(1975)}]{1975Natur.253..251C}%
  \BibitemOpen
  \bibfield  {author} {\bibinfo {author} {\bibfnamefont {G.}~\bibnamefont
  {{Chapline}}},\ }\href {\doibase 10.1038/253251a0} {\bibfield  {journal}
  {\bibinfo  {journal} {Nature (London)}\ }\textbf {\bibinfo {volume} {253}},\
  \bibinfo {pages} {251} (\bibinfo {year} {1975})}\BibitemShut {NoStop}%
\bibitem [{\citenamefont {M{\'e}sz{\'a}ros}(1975)}]{Meszaros:1975ef}%
  \BibitemOpen
  \bibfield  {author} {\bibinfo {author} {\bibfnamefont {P.}~\bibnamefont
  {M{\'e}sz{\'a}ros}},\ }\href@noop {} {\bibfield  {journal} {\bibinfo
  {journal} {Astron. astrophys.}\ }\textbf {\bibinfo {volume} {38}},\ \bibinfo
  {pages} {5} (\bibinfo {year} {1975})}\BibitemShut {NoStop}%
\bibitem [{\citenamefont {Carr}\ \emph {et~al.}(1994)\citenamefont {Carr},
  \citenamefont {Gilbert},\ and\ \citenamefont {Lidsey}}]{Carr:1994ar}%
  \BibitemOpen
  \bibfield  {author} {\bibinfo {author} {\bibfnamefont {B.}~\bibnamefont
  {Carr}}, \bibinfo {author} {\bibfnamefont {J.}~\bibnamefont {Gilbert}}, \
  and\ \bibinfo {author} {\bibfnamefont {J.}~\bibnamefont {Lidsey}},\ }\href
  {\doibase 10.1103/PhysRevD.50.4853} {\bibfield  {journal} {\bibinfo
  {journal} {Phys.~Rev.}\ }\textbf {\bibinfo {volume} {D50}},\ \bibinfo {pages}
  {4853} (\bibinfo {year} {1994})},\ \Eprint
  {http://arxiv.org/abs/astro-ph/9405027} {arXiv:astro-ph/9405027} \BibitemShut
  {NoStop}%
\bibitem [{\citenamefont {Starobinsky}(1980)}]{Starobinsky:1980te}%
  \BibitemOpen
  \bibfield  {author} {\bibinfo {author} {\bibfnamefont {A.~A.}\ \bibnamefont
  {Starobinsky}},\ }\href {\doibase 10.1016/0370-2693(80)90670-X} {\bibfield
  {journal} {\bibinfo  {journal} {Phys. Lett. B}\ }\textbf {\bibinfo {volume}
  {91}},\ \bibinfo {pages} {99} (\bibinfo {year} {1980})}\BibitemShut {NoStop}%
\bibitem [{\citenamefont {Guth}(1981)}]{Guth:1980zm}%
  \BibitemOpen
  \bibfield  {author} {\bibinfo {author} {\bibfnamefont {A.~H.}\ \bibnamefont
  {Guth}},\ }\href {\doibase 10.1103/PhysRevD.23.347} {\bibfield  {journal}
  {\bibinfo  {journal} {Phys. Rev. D}\ }\textbf {\bibinfo {volume} {23}},\
  \bibinfo {pages} {347} (\bibinfo {year} {1981})}\BibitemShut {NoStop}%
\bibitem [{\citenamefont {Sato}(1981)}]{Sato:1980yn}%
  \BibitemOpen
  \bibfield  {author} {\bibinfo {author} {\bibfnamefont {K.}~\bibnamefont
  {Sato}},\ }\href@noop {} {\bibfield  {journal} {\bibinfo  {journal}
  {Mon.~Not.~Roy. astron.~Soc.}\ }\textbf {\bibinfo {volume} {195}},\ \bibinfo
  {pages} {467} (\bibinfo {year} {1981})}\BibitemShut {NoStop}%
\bibitem [{\citenamefont {Albrecht}\ and\ \citenamefont
  {Steinhardt}(1982)}]{Albrecht:1982wi}%
  \BibitemOpen
  \bibfield  {author} {\bibinfo {author} {\bibfnamefont {A.}~\bibnamefont
  {Albrecht}}\ and\ \bibinfo {author} {\bibfnamefont {P.}~\bibnamefont
  {Steinhardt}},\ }\href {\doibase 10.1103/PhysRevLett.48.1220} {\bibfield
  {journal} {\bibinfo  {journal} {Phys.~Rev.~Lett.}\ }\textbf {\bibinfo
  {volume} {48}},\ \bibinfo {pages} {1220} (\bibinfo {year}
  {1982})}\BibitemShut {NoStop}%
\bibitem [{\citenamefont {Guth}\ and\ \citenamefont {Pi}(1982)}]{Guth:1982ec}%
  \BibitemOpen
  \bibfield  {author} {\bibinfo {author} {\bibfnamefont {A.~H.}\ \bibnamefont
  {Guth}}\ and\ \bibinfo {author} {\bibfnamefont {S.~Y.}\ \bibnamefont {Pi}},\
  }\href {\doibase 10.1103/PhysRevLett.49.1110} {\bibfield  {journal} {\bibinfo
   {journal} {Phys. Rev. Lett.}\ }\textbf {\bibinfo {volume} {49}},\ \bibinfo
  {pages} {1110} (\bibinfo {year} {1982})}\BibitemShut {NoStop}%
\bibitem [{\citenamefont {Hawking}(1982)}]{Hawking:1982cz}%
  \BibitemOpen
  \bibfield  {author} {\bibinfo {author} {\bibfnamefont {S.~W.}\ \bibnamefont
  {Hawking}},\ }\href {\doibase 10.1016/0370-2693(82)90373-2} {\bibfield
  {journal} {\bibinfo  {journal} {Phys. Lett. B}\ }\textbf {\bibinfo {volume}
  {115}},\ \bibinfo {pages} {295} (\bibinfo {year} {1982})}\BibitemShut
  {NoStop}%
\bibitem [{\citenamefont {Linde}(1982)}]{Linde:1982uu}%
  \BibitemOpen
  \bibfield  {author} {\bibinfo {author} {\bibfnamefont {A.~D.}\ \bibnamefont
  {Linde}},\ }\href {\doibase 10.1016/0370-2693(82)90293-3} {\bibfield
  {journal} {\bibinfo  {journal} {Phys. Lett. B}\ }\textbf {\bibinfo {volume}
  {116}},\ \bibinfo {pages} {335} (\bibinfo {year} {1982})}\BibitemShut
  {NoStop}%
\bibitem [{\citenamefont {Starobinsky}(1982)}]{Starobinsky:1982ee}%
  \BibitemOpen
  \bibfield  {author} {\bibinfo {author} {\bibfnamefont {A.~A.}\ \bibnamefont
  {Starobinsky}},\ }\href {\doibase 10.1016/0370-2693(82)90541-X} {\bibfield
  {journal} {\bibinfo  {journal} {Phys. Lett. B}\ }\textbf {\bibinfo {volume}
  {117}},\ \bibinfo {pages} {175} (\bibinfo {year} {1982})}\BibitemShut
  {NoStop}%
\bibitem [{\citenamefont {Carr}\ and\ \citenamefont
  {Lidsey}(1993)}]{Carr:1993aq}%
  \BibitemOpen
  \bibfield  {author} {\bibinfo {author} {\bibfnamefont {B.~J.}\ \bibnamefont
  {Carr}}\ and\ \bibinfo {author} {\bibfnamefont {J.~E.}\ \bibnamefont
  {Lidsey}},\ }\href {\doibase 10.1103/PhysRevD.48.543} {\bibfield  {journal}
  {\bibinfo  {journal} {Phys. Rev. D}\ }\textbf {\bibinfo {volume} {48}},\
  \bibinfo {pages} {543} (\bibinfo {year} {1993})}\BibitemShut {NoStop}%
\bibitem [{\citenamefont {Smoot}\ \emph {et~al.}(1992)\citenamefont {Smoot}
  \emph {et~al.}}]{COBE:1992syq}%
  \BibitemOpen
  \bibfield  {author} {\bibinfo {author} {\bibfnamefont {G.~F.}\ \bibnamefont
  {Smoot}} \emph {et~al.} (\bibinfo {collaboration} {COBE}),\ }\href {\doibase
  10.1086/186504} {\bibfield  {journal} {\bibinfo  {journal} {Astrophys. J.
  Lett.}\ }\textbf {\bibinfo {volume} {396}},\ \bibinfo {pages} {L1} (\bibinfo
  {year} {1992})}\BibitemShut {NoStop}%
\bibitem [{\citenamefont {Wright}\ \emph {et~al.}(1992)\citenamefont {Wright}
  \emph {et~al.}}]{Wright:1992tf}%
  \BibitemOpen
  \bibfield  {author} {\bibinfo {author} {\bibfnamefont {E.~L.}\ \bibnamefont
  {Wright}} \emph {et~al.},\ }\href {\doibase 10.1086/186506} {\bibfield
  {journal} {\bibinfo  {journal} {Astrophys. J. Lett.}\ }\textbf {\bibinfo
  {volume} {396}},\ \bibinfo {pages} {L13} (\bibinfo {year}
  {1992})}\BibitemShut {NoStop}%
\bibitem [{\citenamefont {Ivanov}\ \emph {et~al.}(1994)\citenamefont {Ivanov},
  \citenamefont {Naselsky},\ and\ \citenamefont {Novikov}}]{Ivanov:1994pa}%
  \BibitemOpen
  \bibfield  {author} {\bibinfo {author} {\bibfnamefont {P.}~\bibnamefont
  {Ivanov}}, \bibinfo {author} {\bibfnamefont {P.}~\bibnamefont {Naselsky}}, \
  and\ \bibinfo {author} {\bibfnamefont {I.}~\bibnamefont {Novikov}},\ }\href
  {\doibase 10.1103/PhysRevD.50.7173} {\bibfield  {journal} {\bibinfo
  {journal} {Phys. Rev. D}\ }\textbf {\bibinfo {volume} {50}},\ \bibinfo
  {pages} {7173} (\bibinfo {year} {1994})}\BibitemShut {NoStop}%
\bibitem [{\citenamefont {Garcia-Bellido}\ \emph {et~al.}(1996)\citenamefont
  {Garcia-Bellido}, \citenamefont {Linde},\ and\ \citenamefont
  {Wands}}]{Garcia-Bellido:1996mdl}%
  \BibitemOpen
  \bibfield  {author} {\bibinfo {author} {\bibfnamefont {J.}~\bibnamefont
  {Garcia-Bellido}}, \bibinfo {author} {\bibfnamefont {A.~D.}\ \bibnamefont
  {Linde}}, \ and\ \bibinfo {author} {\bibfnamefont {D.}~\bibnamefont
  {Wands}},\ }\href {\doibase 10.1103/PhysRevD.54.6040} {\bibfield  {journal}
  {\bibinfo  {journal} {Phys. Rev. D}\ }\textbf {\bibinfo {volume} {54}},\
  \bibinfo {pages} {6040} (\bibinfo {year} {1996})},\ \Eprint
  {http://arxiv.org/abs/astro-ph/9605094} {arXiv:astro-ph/9605094} \BibitemShut
  {NoStop}%
\bibitem [{\citenamefont {Dolgov}\ and\ \citenamefont
  {Silk}(1993)}]{PhysRevD.47.4244}%
  \BibitemOpen
  \bibfield  {author} {\bibinfo {author} {\bibfnamefont {A.}~\bibnamefont
  {Dolgov}}\ and\ \bibinfo {author} {\bibfnamefont {J.}~\bibnamefont {Silk}},\
  }\href {\doibase 10.1103/PhysRevD.47.4244} {\bibfield  {journal} {\bibinfo
  {journal} {Phys. Rev. D}\ }\textbf {\bibinfo {volume} {47}},\ \bibinfo
  {pages} {4244} (\bibinfo {year} {1993})}\BibitemShut {NoStop}%
\bibitem [{\citenamefont {Khlopov}\ and\ \citenamefont
  {Polnarev}(1980)}]{Khlopov:1980mg}%
  \BibitemOpen
  \bibfield  {author} {\bibinfo {author} {\bibfnamefont {M.~Y.}\ \bibnamefont
  {Khlopov}}\ and\ \bibinfo {author} {\bibfnamefont {A.~G.}\ \bibnamefont
  {Polnarev}},\ }\href {\doibase 10.1016/0370-2693(80)90624-3} {\bibfield
  {journal} {\bibinfo  {journal} {Phys. Lett. B}\ }\textbf {\bibinfo {volume}
  {97}},\ \bibinfo {pages} {383} (\bibinfo {year} {1980})}\BibitemShut
  {NoStop}%
\bibitem [{\citenamefont {{Polnarev}}\ and\ \citenamefont
  {{Khlopov}}(1981)}]{1981AZh....58..706P}%
  \BibitemOpen
  \bibfield  {author} {\bibinfo {author} {\bibfnamefont {A.~G.}\ \bibnamefont
  {{Polnarev}}}\ and\ \bibinfo {author} {\bibfnamefont {M.~Y.}\ \bibnamefont
  {{Khlopov}}},\ }\href@noop {} {\bibfield  {journal} {\bibinfo  {journal}
  {Astron. Zh.}\ }\textbf {\bibinfo {volume} {58}},\ \bibinfo {pages} {706}
  (\bibinfo {year} {1981})}\BibitemShut {NoStop}%
\bibitem [{\citenamefont {Crawford}\ and\ \citenamefont
  {Schramm}(1982)}]{Crawford:1982yz}%
  \BibitemOpen
  \bibfield  {author} {\bibinfo {author} {\bibfnamefont {M.}~\bibnamefont
  {Crawford}}\ and\ \bibinfo {author} {\bibfnamefont {D.~N.}\ \bibnamefont
  {Schramm}},\ }\href {\doibase 10.1038/298538a0} {\bibfield  {journal}
  {\bibinfo  {journal} {Nature}\ }\textbf {\bibinfo {volume} {298}},\ \bibinfo
  {pages} {538} (\bibinfo {year} {1982})}\BibitemShut {NoStop}%
\bibitem [{\citenamefont {Hawking}\ \emph {et~al.}(1982)\citenamefont
  {Hawking}, \citenamefont {Moss},\ and\ \citenamefont
  {Stewart}}]{Hawking:1982ga}%
  \BibitemOpen
  \bibfield  {author} {\bibinfo {author} {\bibfnamefont {S.~W.}\ \bibnamefont
  {Hawking}}, \bibinfo {author} {\bibfnamefont {I.~G.}\ \bibnamefont {Moss}}, \
  and\ \bibinfo {author} {\bibfnamefont {J.~M.}\ \bibnamefont {Stewart}},\
  }\href {\doibase 10.1103/PhysRevD.26.2681} {\bibfield  {journal} {\bibinfo
  {journal} {Phys. Rev. D}\ }\textbf {\bibinfo {volume} {26}},\ \bibinfo
  {pages} {2681} (\bibinfo {year} {1982})}\BibitemShut {NoStop}%
\bibitem [{\citenamefont {Kodama}\ \emph {et~al.}(1982)\citenamefont {Kodama},
  \citenamefont {Sasaki},\ and\ \citenamefont {Sato}}]{Kodama:1982sf}%
  \BibitemOpen
  \bibfield  {author} {\bibinfo {author} {\bibfnamefont {H.}~\bibnamefont
  {Kodama}}, \bibinfo {author} {\bibfnamefont {M.}~\bibnamefont {Sasaki}}, \
  and\ \bibinfo {author} {\bibfnamefont {K.}~\bibnamefont {Sato}},\ }\href
  {\doibase 10.1143/PTP.68.1979} {\bibfield  {journal} {\bibinfo  {journal}
  {Prog. Theor. Phys.}\ }\textbf {\bibinfo {volume} {68}},\ \bibinfo {pages}
  {1979} (\bibinfo {year} {1982})}\BibitemShut {NoStop}%
\bibitem [{\citenamefont {Kibble}(1976)}]{Kibble:1976sj}%
  \BibitemOpen
  \bibfield  {author} {\bibinfo {author} {\bibfnamefont {T.~W.~B.}\
  \bibnamefont {Kibble}},\ }\href {\doibase 10.1088/0305-4470/9/8/029}
  {\bibfield  {journal} {\bibinfo  {journal} {J. Phys. A}\ }\textbf {\bibinfo
  {volume} {9}},\ \bibinfo {pages} {1387} (\bibinfo {year} {1976})}\BibitemShut
  {NoStop}%
\bibitem [{\citenamefont {Hawking}(1989)}]{Hawking:1987bn}%
  \BibitemOpen
  \bibfield  {author} {\bibinfo {author} {\bibfnamefont {S.~W.}\ \bibnamefont
  {Hawking}},\ }\href {\doibase 10.1016/0370-2693(89)90206-2} {\bibfield
  {journal} {\bibinfo  {journal} {Phys. Lett. B}\ }\textbf {\bibinfo {volume}
  {231}},\ \bibinfo {pages} {237} (\bibinfo {year} {1989})}\BibitemShut
  {NoStop}%
\bibitem [{\citenamefont {Polnarev}\ and\ \citenamefont
  {Zembowicz}(1991)}]{Polnarev:1988dh}%
  \BibitemOpen
  \bibfield  {author} {\bibinfo {author} {\bibfnamefont {A.}~\bibnamefont
  {Polnarev}}\ and\ \bibinfo {author} {\bibfnamefont {R.}~\bibnamefont
  {Zembowicz}},\ }\href {\doibase 10.1103/PhysRevD.43.1106} {\bibfield
  {journal} {\bibinfo  {journal} {Phys. Rev. D}\ }\textbf {\bibinfo {volume}
  {43}},\ \bibinfo {pages} {1106} (\bibinfo {year} {1991})}\BibitemShut
  {NoStop}%
\bibitem [{\citenamefont {MacGibbon}\ \emph {et~al.}(1998)\citenamefont
  {MacGibbon}, \citenamefont {Brandenberger},\ and\ \citenamefont
  {Wichoski}}]{MacGibbon:1997pu}%
  \BibitemOpen
  \bibfield  {author} {\bibinfo {author} {\bibfnamefont {J.~H.}\ \bibnamefont
  {MacGibbon}}, \bibinfo {author} {\bibfnamefont {R.~H.}\ \bibnamefont
  {Brandenberger}}, \ and\ \bibinfo {author} {\bibfnamefont {U.~F.}\
  \bibnamefont {Wichoski}},\ }\href {\doibase 10.1103/PhysRevD.57.2158}
  {\bibfield  {journal} {\bibinfo  {journal} {Phys. Rev. D}\ }\textbf {\bibinfo
  {volume} {57}},\ \bibinfo {pages} {2158} (\bibinfo {year} {1998})},\ \Eprint
  {http://arxiv.org/abs/astro-ph/9707146} {arXiv:astro-ph/9707146} \BibitemShut
  {NoStop}%
\bibitem [{\citenamefont {Albrecht}\ and\ \citenamefont
  {Turok}(1989)}]{PhysRevD.40.973}%
  \BibitemOpen
  \bibfield  {author} {\bibinfo {author} {\bibfnamefont {A.}~\bibnamefont
  {Albrecht}}\ and\ \bibinfo {author} {\bibfnamefont {N.}~\bibnamefont
  {Turok}},\ }\href {\doibase 10.1103/PhysRevD.40.973} {\bibfield  {journal}
  {\bibinfo  {journal} {Phys. Rev. D}\ }\textbf {\bibinfo {volume} {40}},\
  \bibinfo {pages} {973} (\bibinfo {year} {1989})}\BibitemShut {NoStop}%
\bibitem [{\citenamefont {Allen}\ and\ \citenamefont
  {Shellard}(1990)}]{PhysRevLett.64.119}%
  \BibitemOpen
  \bibfield  {author} {\bibinfo {author} {\bibfnamefont {B.}~\bibnamefont
  {Allen}}\ and\ \bibinfo {author} {\bibfnamefont {E.~P.~S.}\ \bibnamefont
  {Shellard}},\ }\href {\doibase 10.1103/PhysRevLett.64.119} {\bibfield
  {journal} {\bibinfo  {journal} {Phys. Rev. Lett.}\ }\textbf {\bibinfo
  {volume} {64}},\ \bibinfo {pages} {119} (\bibinfo {year} {1990})}\BibitemShut
  {NoStop}%
\bibitem [{\citenamefont {Caldwell}\ and\ \citenamefont
  {Allen}(1992)}]{Caldwell:1991jj}%
  \BibitemOpen
  \bibfield  {author} {\bibinfo {author} {\bibfnamefont {R.~R.}\ \bibnamefont
  {Caldwell}}\ and\ \bibinfo {author} {\bibfnamefont {B.}~\bibnamefont
  {Allen}},\ }\href {\doibase 10.1103/PhysRevD.45.3447} {\bibfield  {journal}
  {\bibinfo  {journal} {Phys. Rev. D}\ }\textbf {\bibinfo {volume} {45}},\
  \bibinfo {pages} {3447} (\bibinfo {year} {1992})}\BibitemShut {NoStop}%
\bibitem [{\citenamefont {MacGibbon}\ and\ \citenamefont
  {Webber}(1990)}]{MacGibbon:1990zk}%
  \BibitemOpen
  \bibfield  {author} {\bibinfo {author} {\bibfnamefont {J.~H.}\ \bibnamefont
  {MacGibbon}}\ and\ \bibinfo {author} {\bibfnamefont {B.~R.}\ \bibnamefont
  {Webber}},\ }\href {\doibase 10.1103/PhysRevD.41.3052} {\bibfield  {journal}
  {\bibinfo  {journal} {Phys. Rev. D}\ }\textbf {\bibinfo {volume} {41}},\
  \bibinfo {pages} {3052} (\bibinfo {year} {1990})}\BibitemShut {NoStop}%
\bibitem [{\citenamefont {MacGibbon}(1991)}]{MacGibbon:1991tj}%
  \BibitemOpen
  \bibfield  {author} {\bibinfo {author} {\bibfnamefont {J.~H.}\ \bibnamefont
  {MacGibbon}},\ }\href {\doibase 10.1103/PhysRevD.44.376} {\bibfield
  {journal} {\bibinfo  {journal} {Phys. Rev. D}\ }\textbf {\bibinfo {volume}
  {44}},\ \bibinfo {pages} {376} (\bibinfo {year} {1991})}\BibitemShut
  {NoStop}%
\bibitem [{\citenamefont {MacGibbon}\ and\ \citenamefont
  {Carr}(1991)}]{MacGibbon:1991vc}%
  \BibitemOpen
  \bibfield  {author} {\bibinfo {author} {\bibfnamefont {J.~H.}\ \bibnamefont
  {MacGibbon}}\ and\ \bibinfo {author} {\bibfnamefont {B.~J.}\ \bibnamefont
  {Carr}},\ }\href {\doibase 10.1086/169909} {\bibfield  {journal} {\bibinfo
  {journal} {Astrophys. J.}\ }\textbf {\bibinfo {volume} {371}},\ \bibinfo
  {pages} {447} (\bibinfo {year} {1991})}\BibitemShut {NoStop}%
\bibitem [{\citenamefont {{Lacey}}\ and\ \citenamefont
  {{Ostriker}}(1985)}]{1985ApJ...299..633L}%
  \BibitemOpen
  \bibfield  {author} {\bibinfo {author} {\bibfnamefont {C.}~\bibnamefont
  {{Lacey}}}\ and\ \bibinfo {author} {\bibfnamefont {J.}~\bibnamefont
  {{Ostriker}}},\ }\href {\doibase 10.1086/163729} {\bibfield  {journal}
  {\bibinfo  {journal} {Astrophys.~J.}\ }\textbf {\bibinfo {volume} {299}},\
  \bibinfo {pages} {633} (\bibinfo {year} {1985})}\BibitemShut {NoStop}%
\bibitem [{\citenamefont {{Carr}}\ and\ \citenamefont
  {{Lacey}}(1987)}]{1987ApJ...316...23C}%
  \BibitemOpen
  \bibfield  {author} {\bibinfo {author} {\bibfnamefont {B.}~\bibnamefont
  {{Carr}}}\ and\ \bibinfo {author} {\bibfnamefont {C.}~\bibnamefont
  {{Lacey}}},\ }\href {\doibase 10.1086/165176} {\bibfield  {journal} {\bibinfo
   {journal} {Astrophys.~J.}\ }\textbf {\bibinfo {volume} {316}},\ \bibinfo
  {pages} {23} (\bibinfo {year} {1987})}\BibitemShut {NoStop}%
\bibitem [{\citenamefont {{Lacey}}(1991)}]{1991dodg.conf..257L}%
  \BibitemOpen
  \bibfield  {author} {\bibinfo {author} {\bibfnamefont {C.~G.}\ \bibnamefont
  {{Lacey}}},\ }in\ \href@noop {} {\emph {\bibinfo {booktitle} {Dynamics of
  Disc Galaxies}}},\ \bibinfo {editor} {edited by\ \bibinfo {editor}
  {\bibfnamefont {B.}~\bibnamefont {{Sundelius}}}}\ (\bibinfo {year} {1991})\
  p.\ \bibinfo {pages} {257}\BibitemShut {NoStop}%
\bibitem [{\citenamefont {{Carr}}\ and\ \citenamefont
  {{Silk}}(1983)}]{1983ApJ...268....1C}%
  \BibitemOpen
  \bibfield  {author} {\bibinfo {author} {\bibfnamefont {B.}~\bibnamefont
  {{Carr}}}\ and\ \bibinfo {author} {\bibfnamefont {J.}~\bibnamefont
  {{Silk}}},\ }\href {\doibase 10.1086/160924} {\bibfield  {journal} {\bibinfo
  {journal} {Astrophys. J.}\ }\textbf {\bibinfo {volume} {268}},\ \bibinfo
  {pages} {1} (\bibinfo {year} {1983})}\BibitemShut {NoStop}%
\bibitem [{\citenamefont {{Freese}}\ \emph {et~al.}(1983)\citenamefont
  {{Freese}}, \citenamefont {{Price}},\ and\ \citenamefont
  {{Schramm}}}]{1983ApJ...275..405F}%
  \BibitemOpen
  \bibfield  {author} {\bibinfo {author} {\bibfnamefont {K.}~\bibnamefont
  {{Freese}}}, \bibinfo {author} {\bibfnamefont {R.}~\bibnamefont {{Price}}}, \
  and\ \bibinfo {author} {\bibfnamefont {D.}~\bibnamefont {{Schramm}}},\ }\href
  {\doibase 10.1086/161542} {\bibfield  {journal} {\bibinfo  {journal}
  {Astrophys. J.}\ }\textbf {\bibinfo {volume} {275}},\ \bibinfo {pages} {405}
  (\bibinfo {year} {1983})}\BibitemShut {NoStop}%
\bibitem [{\citenamefont {{Carr}}\ and\ \citenamefont
  {{Rees}}(1984)}]{1984MNRAS.206..801C}%
  \BibitemOpen
  \bibfield  {author} {\bibinfo {author} {\bibfnamefont {B.}~\bibnamefont
  {{Carr}}}\ and\ \bibinfo {author} {\bibfnamefont {M.~J.}\ \bibnamefont
  {{Rees}}},\ }\href {\doibase 10.1093/mnras/206.4.801} {\bibfield  {journal}
  {\bibinfo  {journal} {Mon. Not. Roy. Astron. Soc.}\ }\textbf {\bibinfo
  {volume} {206}},\ \bibinfo {pages} {801} (\bibinfo {year}
  {1984})}\BibitemShut {NoStop}%
\bibitem [{\citenamefont {{Bond}}\ and\ \citenamefont
  {{Carr}}(1984)}]{1984MNRAS.207..585B}%
  \BibitemOpen
  \bibfield  {author} {\bibinfo {author} {\bibfnamefont {R.}~\bibnamefont
  {{Bond}}}\ and\ \bibinfo {author} {\bibfnamefont {B.}~\bibnamefont
  {{Carr}}},\ }\href {\doibase 10.1093/mnras/207.3.585} {\bibfield  {journal}
  {\bibinfo  {journal} {Mon.~Not.~Roy. astron.~Soc.}\ }\textbf {\bibinfo
  {volume} {207}},\ \bibinfo {pages} {585} (\bibinfo {year}
  {1984})}\BibitemShut {NoStop}%
\bibitem [{\citenamefont {Hawkins}(1993)}]{Hawkins:1993yud}%
  \BibitemOpen
  \bibfield  {author} {\bibinfo {author} {\bibfnamefont {M.~R.~S.}\
  \bibnamefont {Hawkins}},\ }\href {\doibase 10.1038/366242a0} {\bibfield
  {journal} {\bibinfo  {journal} {Nature}\ }\textbf {\bibinfo {volume} {366}},\
  \bibinfo {pages} {242} (\bibinfo {year} {1993})}\BibitemShut {NoStop}%
\bibitem [{\citenamefont {Alcock}\ \emph {et~al.}(1997)\citenamefont {Alcock}
  \emph {et~al.}}]{MACHO:1996qam}%
  \BibitemOpen
  \bibfield  {author} {\bibinfo {author} {\bibfnamefont {C.}~\bibnamefont
  {Alcock}} \emph {et~al.} (\bibinfo {collaboration} {MACHO}),\ }\href
  {\doibase 10.1086/304535} {\bibfield  {journal} {\bibinfo  {journal}
  {Astrophys. J.}\ }\textbf {\bibinfo {volume} {486}},\ \bibinfo {pages} {697}
  (\bibinfo {year} {1997})},\ \Eprint {http://arxiv.org/abs/astro-ph/9606165}
  {arXiv:astro-ph/9606165} \BibitemShut {NoStop}%
\bibitem [{\citenamefont {Jedamzik}(1998)}]{Jedamzik:1998hc}%
  \BibitemOpen
  \bibfield  {author} {\bibinfo {author} {\bibfnamefont {K.}~\bibnamefont
  {Jedamzik}},\ }\bibfield  {booktitle} {\emph {\bibinfo {booktitle} {Sources
  and detection of dark matter in the universe.}},\ }\href {\doibase
  10.1016/S0370-1573(98)00067-2} {\bibfield  {journal} {\bibinfo  {journal}
  {Phys.~Rept.}\ }\textbf {\bibinfo {volume} {307}},\ \bibinfo {pages} {155}
  (\bibinfo {year} {1998})},\ \Eprint {http://arxiv.org/abs/astro-ph/9805147}
  {arXiv:astro-ph/9805147 [astro-ph]} \BibitemShut {NoStop}%
\bibitem [{\citenamefont {Alcock}\ \emph
  {et~al.}(2000{\natexlab{a}})\citenamefont {Alcock} \emph
  {et~al.}}]{Alcock:2000ph}%
  \BibitemOpen
  \bibfield  {author} {\bibinfo {author} {\bibfnamefont {C.}~\bibnamefont
  {Alcock}} \emph {et~al.} (\bibinfo {collaboration} {MACHO}),\ }\href
  {\doibase 10.1086/309512} {\bibfield  {journal} {\bibinfo  {journal}
  {Astrophys.~J.}\ }\textbf {\bibinfo {volume} {542}},\ \bibinfo {pages} {281}
  (\bibinfo {year} {2000}{\natexlab{a}})},\ \Eprint
  {http://arxiv.org/abs/astro-ph/0001272} {arXiv:astro-ph/0001272} \BibitemShut
  {NoStop}%
\bibitem [{\citenamefont {Tisserand}\ \emph
  {et~al.}(2007{\natexlab{a}})\citenamefont {Tisserand} \emph
  {et~al.}}]{Tisserand:2006zx}%
  \BibitemOpen
  \bibfield  {author} {\bibinfo {author} {\bibfnamefont {P.}~\bibnamefont
  {Tisserand}} \emph {et~al.} (\bibinfo {collaboration} {EROS-2}),\ }\href
  {\doibase 10.1051/0004-6361:20066017} {\bibfield  {journal} {\bibinfo
  {journal} {Astron. astrophys.}\ }\textbf {\bibinfo {volume} {469}},\ \bibinfo
  {pages} {387} (\bibinfo {year} {2007}{\natexlab{a}})},\ \Eprint
  {http://arxiv.org/abs/astro-ph/0607207} {arXiv:astro-ph/0607207} \BibitemShut
  {NoStop}%
\bibitem [{\citenamefont {Carr}\ \emph {et~al.}(2010)\citenamefont {Carr},
  \citenamefont {Kohri}, \citenamefont {Sendouda},\ and\ \citenamefont
  {Yokoyama}}]{Carr:2009jm}%
  \BibitemOpen
  \bibfield  {author} {\bibinfo {author} {\bibfnamefont {B.}~\bibnamefont
  {Carr}}, \bibinfo {author} {\bibfnamefont {K.}~\bibnamefont {Kohri}},
  \bibinfo {author} {\bibfnamefont {Y.}~\bibnamefont {Sendouda}}, \ and\
  \bibinfo {author} {\bibfnamefont {J.}~\bibnamefont {Yokoyama}},\ }\href
  {\doibase 10.1103/PhysRevD.81.104019} {\bibfield  {journal} {\bibinfo
  {journal} {Phys. Rev. D}\ }\textbf {\bibinfo {volume} {81}},\ \bibinfo
  {pages} {104019} (\bibinfo {year} {2010})},\ \Eprint
  {http://arxiv.org/abs/0912.5297} {arXiv:0912.5297 [astro-ph.CO]} \BibitemShut
  {NoStop}%
\bibitem [{\citenamefont {Carr}\ \emph
  {et~al.}(2016{\natexlab{a}})\citenamefont {Carr}, \citenamefont
  {K{\"u}hnel},\ and\ \citenamefont {Sandstad}}]{Carr:2016drx}%
  \BibitemOpen
  \bibfield  {author} {\bibinfo {author} {\bibfnamefont {B.}~\bibnamefont
  {Carr}}, \bibinfo {author} {\bibfnamefont {F.}~\bibnamefont {K{\"u}hnel}}, \
  and\ \bibinfo {author} {\bibfnamefont {M.}~\bibnamefont {Sandstad}},\ }\href
  {\doibase 10.1103/PhysRevD.94.083504} {\bibfield  {journal} {\bibinfo
  {journal} {Phys.~Rev.}\ }\textbf {\bibinfo {volume} {D94}},\ \bibinfo {pages}
  {083504} (\bibinfo {year} {2016}{\natexlab{a}})}\BibitemShut {NoStop}%
\bibitem [{\citenamefont {Bullock}\ and\ \citenamefont
  {Primack}(1997)}]{Bullock:1996at}%
  \BibitemOpen
  \bibfield  {author} {\bibinfo {author} {\bibfnamefont {J.~S.}\ \bibnamefont
  {Bullock}}\ and\ \bibinfo {author} {\bibfnamefont {J.~R.}\ \bibnamefont
  {Primack}},\ }\href {\doibase 10.1103/PhysRevD.55.7423} {\bibfield  {journal}
  {\bibinfo  {journal} {Phys. Rev. D}\ }\textbf {\bibinfo {volume} {55}},\
  \bibinfo {pages} {7423} (\bibinfo {year} {1997})},\ \Eprint
  {http://arxiv.org/abs/astro-ph/9611106} {arXiv:astro-ph/9611106} \BibitemShut
  {NoStop}%
\bibitem [{\citenamefont {Ivanov}(1998)}]{Ivanov:1997ia}%
  \BibitemOpen
  \bibfield  {author} {\bibinfo {author} {\bibfnamefont {P.}~\bibnamefont
  {Ivanov}},\ }\href {\doibase 10.1103/PhysRevD.57.7145} {\bibfield  {journal}
  {\bibinfo  {journal} {Phys. Rev. D}\ }\textbf {\bibinfo {volume} {57}},\
  \bibinfo {pages} {7145} (\bibinfo {year} {1998})},\ \Eprint
  {http://arxiv.org/abs/astro-ph/9708224} {arXiv:astro-ph/9708224} \BibitemShut
  {NoStop}%
\bibitem [{\citenamefont {Young}\ and\ \citenamefont
  {Byrnes}(2015)}]{Young:2015kda}%
  \BibitemOpen
  \bibfield  {author} {\bibinfo {author} {\bibfnamefont {S.}~\bibnamefont
  {Young}}\ and\ \bibinfo {author} {\bibfnamefont {C.}~\bibnamefont {Byrnes}},\
  }\href {\doibase 10.1088/1475-7516/2015/04/034} {\bibfield  {journal}
  {\bibinfo  {journal} {JCAP}\ }\textbf {\bibinfo {volume} {1504}},\ \bibinfo
  {pages} {034} (\bibinfo {year} {2015})},\ \Eprint
  {http://arxiv.org/abs/1503.01505} {arXiv:1503.01505 [astro-ph.CO]}
  \BibitemShut {NoStop}%
\bibitem [{\citenamefont {Kohri}\ \emph {et~al.}(2008)\citenamefont {Kohri},
  \citenamefont {Lyth},\ and\ \citenamefont {Melchiorri}}]{Kohri:2007qn}%
  \BibitemOpen
  \bibfield  {author} {\bibinfo {author} {\bibfnamefont {K.}~\bibnamefont
  {Kohri}}, \bibinfo {author} {\bibfnamefont {D.~H.}\ \bibnamefont {Lyth}}, \
  and\ \bibinfo {author} {\bibfnamefont {A.}~\bibnamefont {Melchiorri}},\
  }\href {\doibase 10.1088/1475-7516/2008/04/038} {\bibfield  {journal}
  {\bibinfo  {journal} {JCAP}\ }\textbf {\bibinfo {volume} {04}},\ \bibinfo
  {pages} {038} (\bibinfo {year} {2008})},\ \Eprint
  {http://arxiv.org/abs/0711.5006} {arXiv:0711.5006 [hep-ph]} \BibitemShut
  {NoStop}%
\bibitem [{\citenamefont {Kawasaki}\ \emph {et~al.}(1998)\citenamefont
  {Kawasaki}, \citenamefont {Sugiyama},\ and\ \citenamefont
  {Yanagida}}]{Kawasaki:1997ju}%
  \BibitemOpen
  \bibfield  {author} {\bibinfo {author} {\bibfnamefont {M.}~\bibnamefont
  {Kawasaki}}, \bibinfo {author} {\bibfnamefont {N.}~\bibnamefont {Sugiyama}},
  \ and\ \bibinfo {author} {\bibfnamefont {T.}~\bibnamefont {Yanagida}},\
  }\href {\doibase 10.1103/PhysRevD.57.6050} {\bibfield  {journal} {\bibinfo
  {journal} {Phys. Rev. D}\ }\textbf {\bibinfo {volume} {57}},\ \bibinfo
  {pages} {6050} (\bibinfo {year} {1998})},\ \Eprint
  {http://arxiv.org/abs/hep-ph/9710259} {arXiv:hep-ph/9710259} \BibitemShut
  {NoStop}%
\bibitem [{\citenamefont {Green}\ and\ \citenamefont
  {Malik}(2001)}]{Green:2000he}%
  \BibitemOpen
  \bibfield  {author} {\bibinfo {author} {\bibfnamefont {A.~M.}\ \bibnamefont
  {Green}}\ and\ \bibinfo {author} {\bibfnamefont {K.~A.}\ \bibnamefont
  {Malik}},\ }\href {\doibase 10.1103/PhysRevD.64.021301} {\bibfield  {journal}
  {\bibinfo  {journal} {Phys. Rev. D}\ }\textbf {\bibinfo {volume} {64}},\
  \bibinfo {pages} {021301} (\bibinfo {year} {2001})},\ \Eprint
  {http://arxiv.org/abs/hep-ph/0008113} {arXiv:hep-ph/0008113} \BibitemShut
  {NoStop}%
\bibitem [{\citenamefont {Bassett}\ and\ \citenamefont
  {Tsujikawa}(2001)}]{Bassett:2000ha}%
  \BibitemOpen
  \bibfield  {author} {\bibinfo {author} {\bibfnamefont {B.~A.}\ \bibnamefont
  {Bassett}}\ and\ \bibinfo {author} {\bibfnamefont {S.}~\bibnamefont
  {Tsujikawa}},\ }\href {\doibase 10.1103/PhysRevD.63.123503} {\bibfield
  {journal} {\bibinfo  {journal} {Phys. Rev. D}\ }\textbf {\bibinfo {volume}
  {63}},\ \bibinfo {pages} {123503} (\bibinfo {year} {2001})},\ \Eprint
  {http://arxiv.org/abs/hep-ph/0008328} {arXiv:hep-ph/0008328} \BibitemShut
  {NoStop}%
\bibitem [{\citenamefont {Ananda}\ \emph {et~al.}(2007)\citenamefont {Ananda},
  \citenamefont {Clarkson},\ and\ \citenamefont {Wands}}]{Ananda:2006af}%
  \BibitemOpen
  \bibfield  {author} {\bibinfo {author} {\bibfnamefont {K.~N.}\ \bibnamefont
  {Ananda}}, \bibinfo {author} {\bibfnamefont {C.}~\bibnamefont {Clarkson}}, \
  and\ \bibinfo {author} {\bibfnamefont {D.}~\bibnamefont {Wands}},\ }\href
  {\doibase 10.1103/PhysRevD.75.123518} {\bibfield  {journal} {\bibinfo
  {journal} {Phys. Rev. D}\ }\textbf {\bibinfo {volume} {75}},\ \bibinfo
  {pages} {123518} (\bibinfo {year} {2007})},\ \Eprint
  {http://arxiv.org/abs/gr-qc/0612013} {arXiv:gr-qc/0612013} \BibitemShut
  {NoStop}%
\bibitem [{\citenamefont {Baumann}\ \emph {et~al.}(2007)\citenamefont
  {Baumann}, \citenamefont {Steinhardt}, \citenamefont {Takahashi},\ and\
  \citenamefont {Ichiki}}]{Baumann:2007zm}%
  \BibitemOpen
  \bibfield  {author} {\bibinfo {author} {\bibfnamefont {D.}~\bibnamefont
  {Baumann}}, \bibinfo {author} {\bibfnamefont {P.~J.}\ \bibnamefont
  {Steinhardt}}, \bibinfo {author} {\bibfnamefont {K.}~\bibnamefont
  {Takahashi}}, \ and\ \bibinfo {author} {\bibfnamefont {K.}~\bibnamefont
  {Ichiki}},\ }\href {\doibase 10.1103/PhysRevD.76.084019} {\bibfield
  {journal} {\bibinfo  {journal} {Phys. Rev. D}\ }\textbf {\bibinfo {volume}
  {76}},\ \bibinfo {pages} {084019} (\bibinfo {year} {2007})},\ \Eprint
  {http://arxiv.org/abs/hep-th/0703290} {arXiv:hep-th/0703290} \BibitemShut
  {NoStop}%
\bibitem [{\citenamefont {Saito}\ and\ \citenamefont
  {Yokoyama}(2009)}]{Saito:2008jc}%
  \BibitemOpen
  \bibfield  {author} {\bibinfo {author} {\bibfnamefont {R.}~\bibnamefont
  {Saito}}\ and\ \bibinfo {author} {\bibfnamefont {J.}~\bibnamefont
  {Yokoyama}},\ }\href {\doibase 10.1103/PhysRevLett.102.161101} {\bibfield
  {journal} {\bibinfo  {journal} {Phys. Rev. Lett.}\ }\textbf {\bibinfo
  {volume} {102}},\ \bibinfo {pages} {161101} (\bibinfo {year} {2009})},\
  \bibinfo {note} {[Erratum: Phys.Rev.Lett. 107, 069901 (2011)]},\ \Eprint
  {http://arxiv.org/abs/0812.4339} {arXiv:0812.4339 [astro-ph]} \BibitemShut
  {NoStop}%
\bibitem [{\citenamefont {Green}\ \emph {et~al.}(2004)\citenamefont {Green},
  \citenamefont {Liddle}, \citenamefont {Malik},\ and\ \citenamefont
  {Sasaki}}]{Green:2004wb}%
  \BibitemOpen
  \bibfield  {author} {\bibinfo {author} {\bibfnamefont {A.~M.}\ \bibnamefont
  {Green}}, \bibinfo {author} {\bibfnamefont {A.~R.}\ \bibnamefont {Liddle}},
  \bibinfo {author} {\bibfnamefont {K.~A.}\ \bibnamefont {Malik}}, \ and\
  \bibinfo {author} {\bibfnamefont {M.}~\bibnamefont {Sasaki}},\ }\href
  {\doibase 10.1103/PhysRevD.70.041502} {\bibfield  {journal} {\bibinfo
  {journal} {Phys. Rev. D}\ }\textbf {\bibinfo {volume} {70}},\ \bibinfo
  {pages} {041502} (\bibinfo {year} {2004})},\ \Eprint
  {http://arxiv.org/abs/astro-ph/0403181} {arXiv:astro-ph/0403181} \BibitemShut
  {NoStop}%
\bibitem [{\citenamefont {Bardeen}\ \emph {et~al.}(1986)\citenamefont
  {Bardeen}, \citenamefont {Bond}, \citenamefont {Kaiser},\ and\ \citenamefont
  {Szalay}}]{Bardeen:1985tr}%
  \BibitemOpen
  \bibfield  {author} {\bibinfo {author} {\bibfnamefont {J.~M.}\ \bibnamefont
  {Bardeen}}, \bibinfo {author} {\bibfnamefont {J.~R.}\ \bibnamefont {Bond}},
  \bibinfo {author} {\bibfnamefont {N.}~\bibnamefont {Kaiser}}, \ and\ \bibinfo
  {author} {\bibfnamefont {A.~S.}\ \bibnamefont {Szalay}},\ }\href {\doibase
  10.1086/164143} {\bibfield  {journal} {\bibinfo  {journal} {Astrophys. J.}\
  }\textbf {\bibinfo {volume} {304}},\ \bibinfo {pages} {15} (\bibinfo {year}
  {1986})}\BibitemShut {NoStop}%
\bibitem [{\citenamefont {Choptuik}(1993)}]{Choptuik:1992jv}%
  \BibitemOpen
  \bibfield  {author} {\bibinfo {author} {\bibfnamefont {M.~W.}\ \bibnamefont
  {Choptuik}},\ }\href {\doibase 10.1103/PhysRevLett.70.9} {\bibfield
  {journal} {\bibinfo  {journal} {Phys. Rev. Lett.}\ }\textbf {\bibinfo
  {volume} {70}},\ \bibinfo {pages} {9} (\bibinfo {year} {1993})}\BibitemShut
  {NoStop}%
\bibitem [{\citenamefont {Niemeyer}\ and\ \citenamefont
  {Jedamzik}(1998)}]{Niemeyer:1997mt}%
  \BibitemOpen
  \bibfield  {author} {\bibinfo {author} {\bibfnamefont {J.~C.}\ \bibnamefont
  {Niemeyer}}\ and\ \bibinfo {author} {\bibfnamefont {K.}~\bibnamefont
  {Jedamzik}},\ }\href {\doibase 10.1103/PhysRevLett.80.5481} {\bibfield
  {journal} {\bibinfo  {journal} {Phys. Rev. Lett.}\ }\textbf {\bibinfo
  {volume} {80}},\ \bibinfo {pages} {5481} (\bibinfo {year} {1998})},\ \Eprint
  {http://arxiv.org/abs/astro-ph/9709072} {arXiv:astro-ph/9709072} \BibitemShut
  {NoStop}%
\bibitem [{\citenamefont {Niemeyer}\ and\ \citenamefont
  {Jedamzik}(1999)}]{Niemeyer:1999ak}%
  \BibitemOpen
  \bibfield  {author} {\bibinfo {author} {\bibfnamefont {J.~C.}\ \bibnamefont
  {Niemeyer}}\ and\ \bibinfo {author} {\bibfnamefont {K.}~\bibnamefont
  {Jedamzik}},\ }\href {\doibase 10.1103/PhysRevD.59.124013} {\bibfield
  {journal} {\bibinfo  {journal} {Phys. Rev. D}\ }\textbf {\bibinfo {volume}
  {59}},\ \bibinfo {pages} {124013} (\bibinfo {year} {1999})},\ \Eprint
  {http://arxiv.org/abs/astro-ph/9901292} {arXiv:astro-ph/9901292} \BibitemShut
  {NoStop}%
\bibitem [{\citenamefont {Musco}\ \emph {et~al.}(2005)\citenamefont {Musco},
  \citenamefont {Miller},\ and\ \citenamefont {Rezzolla}}]{Musco:2004ak}%
  \BibitemOpen
  \bibfield  {author} {\bibinfo {author} {\bibfnamefont {I.}~\bibnamefont
  {Musco}}, \bibinfo {author} {\bibfnamefont {J.~C.}\ \bibnamefont {Miller}}, \
  and\ \bibinfo {author} {\bibfnamefont {L.}~\bibnamefont {Rezzolla}},\ }\href
  {\doibase 10.1088/0264-9381/22/7/013} {\bibfield  {journal} {\bibinfo
  {journal} {Class. Quant. Grav.}\ }\textbf {\bibinfo {volume} {22}},\ \bibinfo
  {pages} {1405} (\bibinfo {year} {2005})},\ \Eprint
  {http://arxiv.org/abs/gr-qc/0412063} {arXiv:gr-qc/0412063} \BibitemShut
  {NoStop}%
\bibitem [{\citenamefont {Musco}\ \emph {et~al.}(2009)\citenamefont {Musco},
  \citenamefont {Miller},\ and\ \citenamefont {Polnarev}}]{Musco:2008hv}%
  \BibitemOpen
  \bibfield  {author} {\bibinfo {author} {\bibfnamefont {I.}~\bibnamefont
  {Musco}}, \bibinfo {author} {\bibfnamefont {J.~C.}\ \bibnamefont {Miller}}, \
  and\ \bibinfo {author} {\bibfnamefont {A.~G.}\ \bibnamefont {Polnarev}},\
  }\href {\doibase 10.1088/0264-9381/26/23/235001} {\bibfield  {journal}
  {\bibinfo  {journal} {Class. Quant. Grav.}\ }\textbf {\bibinfo {volume}
  {26}},\ \bibinfo {pages} {235001} (\bibinfo {year} {2009})},\ \Eprint
  {http://arxiv.org/abs/0811.1452} {arXiv:0811.1452 [gr-qc]} \BibitemShut
  {NoStop}%
\bibitem [{\citenamefont {Musco}\ and\ \citenamefont
  {Miller}(2013)}]{Musco:2012au}%
  \BibitemOpen
  \bibfield  {author} {\bibinfo {author} {\bibfnamefont {I.}~\bibnamefont
  {Musco}}\ and\ \bibinfo {author} {\bibfnamefont {J.~C.}\ \bibnamefont
  {Miller}},\ }\href {\doibase 10.1088/0264-9381/30/14/145009} {\bibfield
  {journal} {\bibinfo  {journal} {Class. Quant. Grav.}\ }\textbf {\bibinfo
  {volume} {30}},\ \bibinfo {pages} {145009} (\bibinfo {year} {2013})},\
  \Eprint {http://arxiv.org/abs/1201.2379} {arXiv:1201.2379 [gr-qc]}
  \BibitemShut {NoStop}%
\bibitem [{\citenamefont {Jedamzik}(1997)}]{Jedamzik:1996mr}%
  \BibitemOpen
  \bibfield  {author} {\bibinfo {author} {\bibfnamefont {K.}~\bibnamefont
  {Jedamzik}},\ }\href {\doibase 10.1103/PhysRevD.55.R5871} {\bibfield
  {journal} {\bibinfo  {journal} {Phys. Rev. D}\ }\textbf {\bibinfo {volume}
  {55}},\ \bibinfo {pages} {5871} (\bibinfo {year} {1997})},\ \Eprint
  {http://arxiv.org/abs/astro-ph/9605152} {arXiv:astro-ph/9605152} \BibitemShut
  {NoStop}%
\bibitem [{\citenamefont {Nakama}\ \emph {et~al.}(2014)\citenamefont {Nakama},
  \citenamefont {Harada}, \citenamefont {Polnarev},\ and\ \citenamefont
  {Yokoyama}}]{Nakama:2013ica}%
  \BibitemOpen
  \bibfield  {author} {\bibinfo {author} {\bibfnamefont {T.}~\bibnamefont
  {Nakama}}, \bibinfo {author} {\bibfnamefont {T.}~\bibnamefont {Harada}},
  \bibinfo {author} {\bibfnamefont {A.~G.}\ \bibnamefont {Polnarev}}, \ and\
  \bibinfo {author} {\bibfnamefont {J.}~\bibnamefont {Yokoyama}},\ }\href
  {\doibase 10.1088/1475-7516/2014/01/037} {\bibfield  {journal} {\bibinfo
  {journal} {JCAP}\ }\textbf {\bibinfo {volume} {01}},\ \bibinfo {pages} {037}
  (\bibinfo {year} {2014})},\ \Eprint {http://arxiv.org/abs/1310.3007}
  {arXiv:1310.3007 [gr-qc]} \BibitemShut {NoStop}%
\bibitem [{\citenamefont {Harada}\ \emph {et~al.}(2016)\citenamefont {Harada},
  \citenamefont {Yoo}, \citenamefont {Kohri}, \citenamefont {Nakao},\ and\
  \citenamefont {Jhingan}}]{Harada:2016mhb}%
  \BibitemOpen
  \bibfield  {author} {\bibinfo {author} {\bibfnamefont {T.}~\bibnamefont
  {Harada}}, \bibinfo {author} {\bibfnamefont {C.-M.}\ \bibnamefont {Yoo}},
  \bibinfo {author} {\bibfnamefont {K.}~\bibnamefont {Kohri}}, \bibinfo
  {author} {\bibfnamefont {K.-i.}\ \bibnamefont {Nakao}}, \ and\ \bibinfo
  {author} {\bibfnamefont {S.}~\bibnamefont {Jhingan}},\ }\href {\doibase
  10.3847/1538-4357/833/1/61} {\bibfield  {journal} {\bibinfo  {journal}
  {Astrophys. J.}\ }\textbf {\bibinfo {volume} {833}},\ \bibinfo {pages} {61}
  (\bibinfo {year} {2016})},\ \Eprint {http://arxiv.org/abs/1609.01588}
  {arXiv:1609.01588 [astro-ph.CO]} \BibitemShut {NoStop}%
\bibitem [{\citenamefont {Harada}\ \emph {et~al.}(2017)\citenamefont {Harada},
  \citenamefont {Yoo}, \citenamefont {Kohri},\ and\ \citenamefont
  {Nakao}}]{Harada:2017fjm}%
  \BibitemOpen
  \bibfield  {author} {\bibinfo {author} {\bibfnamefont {T.}~\bibnamefont
  {Harada}}, \bibinfo {author} {\bibfnamefont {C.-M.}\ \bibnamefont {Yoo}},
  \bibinfo {author} {\bibfnamefont {K.}~\bibnamefont {Kohri}}, \ and\ \bibinfo
  {author} {\bibfnamefont {K.-I.}\ \bibnamefont {Nakao}},\ }\href {\doibase
  10.1103/PhysRevD.96.083517} {\bibfield  {journal} {\bibinfo  {journal} {Phys.
  Rev. D}\ }\textbf {\bibinfo {volume} {96}},\ \bibinfo {pages} {083517}
  (\bibinfo {year} {2017})},\ \bibinfo {note} {[Erratum: Phys.Rev.D 99, 069904
  (2019)]},\ \Eprint {http://arxiv.org/abs/1707.03595} {arXiv:1707.03595
  [gr-qc]} \BibitemShut {NoStop}%
\bibitem [{\citenamefont {Cline}\ \emph {et~al.}(1997)\citenamefont {Cline},
  \citenamefont {Sanders},\ and\ \citenamefont {Hong}}]{Cline:1996zg}%
  \BibitemOpen
  \bibfield  {author} {\bibinfo {author} {\bibfnamefont {D.~B.}\ \bibnamefont
  {Cline}}, \bibinfo {author} {\bibfnamefont {D.~A.}\ \bibnamefont {Sanders}},
  \ and\ \bibinfo {author} {\bibfnamefont {W.}~\bibnamefont {Hong}},\ }\href
  {\doibase 10.1086/304480} {\bibfield  {journal} {\bibinfo  {journal}
  {Astrophys. J.}\ }\textbf {\bibinfo {volume} {486}},\ \bibinfo {pages} {169}
  (\bibinfo {year} {1997})}\BibitemShut {NoStop}%
\bibitem [{\citenamefont {Heckler}(1997)}]{Heckler:1995qq}%
  \BibitemOpen
  \bibfield  {author} {\bibinfo {author} {\bibfnamefont {A.~F.}\ \bibnamefont
  {Heckler}},\ }\href {\doibase 10.1103/PhysRevD.55.480} {\bibfield  {journal}
  {\bibinfo  {journal} {Phys. Rev. D}\ }\textbf {\bibinfo {volume} {55}},\
  \bibinfo {pages} {480} (\bibinfo {year} {1997})},\ \Eprint
  {http://arxiv.org/abs/astro-ph/9601029} {arXiv:astro-ph/9601029} \BibitemShut
  {NoStop}%
\bibitem [{\citenamefont {MacGibbon}\ \emph {et~al.}(2008)\citenamefont
  {MacGibbon}, \citenamefont {Carr},\ and\ \citenamefont
  {Page}}]{MacGibbon:2007yq}%
  \BibitemOpen
  \bibfield  {author} {\bibinfo {author} {\bibfnamefont {J.~H.}\ \bibnamefont
  {MacGibbon}}, \bibinfo {author} {\bibfnamefont {B.~J.}\ \bibnamefont {Carr}},
  \ and\ \bibinfo {author} {\bibfnamefont {D.~N.}\ \bibnamefont {Page}},\
  }\href {\doibase 10.1103/PhysRevD.78.064043} {\bibfield  {journal} {\bibinfo
  {journal} {Phys. Rev. D}\ }\textbf {\bibinfo {volume} {78}},\ \bibinfo
  {pages} {064043} (\bibinfo {year} {2008})},\ \Eprint
  {http://arxiv.org/abs/0709.2380} {arXiv:0709.2380 [astro-ph]} \BibitemShut
  {NoStop}%
\bibitem [{\citenamefont {{Wright}}(1996)}]{1996ApJ...459..487W}%
  \BibitemOpen
  \bibfield  {author} {\bibinfo {author} {\bibfnamefont {E.~L.}\ \bibnamefont
  {{Wright}}},\ }\href {\doibase 10.1086/176910} {\bibfield  {journal}
  {\bibinfo  {journal} {Astrophys. J.}\ }\textbf {\bibinfo {volume} {459}},\
  \bibinfo {pages} {487} (\bibinfo {year} {1996})},\ \Eprint
  {http://arxiv.org/abs/astro-ph/9509074} {astro-ph/9509074} \BibitemShut
  {NoStop}%
\bibitem [{\citenamefont {Lehoucq}\ \emph {et~al.}(2009)\citenamefont
  {Lehoucq}, \citenamefont {Cass\'e}, \citenamefont {Casandjian},\ and\
  \citenamefont {Grenier}}]{Lehoucq:2009ge}%
  \BibitemOpen
  \bibfield  {author} {\bibinfo {author} {\bibfnamefont {R.}~\bibnamefont
  {Lehoucq}}, \bibinfo {author} {\bibfnamefont {M.}~\bibnamefont {Cass\'e}},
  \bibinfo {author} {\bibfnamefont {J.-M.}\ \bibnamefont {Casandjian}}, \ and\
  \bibinfo {author} {\bibfnamefont {I.}~\bibnamefont {Grenier}},\ }\href
  {\doibase 10.1051/0004-6361/200911961} {\bibfield  {journal} {\bibinfo
  {journal} {Astron. Astrophys.}\ }\textbf {\bibinfo {volume} {502}},\ \bibinfo
  {pages} {37} (\bibinfo {year} {2009})},\ \Eprint
  {http://arxiv.org/abs/0906.1648} {arXiv:0906.1648 [astro-ph.HE]} \BibitemShut
  {NoStop}%
\bibitem [{\citenamefont {Carr}\ \emph
  {et~al.}(2016{\natexlab{b}})\citenamefont {Carr}, \citenamefont {Kohri},
  \citenamefont {Sendouda},\ and\ \citenamefont {Yokoyama}}]{Carr:2016hva}%
  \BibitemOpen
  \bibfield  {author} {\bibinfo {author} {\bibfnamefont {B.~J.}\ \bibnamefont
  {Carr}}, \bibinfo {author} {\bibfnamefont {K.}~\bibnamefont {Kohri}},
  \bibinfo {author} {\bibfnamefont {Y.}~\bibnamefont {Sendouda}}, \ and\
  \bibinfo {author} {\bibfnamefont {J.}~\bibnamefont {Yokoyama}},\ }\href
  {\doibase 10.1103/PhysRevD.94.044029} {\bibfield  {journal} {\bibinfo
  {journal} {Phys. Rev. D}\ }\textbf {\bibinfo {volume} {94}},\ \bibinfo
  {pages} {044029} (\bibinfo {year} {2016}{\natexlab{b}})},\ \Eprint
  {http://arxiv.org/abs/1604.05349} {arXiv:1604.05349 [astro-ph.CO]}
  \BibitemShut {NoStop}%
\bibitem [{\citenamefont {Paczynski}(1986)}]{Paczynski:1985jf}%
  \BibitemOpen
  \bibfield  {author} {\bibinfo {author} {\bibfnamefont {B.}~\bibnamefont
  {Paczynski}},\ }\href {\doibase 10.1086/164140} {\bibfield  {journal}
  {\bibinfo  {journal} {Astrophys. J.}\ }\textbf {\bibinfo {volume} {304}},\
  \bibinfo {pages} {1} (\bibinfo {year} {1986})}\BibitemShut {NoStop}%
\bibitem [{\citenamefont {Fields}\ \emph {et~al.}(2000)\citenamefont {Fields},
  \citenamefont {Freese},\ and\ \citenamefont {Graff}}]{Fields:1999ar}%
  \BibitemOpen
  \bibfield  {author} {\bibinfo {author} {\bibfnamefont {B.~D.}\ \bibnamefont
  {Fields}}, \bibinfo {author} {\bibfnamefont {K.}~\bibnamefont {Freese}}, \
  and\ \bibinfo {author} {\bibfnamefont {D.~S.}\ \bibnamefont {Graff}},\ }\href
  {\doibase 10.1086/308727} {\bibfield  {journal} {\bibinfo  {journal}
  {Astrophys. J.}\ }\textbf {\bibinfo {volume} {534}},\ \bibinfo {pages} {265}
  (\bibinfo {year} {2000})},\ \Eprint {http://arxiv.org/abs/astro-ph/9904291}
  {arXiv:astro-ph/9904291} \BibitemShut {NoStop}%
\bibitem [{\citenamefont {Alcock}\ \emph
  {et~al.}(2000{\natexlab{b}})\citenamefont {Alcock} \emph
  {et~al.}}]{MACHO:2000qbb}%
  \BibitemOpen
  \bibfield  {author} {\bibinfo {author} {\bibfnamefont {C.}~\bibnamefont
  {Alcock}} \emph {et~al.} (\bibinfo {collaboration} {MACHO}),\ }\href
  {\doibase 10.1086/309512} {\bibfield  {journal} {\bibinfo  {journal}
  {Astrophys. J.}\ }\textbf {\bibinfo {volume} {542}},\ \bibinfo {pages} {281}
  (\bibinfo {year} {2000}{\natexlab{b}})},\ \Eprint
  {http://arxiv.org/abs/astro-ph/0001272} {arXiv:astro-ph/0001272} \BibitemShut
  {NoStop}%
\bibitem [{\citenamefont {Tisserand}\ \emph
  {et~al.}(2007{\natexlab{b}})\citenamefont {Tisserand} \emph
  {et~al.}}]{EROS-2:2006ryy}%
  \BibitemOpen
  \bibfield  {author} {\bibinfo {author} {\bibfnamefont {P.}~\bibnamefont
  {Tisserand}} \emph {et~al.} (\bibinfo {collaboration} {EROS-2}),\ }\href
  {\doibase 10.1051/0004-6361:20066017} {\bibfield  {journal} {\bibinfo
  {journal} {Astron. Astrophys.}\ }\textbf {\bibinfo {volume} {469}},\ \bibinfo
  {pages} {387} (\bibinfo {year} {2007}{\natexlab{b}})},\ \Eprint
  {http://arxiv.org/abs/astro-ph/0607207} {arXiv:astro-ph/0607207} \BibitemShut
  {NoStop}%
\bibitem [{\citenamefont {Wyrzykowski}\ \emph {et~al.}(2011)\citenamefont
  {Wyrzykowski} \emph {et~al.}}]{Wyrzykowski:2011tr}%
  \BibitemOpen
  \bibfield  {author} {\bibinfo {author} {\bibfnamefont {L.}~\bibnamefont
  {Wyrzykowski}} \emph {et~al.},\ }\href {\doibase
  10.1111/j.1365-2966.2011.19243.x} {\bibfield  {journal} {\bibinfo  {journal}
  {Mon. Not. Roy. Astron. Soc.}\ }\textbf {\bibinfo {volume} {416}},\ \bibinfo
  {pages} {2949} (\bibinfo {year} {2011})},\ \Eprint
  {http://arxiv.org/abs/1106.2925} {arXiv:1106.2925 [astro-ph.GA]} \BibitemShut
  {NoStop}%
\bibitem [{\citenamefont {Nakamura}\ \emph {et~al.}(1997)\citenamefont
  {Nakamura}, \citenamefont {Sasaki}, \citenamefont {Tanaka},\ and\
  \citenamefont {Thorne}}]{Nakamura:1997sm}%
  \BibitemOpen
  \bibfield  {author} {\bibinfo {author} {\bibfnamefont {T.}~\bibnamefont
  {Nakamura}}, \bibinfo {author} {\bibfnamefont {M.}~\bibnamefont {Sasaki}},
  \bibinfo {author} {\bibfnamefont {T.}~\bibnamefont {Tanaka}}, \ and\ \bibinfo
  {author} {\bibfnamefont {K.~S.}\ \bibnamefont {Thorne}},\ }\href {\doibase
  10.1086/310886} {\bibfield  {journal} {\bibinfo  {journal} {Astrophys. J.
  Lett.}\ }\textbf {\bibinfo {volume} {487}},\ \bibinfo {pages} {L139}
  (\bibinfo {year} {1997})},\ \Eprint {http://arxiv.org/abs/astro-ph/9708060}
  {arXiv:astro-ph/9708060} \BibitemShut {NoStop}%
\bibitem [{\citenamefont {Ioka}\ \emph {et~al.}(1998)\citenamefont {Ioka},
  \citenamefont {Chiba}, \citenamefont {Tanaka},\ and\ \citenamefont
  {Nakamura}}]{Ioka:1998nz}%
  \BibitemOpen
  \bibfield  {author} {\bibinfo {author} {\bibfnamefont {K.}~\bibnamefont
  {Ioka}}, \bibinfo {author} {\bibfnamefont {T.}~\bibnamefont {Chiba}},
  \bibinfo {author} {\bibfnamefont {T.}~\bibnamefont {Tanaka}}, \ and\ \bibinfo
  {author} {\bibfnamefont {T.}~\bibnamefont {Nakamura}},\ }\href {\doibase
  10.1103/PhysRevD.58.063003} {\bibfield  {journal} {\bibinfo  {journal} {Phys.
  Rev. D}\ }\textbf {\bibinfo {volume} {58}},\ \bibinfo {pages} {063003}
  (\bibinfo {year} {1998})},\ \Eprint {http://arxiv.org/abs/astro-ph/9807018}
  {arXiv:astro-ph/9807018} \BibitemShut {NoStop}%
\bibitem [{\citenamefont {Carr}\ and\ \citenamefont
  {Sakellariadou}(1999)}]{Carr:1997cn}%
  \BibitemOpen
  \bibfield  {author} {\bibinfo {author} {\bibfnamefont {B.~J.}\ \bibnamefont
  {Carr}}\ and\ \bibinfo {author} {\bibfnamefont {M.}~\bibnamefont
  {Sakellariadou}},\ }\href {\doibase 10.1086/307071} {\bibfield  {journal}
  {\bibinfo  {journal} {Astrophys. J.}\ }\textbf {\bibinfo {volume} {516}},\
  \bibinfo {pages} {195} (\bibinfo {year} {1999})}\BibitemShut {NoStop}%
\bibitem [{\citenamefont {{Bahcall}}\ \emph {et~al.}(1985)\citenamefont
  {{Bahcall}}, \citenamefont {{Hut}},\ and\ \citenamefont
  {{Tremaine}}}]{1985ApJ...290...15B}%
  \BibitemOpen
  \bibfield  {author} {\bibinfo {author} {\bibfnamefont {J.~N.}\ \bibnamefont
  {{Bahcall}}}, \bibinfo {author} {\bibfnamefont {P.}~\bibnamefont {{Hut}}}, \
  and\ \bibinfo {author} {\bibfnamefont {S.}~\bibnamefont {{Tremaine}}},\
  }\href {\doibase 10.1086/162953} {\bibfield  {journal} {\bibinfo  {journal}
  {Astrophys. J,}\ }\textbf {\bibinfo {volume} {290}},\ \bibinfo {pages} {15}
  (\bibinfo {year} {1985})}\BibitemShut {NoStop}%
\bibitem [{\citenamefont {Yoo}\ \emph {et~al.}(2004)\citenamefont {Yoo},
  \citenamefont {Chaname},\ and\ \citenamefont {Gould}}]{Yoo:2003fr}%
  \BibitemOpen
  \bibfield  {author} {\bibinfo {author} {\bibfnamefont {J.}~\bibnamefont
  {Yoo}}, \bibinfo {author} {\bibfnamefont {J.}~\bibnamefont {Chaname}}, \ and\
  \bibinfo {author} {\bibfnamefont {A.}~\bibnamefont {Gould}},\ }\href
  {\doibase 10.1086/380562} {\bibfield  {journal} {\bibinfo  {journal}
  {Astrophys. J.}\ }\textbf {\bibinfo {volume} {601}},\ \bibinfo {pages} {311}
  (\bibinfo {year} {2004})},\ \Eprint {http://arxiv.org/abs/astro-ph/0307437}
  {arXiv:astro-ph/0307437} \BibitemShut {NoStop}%
\bibitem [{\citenamefont {{Monroy-Rodr{\'\i}guez}}\ and\ \citenamefont
  {{Allen}}(2014)}]{2014ApJ...790..159M}%
  \BibitemOpen
  \bibfield  {author} {\bibinfo {author} {\bibfnamefont {M.~A.}\ \bibnamefont
  {{Monroy-Rodr{\'\i}guez}}}\ and\ \bibinfo {author} {\bibfnamefont
  {C.}~\bibnamefont {{Allen}}},\ }\href {\doibase 10.1088/0004-637X/790/2/159}
  {\bibfield  {journal} {\bibinfo  {journal} {Astrophys. J.}\ }\textbf
  {\bibinfo {volume} {790}},\ \bibinfo {eid} {159} (\bibinfo {year} {2014})},\
  \Eprint {http://arxiv.org/abs/1406.5169} {arXiv:1406.5169 [astro-ph.GA]}
  \BibitemShut {NoStop}%
\bibitem [{\citenamefont {{Quinn}}\ \emph {et~al.}(2009)\citenamefont
  {{Quinn}}, \citenamefont {{Wilkinson}}, \citenamefont {{Irwin}},
  \citenamefont {{Marshall}}, \citenamefont {{Koch}},\ and\ \citenamefont
  {{Belokurov}}}]{2009MNRAS.396L..11Q}%
  \BibitemOpen
  \bibfield  {author} {\bibinfo {author} {\bibfnamefont {D.~P.}\ \bibnamefont
  {{Quinn}}}, \bibinfo {author} {\bibfnamefont {M.~I.}\ \bibnamefont
  {{Wilkinson}}}, \bibinfo {author} {\bibfnamefont {M.~J.}\ \bibnamefont
  {{Irwin}}}, \bibinfo {author} {\bibfnamefont {J.}~\bibnamefont {{Marshall}}},
  \bibinfo {author} {\bibfnamefont {A.}~\bibnamefont {{Koch}}}, \ and\ \bibinfo
  {author} {\bibfnamefont {V.}~\bibnamefont {{Belokurov}}},\ }\href {\doibase
  10.1111/j.1745-3933.2009.00652.x} {\bibfield  {journal} {\bibinfo  {journal}
  {Mon. Not. Roy. Astron. Soc.}\ }\textbf {\bibinfo {volume} {396}},\ \bibinfo
  {pages} {L11} (\bibinfo {year} {2009})},\ \Eprint
  {http://arxiv.org/abs/0903.1644} {arXiv:0903.1644 [astro-ph.GA]} \BibitemShut
  {NoStop}%
\bibitem [{\citenamefont {Afshordi}\ \emph {et~al.}(2003)\citenamefont
  {Afshordi}, \citenamefont {McDonald},\ and\ \citenamefont
  {Spergel}}]{Afshordi:2003zb}%
  \BibitemOpen
  \bibfield  {author} {\bibinfo {author} {\bibfnamefont {N.}~\bibnamefont
  {Afshordi}}, \bibinfo {author} {\bibfnamefont {P.}~\bibnamefont {McDonald}},
  \ and\ \bibinfo {author} {\bibfnamefont {D.~N.}\ \bibnamefont {Spergel}},\
  }\href {\doibase 10.1086/378763} {\bibfield  {journal} {\bibinfo  {journal}
  {Astrophys.~J.~Lett.}\ }\textbf {\bibinfo {volume} {594}},\ \bibinfo {pages}
  {L71} (\bibinfo {year} {2003})},\ \Eprint
  {http://arxiv.org/abs/astro-ph/0302035} {arXiv:astro-ph/0302035} \BibitemShut
  {NoStop}%
\bibitem [{\citenamefont {Capela}\ \emph {et~al.}(2013)\citenamefont {Capela},
  \citenamefont {Pshirkov},\ and\ \citenamefont {Tinyakov}}]{Capela:2013yf}%
  \BibitemOpen
  \bibfield  {author} {\bibinfo {author} {\bibfnamefont {F.}~\bibnamefont
  {Capela}}, \bibinfo {author} {\bibfnamefont {M.}~\bibnamefont {Pshirkov}}, \
  and\ \bibinfo {author} {\bibfnamefont {P.}~\bibnamefont {Tinyakov}},\ }\href
  {\doibase 10.1103/PhysRevD.87.123524} {\bibfield  {journal} {\bibinfo
  {journal} {Phys. Rev. D}\ }\textbf {\bibinfo {volume} {87}},\ \bibinfo
  {pages} {123524} (\bibinfo {year} {2013})},\ \Eprint
  {http://arxiv.org/abs/1301.4984} {arXiv:1301.4984 [astro-ph.CO]} \BibitemShut
  {NoStop}%
\bibitem [{\citenamefont {Pani}\ and\ \citenamefont
  {Loeb}(2014)}]{Pani:2014rca}%
  \BibitemOpen
  \bibfield  {author} {\bibinfo {author} {\bibfnamefont {P.}~\bibnamefont
  {Pani}}\ and\ \bibinfo {author} {\bibfnamefont {A.}~\bibnamefont {Loeb}},\
  }\href {\doibase 10.1088/1475-7516/2014/06/026} {\bibfield  {journal}
  {\bibinfo  {journal} {JCAP}\ }\textbf {\bibinfo {volume} {06}},\ \bibinfo
  {pages} {026} (\bibinfo {year} {2014})},\ \Eprint
  {http://arxiv.org/abs/1401.3025} {arXiv:1401.3025 [astro-ph.CO]} \BibitemShut
  {NoStop}%
\bibitem [{\citenamefont {Graham}\ \emph {et~al.}(2015)\citenamefont {Graham},
  \citenamefont {Rajendran},\ and\ \citenamefont {Varela}}]{Graham:2015apa}%
  \BibitemOpen
  \bibfield  {author} {\bibinfo {author} {\bibfnamefont {P.~W.}\ \bibnamefont
  {Graham}}, \bibinfo {author} {\bibfnamefont {S.}~\bibnamefont {Rajendran}}, \
  and\ \bibinfo {author} {\bibfnamefont {J.}~\bibnamefont {Varela}},\ }\href
  {\doibase 10.1103/PhysRevD.92.063007} {\bibfield  {journal} {\bibinfo
  {journal} {Phys. Rev. D}\ }\textbf {\bibinfo {volume} {92}},\ \bibinfo
  {pages} {063007} (\bibinfo {year} {2015})},\ \Eprint
  {http://arxiv.org/abs/1505.04444} {arXiv:1505.04444 [hep-ph]} \BibitemShut
  {NoStop}%
\bibitem [{\citenamefont {{Carr}}(1981)}]{1981MNRAS.194..639C}%
  \BibitemOpen
  \bibfield  {author} {\bibinfo {author} {\bibfnamefont {B.~J.}\ \bibnamefont
  {{Carr}}},\ }\href {\doibase 10.1093/mnras/194.3.639} {\bibfield  {journal}
  {\bibinfo  {journal} {Mon. Not. Roy. Astron. Soc.}\ }\textbf {\bibinfo
  {volume} {194}},\ \bibinfo {pages} {639} (\bibinfo {year}
  {1981})}\BibitemShut {NoStop}%
\bibitem [{\citenamefont {Ricotti}\ \emph {et~al.}(2008)\citenamefont
  {Ricotti}, \citenamefont {Ostriker},\ and\ \citenamefont
  {Mack}}]{Ricotti:2007au}%
  \BibitemOpen
  \bibfield  {author} {\bibinfo {author} {\bibfnamefont {M.}~\bibnamefont
  {Ricotti}}, \bibinfo {author} {\bibfnamefont {J.~P.}\ \bibnamefont
  {Ostriker}}, \ and\ \bibinfo {author} {\bibfnamefont {K.~J.}\ \bibnamefont
  {Mack}},\ }\href {\doibase 10.1086/587831} {\bibfield  {journal} {\bibinfo
  {journal} {Astrophys. J.}\ }\textbf {\bibinfo {volume} {680}},\ \bibinfo
  {pages} {829} (\bibinfo {year} {2008})},\ \Eprint
  {http://arxiv.org/abs/0709.0524} {arXiv:0709.0524 [astro-ph]} \BibitemShut
  {NoStop}%
\bibitem [{\citenamefont {Montero-Camacho}\ \emph {et~al.}(2019)\citenamefont
  {Montero-Camacho}, \citenamefont {Fang}, \citenamefont {Vasquez},
  \citenamefont {Silva},\ and\ \citenamefont
  {Hirata}}]{Montero-Camacho:2019jte}%
  \BibitemOpen
  \bibfield  {author} {\bibinfo {author} {\bibfnamefont {P.}~\bibnamefont
  {Montero-Camacho}}, \bibinfo {author} {\bibfnamefont {X.}~\bibnamefont
  {Fang}}, \bibinfo {author} {\bibfnamefont {G.}~\bibnamefont {Vasquez}},
  \bibinfo {author} {\bibfnamefont {M.}~\bibnamefont {Silva}}, \ and\ \bibinfo
  {author} {\bibfnamefont {C.~M.}\ \bibnamefont {Hirata}},\ }\href {\doibase
  10.1088/1475-7516/2019/08/031} {\bibfield  {journal} {\bibinfo  {journal}
  {JCAP}\ }\textbf {\bibinfo {volume} {08}},\ \bibinfo {pages} {031} (\bibinfo
  {year} {2019})},\ \Eprint {http://arxiv.org/abs/1906.05950} {arXiv:1906.05950
  [astro-ph.CO]} \BibitemShut {NoStop}%
\bibitem [{\citenamefont {Mack}\ \emph {et~al.}(2007)\citenamefont {Mack},
  \citenamefont {Ostriker},\ and\ \citenamefont {Ricotti}}]{Mack:2006gz}%
  \BibitemOpen
  \bibfield  {author} {\bibinfo {author} {\bibfnamefont {K.~J.}\ \bibnamefont
  {Mack}}, \bibinfo {author} {\bibfnamefont {J.~P.}\ \bibnamefont {Ostriker}},
  \ and\ \bibinfo {author} {\bibfnamefont {M.}~\bibnamefont {Ricotti}},\ }\href
  {\doibase 10.1086/518998} {\bibfield  {journal} {\bibinfo  {journal}
  {Astrophys. J.}\ }\textbf {\bibinfo {volume} {665}},\ \bibinfo {pages} {1277}
  (\bibinfo {year} {2007})},\ \Eprint {http://arxiv.org/abs/astro-ph/0608642}
  {arXiv:astro-ph/0608642} \BibitemShut {NoStop}%
\bibitem [{\citenamefont {Lacki}\ and\ \citenamefont
  {Beacom}(2010)}]{Lacki:2010zf}%
  \BibitemOpen
  \bibfield  {author} {\bibinfo {author} {\bibfnamefont {B.~C.}\ \bibnamefont
  {Lacki}}\ and\ \bibinfo {author} {\bibfnamefont {J.~F.}\ \bibnamefont
  {Beacom}},\ }\href {\doibase 10.1088/2041-8205/720/1/L67} {\bibfield
  {journal} {\bibinfo  {journal} {Astrophys. J. Lett.}\ }\textbf {\bibinfo
  {volume} {720}},\ \bibinfo {pages} {L67} (\bibinfo {year} {2010})},\ \Eprint
  {http://arxiv.org/abs/1003.3466} {arXiv:1003.3466 [astro-ph.CO]} \BibitemShut
  {NoStop}%
\bibitem [{\citenamefont {Carr}\ \emph
  {et~al.}(2021{\natexlab{a}})\citenamefont {Carr}, \citenamefont {Kohri},
  \citenamefont {Sendouda},\ and\ \citenamefont {Yokoyama}}]{Carr:2020gox}%
  \BibitemOpen
  \bibfield  {author} {\bibinfo {author} {\bibfnamefont {B.}~\bibnamefont
  {Carr}}, \bibinfo {author} {\bibfnamefont {K.}~\bibnamefont {Kohri}},
  \bibinfo {author} {\bibfnamefont {Y.}~\bibnamefont {Sendouda}}, \ and\
  \bibinfo {author} {\bibfnamefont {J.}~\bibnamefont {Yokoyama}},\ }\href
  {\doibase 10.1088/1361-6633/ac1e31} {\bibfield  {journal} {\bibinfo
  {journal} {Rept. Prog. Phys.}\ }\textbf {\bibinfo {volume} {84}},\ \bibinfo
  {pages} {116902} (\bibinfo {year} {2021}{\natexlab{a}})},\ \Eprint
  {http://arxiv.org/abs/2002.12778} {arXiv:2002.12778 [astro-ph.CO]}
  \BibitemShut {NoStop}%
\bibitem [{\citenamefont {Green}\ and\ \citenamefont
  {Kavanagh}(2021)}]{Green:2020jor}%
  \BibitemOpen
  \bibfield  {author} {\bibinfo {author} {\bibfnamefont {A.~M.}\ \bibnamefont
  {Green}}\ and\ \bibinfo {author} {\bibfnamefont {B.~J.}\ \bibnamefont
  {Kavanagh}},\ }\href {\doibase 10.1088/1361-6471/abc534} {\bibfield
  {journal} {\bibinfo  {journal} {J. Phys. G}\ }\textbf {\bibinfo {volume}
  {48}},\ \bibinfo {pages} {043001} (\bibinfo {year} {2021})},\ \Eprint
  {http://arxiv.org/abs/2007.10722} {arXiv:2007.10722 [astro-ph.CO]}
  \BibitemShut {NoStop}%
\bibitem [{\citenamefont {Clesse}\ and\ \citenamefont
  {Garc\'\i{}a-Bellido}(2018)}]{Clesse:2017bsw}%
  \BibitemOpen
  \bibfield  {author} {\bibinfo {author} {\bibfnamefont {S.}~\bibnamefont
  {Clesse}}\ and\ \bibinfo {author} {\bibfnamefont {J.}~\bibnamefont
  {Garc\'\i{}a-Bellido}},\ }\href {\doibase 10.1016/j.dark.2018.08.004}
  {\bibfield  {journal} {\bibinfo  {journal} {Phys. Dark Univ.}\ }\textbf
  {\bibinfo {volume} {22}},\ \bibinfo {pages} {137} (\bibinfo {year} {2018})},\
  \Eprint {http://arxiv.org/abs/1711.10458} {arXiv:1711.10458 [astro-ph.CO]}
  \BibitemShut {NoStop}%
\bibitem [{\citenamefont {Carr}\ \emph
  {et~al.}(2021{\natexlab{b}})\citenamefont {Carr}, \citenamefont {Clesse},
  \citenamefont {Garc\'\i{}a-Bellido},\ and\ \citenamefont
  {K\"uhnel}}]{Carr:2019kxo}%
  \BibitemOpen
  \bibfield  {author} {\bibinfo {author} {\bibfnamefont {B.}~\bibnamefont
  {Carr}}, \bibinfo {author} {\bibfnamefont {S.}~\bibnamefont {Clesse}},
  \bibinfo {author} {\bibfnamefont {J.}~\bibnamefont {Garc\'\i{}a-Bellido}}, \
  and\ \bibinfo {author} {\bibfnamefont {F.}~\bibnamefont {K\"uhnel}},\ }\href
  {\doibase 10.1016/j.dark.2020.100755} {\bibfield  {journal} {\bibinfo
  {journal} {Phys. Dark Univ.}\ }\textbf {\bibinfo {volume} {31}},\ \bibinfo
  {pages} {100755} (\bibinfo {year} {2021}{\natexlab{b}})},\ \Eprint
  {http://arxiv.org/abs/1906.08217} {arXiv:1906.08217 [astro-ph.CO]}
  \BibitemShut {NoStop}%
\bibitem [{\citenamefont {Carr}\ \emph {et~al.}(2024)\citenamefont {Carr},
  \citenamefont {Clesse}, \citenamefont {Garcia-Bellido}, \citenamefont
  {Hawkins},\ and\ \citenamefont {Kuhnel}}]{Carr:2023tpt}%
  \BibitemOpen
  \bibfield  {author} {\bibinfo {author} {\bibfnamefont {B.}~\bibnamefont
  {Carr}}, \bibinfo {author} {\bibfnamefont {S.}~\bibnamefont {Clesse}},
  \bibinfo {author} {\bibfnamefont {J.}~\bibnamefont {Garcia-Bellido}},
  \bibinfo {author} {\bibfnamefont {M.}~\bibnamefont {Hawkins}}, \ and\
  \bibinfo {author} {\bibfnamefont {F.}~\bibnamefont {Kuhnel}},\ }\href
  {\doibase 10.1016/j.physrep.2023.11.005} {\bibfield  {journal} {\bibinfo
  {journal} {Phys. Rept.}\ }\textbf {\bibinfo {volume} {1054}},\ \bibinfo
  {pages} {1} (\bibinfo {year} {2024})},\ \Eprint
  {http://arxiv.org/abs/2306.03903} {arXiv:2306.03903 [astro-ph.CO]}
  \BibitemShut {NoStop}%
\bibitem [{\citenamefont {Kavanagh}()}]{PBHbounds}%
  \BibitemOpen
  \bibfield  {author} {\bibinfo {author} {\bibfnamefont {B.~J.}\ \bibnamefont
  {Kavanagh}},\ }\href {\doibase 10.5281/zenodo.3538998} {\enquote {\bibinfo
  {title} {{PBHbounds, 10.5281/zenodo.3538998}},}\ }\BibitemShut {NoStop}%
\bibitem [{\citenamefont {Byrnes}\ \emph {et~al.}(2018)\citenamefont {Byrnes},
  \citenamefont {Hindmarsh}, \citenamefont {Young},\ and\ \citenamefont
  {Hawkins}}]{Byrnes:2018clq}%
  \BibitemOpen
  \bibfield  {author} {\bibinfo {author} {\bibfnamefont {C.~T.}\ \bibnamefont
  {Byrnes}}, \bibinfo {author} {\bibfnamefont {M.}~\bibnamefont {Hindmarsh}},
  \bibinfo {author} {\bibfnamefont {S.}~\bibnamefont {Young}}, \ and\ \bibinfo
  {author} {\bibfnamefont {M.~R.~S.}\ \bibnamefont {Hawkins}},\ }\href
  {\doibase 10.1088/1475-7516/2018/08/041} {\bibfield  {journal} {\bibinfo
  {journal} {JCAP}\ }\textbf {\bibinfo {volume} {08}},\ \bibinfo {pages} {041}
  (\bibinfo {year} {2018})},\ \Eprint {http://arxiv.org/abs/1801.06138}
  {arXiv:1801.06138 [astro-ph.CO]} \BibitemShut {NoStop}%
\bibitem [{\citenamefont {Akrami}\ \emph {et~al.}(2020)\citenamefont {Akrami}
  \emph {et~al.}}]{Planck:2018jri}%
  \BibitemOpen
  \bibfield  {author} {\bibinfo {author} {\bibfnamefont {Y.}~\bibnamefont
  {Akrami}} \emph {et~al.} (\bibinfo {collaboration} {Planck}),\ }\href
  {\doibase 10.1051/0004-6361/201833887} {\bibfield  {journal} {\bibinfo
  {journal} {Astron. Astrophys.}\ }\textbf {\bibinfo {volume} {641}},\ \bibinfo
  {pages} {A10} (\bibinfo {year} {2020})},\ \Eprint
  {http://arxiv.org/abs/1807.06211} {arXiv:1807.06211 [astro-ph.CO]}
  \BibitemShut {NoStop}%
\bibitem [{\citenamefont {Ballesteros}\ and\ \citenamefont
  {Taoso}(2018)}]{Ballesteros:2017fsr}%
  \BibitemOpen
  \bibfield  {author} {\bibinfo {author} {\bibfnamefont {G.}~\bibnamefont
  {Ballesteros}}\ and\ \bibinfo {author} {\bibfnamefont {M.}~\bibnamefont
  {Taoso}},\ }\href {\doibase 10.1103/PhysRevD.97.023501} {\bibfield  {journal}
  {\bibinfo  {journal} {Phys. Rev. D}\ }\textbf {\bibinfo {volume} {97}},\
  \bibinfo {pages} {023501} (\bibinfo {year} {2018})},\ \Eprint
  {http://arxiv.org/abs/1709.05565} {arXiv:1709.05565 [hep-ph]} \BibitemShut
  {NoStop}%
\bibitem [{\citenamefont {Hertzberg}\ and\ \citenamefont
  {Yamada}(2018)}]{Hertzberg:2017dkh}%
  \BibitemOpen
  \bibfield  {author} {\bibinfo {author} {\bibfnamefont {M.~P.}\ \bibnamefont
  {Hertzberg}}\ and\ \bibinfo {author} {\bibfnamefont {M.}~\bibnamefont
  {Yamada}},\ }\href {\doibase 10.1103/PhysRevD.97.083509} {\bibfield
  {journal} {\bibinfo  {journal} {Phys. Rev. D}\ }\textbf {\bibinfo {volume}
  {97}},\ \bibinfo {pages} {083509} (\bibinfo {year} {2018})},\ \Eprint
  {http://arxiv.org/abs/1712.09750} {arXiv:1712.09750 [astro-ph.CO]}
  \BibitemShut {NoStop}%
\bibitem [{\citenamefont {Garcia-Bellido}\ and\ \citenamefont
  {Ruiz~Morales}(2017)}]{Garcia-Bellido:2017mdw}%
  \BibitemOpen
  \bibfield  {author} {\bibinfo {author} {\bibfnamefont {J.}~\bibnamefont
  {Garcia-Bellido}}\ and\ \bibinfo {author} {\bibfnamefont {E.}~\bibnamefont
  {Ruiz~Morales}},\ }\href {\doibase 10.1016/j.dark.2017.09.007} {\bibfield
  {journal} {\bibinfo  {journal} {Phys. Dark Univ.}\ }\textbf {\bibinfo
  {volume} {18}},\ \bibinfo {pages} {47} (\bibinfo {year} {2017})},\ \Eprint
  {http://arxiv.org/abs/1702.03901} {arXiv:1702.03901 [astro-ph.CO]}
  \BibitemShut {NoStop}%
\bibitem [{\citenamefont {Mishra}\ and\ \citenamefont
  {Sahni}(2020)}]{Mishra:2019pzq}%
  \BibitemOpen
  \bibfield  {author} {\bibinfo {author} {\bibfnamefont {S.~S.}\ \bibnamefont
  {Mishra}}\ and\ \bibinfo {author} {\bibfnamefont {V.}~\bibnamefont {Sahni}},\
  }\href {\doibase 10.1088/1475-7516/2020/04/007} {\bibfield  {journal}
  {\bibinfo  {journal} {JCAP}\ }\textbf {\bibinfo {volume} {04}},\ \bibinfo
  {pages} {007} (\bibinfo {year} {2020})},\ \Eprint
  {http://arxiv.org/abs/1911.00057} {arXiv:1911.00057 [gr-qc]} \BibitemShut
  {NoStop}%
\bibitem [{\citenamefont {Palma}\ \emph {et~al.}(2020)\citenamefont {Palma},
  \citenamefont {Sypsas},\ and\ \citenamefont {Zenteno}}]{Palma:2020ejf}%
  \BibitemOpen
  \bibfield  {author} {\bibinfo {author} {\bibfnamefont {G.~A.}\ \bibnamefont
  {Palma}}, \bibinfo {author} {\bibfnamefont {S.}~\bibnamefont {Sypsas}}, \
  and\ \bibinfo {author} {\bibfnamefont {C.}~\bibnamefont {Zenteno}},\ }\href
  {\doibase 10.1103/PhysRevLett.125.121301} {\bibfield  {journal} {\bibinfo
  {journal} {Phys. Rev. Lett.}\ }\textbf {\bibinfo {volume} {125}},\ \bibinfo
  {pages} {121301} (\bibinfo {year} {2020})},\ \Eprint
  {http://arxiv.org/abs/2004.06106} {arXiv:2004.06106 [astro-ph.CO]}
  \BibitemShut {NoStop}%
\bibitem [{\citenamefont {Fumagalli}\ \emph {et~al.}(2023)\citenamefont
  {Fumagalli}, \citenamefont {Renaux-Petel}, \citenamefont {Ronayne},\ and\
  \citenamefont {Witkowski}}]{Fumagalli:2020adf}%
  \BibitemOpen
  \bibfield  {author} {\bibinfo {author} {\bibfnamefont {J.}~\bibnamefont
  {Fumagalli}}, \bibinfo {author} {\bibfnamefont {S.}~\bibnamefont
  {Renaux-Petel}}, \bibinfo {author} {\bibfnamefont {J.~W.}\ \bibnamefont
  {Ronayne}}, \ and\ \bibinfo {author} {\bibfnamefont {L.~T.}\ \bibnamefont
  {Witkowski}},\ }\href {\doibase 10.1016/j.physletb.2023.137921} {\bibfield
  {journal} {\bibinfo  {journal} {Phys. Lett. B}\ }\textbf {\bibinfo {volume}
  {841}},\ \bibinfo {pages} {137921} (\bibinfo {year} {2023})},\ \Eprint
  {http://arxiv.org/abs/2004.08369} {arXiv:2004.08369 [hep-th]} \BibitemShut
  {NoStop}%
\bibitem [{\citenamefont {\"Ozsoy}\ and\ \citenamefont
  {Tasinato}(2023)}]{Ozsoy:2023ryl}%
  \BibitemOpen
  \bibfield  {author} {\bibinfo {author} {\bibfnamefont {O.}~\bibnamefont
  {\"Ozsoy}}\ and\ \bibinfo {author} {\bibfnamefont {G.}~\bibnamefont
  {Tasinato}},\ }\href {\doibase 10.3390/universe9050203} {\bibfield  {journal}
  {\bibinfo  {journal} {Universe}\ }\textbf {\bibinfo {volume} {9}},\ \bibinfo
  {pages} {203} (\bibinfo {year} {2023})},\ \Eprint
  {http://arxiv.org/abs/2301.03600} {arXiv:2301.03600 [astro-ph.CO]}
  \BibitemShut {NoStop}%
\bibitem [{\citenamefont {Cotner}\ and\ \citenamefont
  {Kusenko}(2017)}]{Cotner:2016cvr}%
  \BibitemOpen
  \bibfield  {author} {\bibinfo {author} {\bibfnamefont {E.}~\bibnamefont
  {Cotner}}\ and\ \bibinfo {author} {\bibfnamefont {A.}~\bibnamefont
  {Kusenko}},\ }\href {\doibase 10.1103/PhysRevLett.119.031103} {\bibfield
  {journal} {\bibinfo  {journal} {Phys. Rev. Lett.}\ }\textbf {\bibinfo
  {volume} {119}},\ \bibinfo {pages} {031103} (\bibinfo {year} {2017})},\
  \Eprint {http://arxiv.org/abs/1612.02529} {arXiv:1612.02529 [astro-ph.CO]}
  \BibitemShut {NoStop}%
\bibitem [{\citenamefont {{Dvali}}\ \emph {et~al.}(2022)\citenamefont
  {{Dvali}}, \citenamefont {{K{\"u}hnel}},\ and\ \citenamefont
  {{Zantedeschi}}}]{2022PhRvL.129f1302D}%
  \BibitemOpen
  \bibfield  {author} {\bibinfo {author} {\bibfnamefont {G.}~\bibnamefont
  {{Dvali}}}, \bibinfo {author} {\bibfnamefont {F.}~\bibnamefont
  {{K{\"u}hnel}}}, \ and\ \bibinfo {author} {\bibfnamefont {M.}~\bibnamefont
  {{Zantedeschi}}},\ }\href {\doibase 10.1103/PhysRevLett.129.061302}
  {\bibfield  {journal} {\bibinfo  {journal} {\prl}\ }\textbf {\bibinfo
  {volume} {129}},\ \bibinfo {eid} {061302} (\bibinfo {year} {2022})},\ \Eprint
  {http://arxiv.org/abs/2112.08354} {arXiv:2112.08354 [hep-th]} \BibitemShut
  {NoStop}%
\bibitem [{\citenamefont {Amendola}\ \emph {et~al.}(2018)\citenamefont
  {Amendola}, \citenamefont {Rubio},\ and\ \citenamefont
  {Wetterich}}]{Amendola:2017xhl}%
  \BibitemOpen
  \bibfield  {author} {\bibinfo {author} {\bibfnamefont {L.}~\bibnamefont
  {Amendola}}, \bibinfo {author} {\bibfnamefont {J.}~\bibnamefont {Rubio}}, \
  and\ \bibinfo {author} {\bibfnamefont {C.}~\bibnamefont {Wetterich}},\ }\href
  {\doibase 10.1103/PhysRevD.97.081302} {\bibfield  {journal} {\bibinfo
  {journal} {Phys. Rev. D}\ }\textbf {\bibinfo {volume} {97}},\ \bibinfo
  {pages} {081302} (\bibinfo {year} {2018})},\ \Eprint
  {http://arxiv.org/abs/1711.09915} {arXiv:1711.09915 [astro-ph.CO]}
  \BibitemShut {NoStop}%
\bibitem [{\citenamefont {Flores}\ and\ \citenamefont
  {Kusenko}(2021)}]{Flores:2020drq}%
  \BibitemOpen
  \bibfield  {author} {\bibinfo {author} {\bibfnamefont {M.~M.}\ \bibnamefont
  {Flores}}\ and\ \bibinfo {author} {\bibfnamefont {A.}~\bibnamefont
  {Kusenko}},\ }\href {\doibase 10.1103/PhysRevLett.126.041101} {\bibfield
  {journal} {\bibinfo  {journal} {Phys. Rev. Lett.}\ }\textbf {\bibinfo
  {volume} {126}},\ \bibinfo {pages} {041101} (\bibinfo {year} {2021})},\
  \Eprint {http://arxiv.org/abs/2008.12456} {arXiv:2008.12456 [astro-ph.CO]}
  \BibitemShut {NoStop}%
\bibitem [{\citenamefont {Jenkins}\ and\ \citenamefont
  {Sakellariadou}(2020)}]{Jenkins:2020ctp}%
  \BibitemOpen
  \bibfield  {author} {\bibinfo {author} {\bibfnamefont {A.~C.}\ \bibnamefont
  {Jenkins}}\ and\ \bibinfo {author} {\bibfnamefont {M.}~\bibnamefont
  {Sakellariadou}},\ }\href@noop {} {\  (\bibinfo {year} {2020})},\ \Eprint
  {http://arxiv.org/abs/2006.16249} {arXiv:2006.16249 [astro-ph.CO]}
  \BibitemShut {NoStop}%
\bibitem [{\citenamefont {Germani}\ and\ \citenamefont
  {Musco}(2019)}]{Germani:2018jgr}%
  \BibitemOpen
  \bibfield  {author} {\bibinfo {author} {\bibfnamefont {C.}~\bibnamefont
  {Germani}}\ and\ \bibinfo {author} {\bibfnamefont {I.}~\bibnamefont
  {Musco}},\ }\href {\doibase 10.1103/PhysRevLett.122.141302} {\bibfield
  {journal} {\bibinfo  {journal} {Phys. Rev. Lett.}\ }\textbf {\bibinfo
  {volume} {122}},\ \bibinfo {pages} {141302} (\bibinfo {year} {2019})},\
  \Eprint {http://arxiv.org/abs/1805.04087} {arXiv:1805.04087 [astro-ph.CO]}
  \BibitemShut {NoStop}%
\bibitem [{\citenamefont {Escriv\`a}\ \emph {et~al.}(2020)\citenamefont
  {Escriv\`a}, \citenamefont {Germani},\ and\ \citenamefont
  {Sheth}}]{Escriva:2019phb}%
  \BibitemOpen
  \bibfield  {author} {\bibinfo {author} {\bibfnamefont {A.}~\bibnamefont
  {Escriv\`a}}, \bibinfo {author} {\bibfnamefont {C.}~\bibnamefont {Germani}},
  \ and\ \bibinfo {author} {\bibfnamefont {R.~K.}\ \bibnamefont {Sheth}},\
  }\href {\doibase 10.1103/PhysRevD.101.044022} {\bibfield  {journal} {\bibinfo
   {journal} {Phys. Rev. D}\ }\textbf {\bibinfo {volume} {101}},\ \bibinfo
  {pages} {044022} (\bibinfo {year} {2020})},\ \Eprint
  {http://arxiv.org/abs/1907.13311} {arXiv:1907.13311 [gr-qc]} \BibitemShut
  {NoStop}%
\bibitem [{\citenamefont {Kawasaki}\ and\ \citenamefont
  {Nakatsuka}(2019)}]{Kawasaki:2019mbl}%
  \BibitemOpen
  \bibfield  {author} {\bibinfo {author} {\bibfnamefont {M.}~\bibnamefont
  {Kawasaki}}\ and\ \bibinfo {author} {\bibfnamefont {H.}~\bibnamefont
  {Nakatsuka}},\ }\href {\doibase 10.1103/PhysRevD.99.123501} {\bibfield
  {journal} {\bibinfo  {journal} {Phys. Rev. D}\ }\textbf {\bibinfo {volume}
  {99}},\ \bibinfo {pages} {123501} (\bibinfo {year} {2019})},\ \Eprint
  {http://arxiv.org/abs/1903.02994} {arXiv:1903.02994 [astro-ph.CO]}
  \BibitemShut {NoStop}%
\bibitem [{\citenamefont {De~Luca}\ \emph {et~al.}(2019)\citenamefont
  {De~Luca}, \citenamefont {Franciolini}, \citenamefont {Kehagias},
  \citenamefont {Peloso}, \citenamefont {Riotto},\ and\ \citenamefont
  {\"Unal}}]{DeLuca:2019qsy}%
  \BibitemOpen
  \bibfield  {author} {\bibinfo {author} {\bibfnamefont {V.}~\bibnamefont
  {De~Luca}}, \bibinfo {author} {\bibfnamefont {G.}~\bibnamefont
  {Franciolini}}, \bibinfo {author} {\bibfnamefont {A.}~\bibnamefont
  {Kehagias}}, \bibinfo {author} {\bibfnamefont {M.}~\bibnamefont {Peloso}},
  \bibinfo {author} {\bibfnamefont {A.}~\bibnamefont {Riotto}}, \ and\ \bibinfo
  {author} {\bibfnamefont {C.}~\bibnamefont {\"Unal}},\ }\href {\doibase
  10.1088/1475-7516/2019/07/048} {\bibfield  {journal} {\bibinfo  {journal}
  {JCAP}\ }\textbf {\bibinfo {volume} {07}},\ \bibinfo {pages} {048} (\bibinfo
  {year} {2019})},\ \Eprint {http://arxiv.org/abs/1904.00970} {arXiv:1904.00970
  [astro-ph.CO]} \BibitemShut {NoStop}%
\bibitem [{\citenamefont {Young}\ \emph {et~al.}(2019)\citenamefont {Young},
  \citenamefont {Musco},\ and\ \citenamefont {Byrnes}}]{Young:2019yug}%
  \BibitemOpen
  \bibfield  {author} {\bibinfo {author} {\bibfnamefont {S.}~\bibnamefont
  {Young}}, \bibinfo {author} {\bibfnamefont {I.}~\bibnamefont {Musco}}, \ and\
  \bibinfo {author} {\bibfnamefont {C.~T.}\ \bibnamefont {Byrnes}},\ }\href
  {\doibase 10.1088/1475-7516/2019/11/012} {\bibfield  {journal} {\bibinfo
  {journal} {JCAP}\ }\textbf {\bibinfo {volume} {11}},\ \bibinfo {pages} {012}
  (\bibinfo {year} {2019})},\ \Eprint {http://arxiv.org/abs/1904.00984}
  {arXiv:1904.00984 [astro-ph.CO]} \BibitemShut {NoStop}%
\bibitem [{\citenamefont {Garc\'\i{}a-Bellido}\ \emph
  {et~al.}(2021)\citenamefont {Garc\'\i{}a-Bellido}, \citenamefont {Carr},\
  and\ \citenamefont {Clesse}}]{Garcia-Bellido:2019vlf}%
  \BibitemOpen
  \bibfield  {author} {\bibinfo {author} {\bibfnamefont {J.}~\bibnamefont
  {Garc\'\i{}a-Bellido}}, \bibinfo {author} {\bibfnamefont {B.}~\bibnamefont
  {Carr}}, \ and\ \bibinfo {author} {\bibfnamefont {S.}~\bibnamefont
  {Clesse}},\ }\href {\doibase 10.3390/universe8010012} {\bibfield  {journal}
  {\bibinfo  {journal} {Universe}\ }\textbf {\bibinfo {volume} {8}},\ \bibinfo
  {pages} {12} (\bibinfo {year} {2021})},\ \Eprint
  {http://arxiv.org/abs/1904.11482} {arXiv:1904.11482 [astro-ph.CO]}
  \BibitemShut {NoStop}%
\bibitem [{\citenamefont {Cole}\ \emph {et~al.}(2023)\citenamefont {Cole},
  \citenamefont {Gow}, \citenamefont {Byrnes},\ and\ \citenamefont
  {Patil}}]{Cole:2023wyx}%
  \BibitemOpen
  \bibfield  {author} {\bibinfo {author} {\bibfnamefont {P.~S.}\ \bibnamefont
  {Cole}}, \bibinfo {author} {\bibfnamefont {A.~D.}\ \bibnamefont {Gow}},
  \bibinfo {author} {\bibfnamefont {C.~T.}\ \bibnamefont {Byrnes}}, \ and\
  \bibinfo {author} {\bibfnamefont {S.~P.}\ \bibnamefont {Patil}},\ }\href
  {\doibase 10.1088/1475-7516/2023/08/031} {\bibfield  {journal} {\bibinfo
  {journal} {JCAP}\ }\textbf {\bibinfo {volume} {08}},\ \bibinfo {pages} {031}
  (\bibinfo {year} {2023})},\ \Eprint {http://arxiv.org/abs/2304.01997}
  {arXiv:2304.01997 [astro-ph.CO]} \BibitemShut {NoStop}%
\bibitem [{\citenamefont {Green}(2016)}]{Green:2016xgy}%
  \BibitemOpen
  \bibfield  {author} {\bibinfo {author} {\bibfnamefont {A.~M.}\ \bibnamefont
  {Green}},\ }\href {\doibase 10.1103/PhysRevD.94.063530} {\bibfield  {journal}
  {\bibinfo  {journal} {Phys. Rev. D}\ }\textbf {\bibinfo {volume} {94}},\
  \bibinfo {pages} {063530} (\bibinfo {year} {2016})},\ \Eprint
  {http://arxiv.org/abs/1609.01143} {arXiv:1609.01143 [astro-ph.CO]}
  \BibitemShut {NoStop}%
\bibitem [{\citenamefont {Carr}\ \emph {et~al.}(2017)\citenamefont {Carr},
  \citenamefont {Raidal}, \citenamefont {Tenkanen}, \citenamefont {Vaskonen},\
  and\ \citenamefont {Veerm\"ae}}]{Carr:2017jsz}%
  \BibitemOpen
  \bibfield  {author} {\bibinfo {author} {\bibfnamefont {B.}~\bibnamefont
  {Carr}}, \bibinfo {author} {\bibfnamefont {M.}~\bibnamefont {Raidal}},
  \bibinfo {author} {\bibfnamefont {T.}~\bibnamefont {Tenkanen}}, \bibinfo
  {author} {\bibfnamefont {V.}~\bibnamefont {Vaskonen}}, \ and\ \bibinfo
  {author} {\bibfnamefont {H.}~\bibnamefont {Veerm\"ae}},\ }\href {\doibase
  10.1103/PhysRevD.96.023514} {\bibfield  {journal} {\bibinfo  {journal} {Phys.
  Rev. D}\ }\textbf {\bibinfo {volume} {96}},\ \bibinfo {pages} {023514}
  (\bibinfo {year} {2017})},\ \Eprint {http://arxiv.org/abs/1705.05567}
  {arXiv:1705.05567 [astro-ph.CO]} \BibitemShut {NoStop}%
\bibitem [{\citenamefont {{Kashlinsky}}(2016)}]{2016ApJ...823L..25K}%
  \BibitemOpen
  \bibfield  {author} {\bibinfo {author} {\bibfnamefont {A.}~\bibnamefont
  {{Kashlinsky}}},\ }\href {\doibase 10.3847/2041-8205/823/2/L25} {\bibfield
  {journal} {\bibinfo  {journal} {Astrophys. J. Lett.}\ }\textbf {\bibinfo
  {volume} {823}},\ \bibinfo {eid} {L25} (\bibinfo {year} {2016})},\ \Eprint
  {http://arxiv.org/abs/1605.04023} {arXiv:1605.04023 [astro-ph.CO]}
  \BibitemShut {NoStop}%
\bibitem [{\citenamefont {Inman}\ and\ \citenamefont
  {Ali-Ha\"\i{}moud}(2019)}]{Inman:2019wvr}%
  \BibitemOpen
  \bibfield  {author} {\bibinfo {author} {\bibfnamefont {D.}~\bibnamefont
  {Inman}}\ and\ \bibinfo {author} {\bibfnamefont {Y.}~\bibnamefont
  {Ali-Ha\"\i{}moud}},\ }\href {\doibase 10.1103/PhysRevD.100.083528}
  {\bibfield  {journal} {\bibinfo  {journal} {Phys. Rev. D}\ }\textbf {\bibinfo
  {volume} {100}},\ \bibinfo {pages} {083528} (\bibinfo {year} {2019})},\
  \Eprint {http://arxiv.org/abs/1907.08129} {arXiv:1907.08129 [astro-ph.CO]}
  \BibitemShut {NoStop}%
\bibitem [{\citenamefont {Gorton}\ and\ \citenamefont
  {Green}(2022)}]{Gorton:2022fyb}%
  \BibitemOpen
  \bibfield  {author} {\bibinfo {author} {\bibfnamefont {M.}~\bibnamefont
  {Gorton}}\ and\ \bibinfo {author} {\bibfnamefont {A.~M.}\ \bibnamefont
  {Green}},\ }\href {\doibase 10.1088/1475-7516/2022/08/035} {\bibfield
  {journal} {\bibinfo  {journal} {JCAP}\ }\textbf {\bibinfo {volume} {08}},\
  \bibinfo {pages} {035} (\bibinfo {year} {2022})},\ \Eprint
  {http://arxiv.org/abs/2203.04209} {arXiv:2203.04209 [astro-ph.CO]}
  \BibitemShut {NoStop}%
\bibitem [{\citenamefont {Peta\v{c}}\ \emph {et~al.}(2022)\citenamefont
  {Peta\v{c}}, \citenamefont {Lavalle},\ and\ \citenamefont
  {Jedamzik}}]{Petac:2022rio}%
  \BibitemOpen
  \bibfield  {author} {\bibinfo {author} {\bibfnamefont {M.}~\bibnamefont
  {Peta\v{c}}}, \bibinfo {author} {\bibfnamefont {J.}~\bibnamefont {Lavalle}},
  \ and\ \bibinfo {author} {\bibfnamefont {K.}~\bibnamefont {Jedamzik}},\
  }\href {\doibase 10.1103/PhysRevD.105.083520} {\bibfield  {journal} {\bibinfo
   {journal} {Phys. Rev. D}\ }\textbf {\bibinfo {volume} {105}},\ \bibinfo
  {pages} {083520} (\bibinfo {year} {2022})},\ \Eprint
  {http://arxiv.org/abs/2201.02521} {arXiv:2201.02521 [astro-ph.CO]}
  \BibitemShut {NoStop}%
\bibitem [{\citenamefont {Calcino}\ \emph {et~al.}(2018)\citenamefont
  {Calcino}, \citenamefont {Garcia-Bellido},\ and\ \citenamefont
  {Davis}}]{Calcino:2018mwh}%
  \BibitemOpen
  \bibfield  {author} {\bibinfo {author} {\bibfnamefont {J.}~\bibnamefont
  {Calcino}}, \bibinfo {author} {\bibfnamefont {J.}~\bibnamefont
  {Garcia-Bellido}}, \ and\ \bibinfo {author} {\bibfnamefont {T.~M.}\
  \bibnamefont {Davis}},\ }\href {\doibase 10.1093/mnras/sty1368} {\bibfield
  {journal} {\bibinfo  {journal} {Mon. Not. Roy. Astron. Soc.}\ }\textbf
  {\bibinfo {volume} {479}},\ \bibinfo {pages} {2889} (\bibinfo {year}
  {2018})},\ \Eprint {http://arxiv.org/abs/1803.09205} {arXiv:1803.09205
  [astro-ph.CO]} \BibitemShut {NoStop}%
\bibitem [{\citenamefont {Young}\ and\ \citenamefont
  {Byrnes}(2020)}]{Young:2019gfc}%
  \BibitemOpen
  \bibfield  {author} {\bibinfo {author} {\bibfnamefont {S.}~\bibnamefont
  {Young}}\ and\ \bibinfo {author} {\bibfnamefont {C.~T.}\ \bibnamefont
  {Byrnes}},\ }\href {\doibase 10.1088/1475-7516/2020/03/004} {\bibfield
  {journal} {\bibinfo  {journal} {JCAP}\ }\textbf {\bibinfo {volume} {03}},\
  \bibinfo {pages} {004} (\bibinfo {year} {2020})},\ \Eprint
  {http://arxiv.org/abs/1910.06077} {arXiv:1910.06077 [astro-ph.CO]}
  \BibitemShut {NoStop}%
\bibitem [{\citenamefont {De~Luca}\ \emph {et~al.}(2022)\citenamefont
  {De~Luca}, \citenamefont {Franciolini}, \citenamefont {Riotto},\ and\
  \citenamefont {Veerm\"ae}}]{DeLuca:2022uvz}%
  \BibitemOpen
  \bibfield  {author} {\bibinfo {author} {\bibfnamefont {V.}~\bibnamefont
  {De~Luca}}, \bibinfo {author} {\bibfnamefont {G.}~\bibnamefont
  {Franciolini}}, \bibinfo {author} {\bibfnamefont {A.}~\bibnamefont {Riotto}},
  \ and\ \bibinfo {author} {\bibfnamefont {H.}~\bibnamefont {Veerm\"ae}},\
  }\href {\doibase 10.1103/PhysRevLett.129.191302} {\bibfield  {journal}
  {\bibinfo  {journal} {Phys. Rev. Lett.}\ }\textbf {\bibinfo {volume} {129}},\
  \bibinfo {pages} {191302} (\bibinfo {year} {2022})},\ \Eprint
  {http://arxiv.org/abs/2208.01683} {arXiv:2208.01683 [astro-ph.CO]}
  \BibitemShut {NoStop}%
\bibitem [{\citenamefont {Adamek}\ \emph {et~al.}(2019)\citenamefont {Adamek},
  \citenamefont {Byrnes}, \citenamefont {Gosenca},\ and\ \citenamefont
  {Hotchkiss}}]{Adamek:2019gns}%
  \BibitemOpen
  \bibfield  {author} {\bibinfo {author} {\bibfnamefont {J.}~\bibnamefont
  {Adamek}}, \bibinfo {author} {\bibfnamefont {C.~T.}\ \bibnamefont {Byrnes}},
  \bibinfo {author} {\bibfnamefont {M.}~\bibnamefont {Gosenca}}, \ and\
  \bibinfo {author} {\bibfnamefont {S.}~\bibnamefont {Hotchkiss}},\ }\href
  {\doibase 10.1103/PhysRevD.100.023506} {\bibfield  {journal} {\bibinfo
  {journal} {Phys. Rev. D}\ }\textbf {\bibinfo {volume} {100}},\ \bibinfo
  {pages} {023506} (\bibinfo {year} {2019})},\ \Eprint
  {http://arxiv.org/abs/1901.08528} {arXiv:1901.08528 [astro-ph.CO]}
  \BibitemShut {NoStop}%
\bibitem [{\citenamefont {Carr}\ \emph
  {et~al.}(2021{\natexlab{c}})\citenamefont {Carr}, \citenamefont {Kuhnel},\
  and\ \citenamefont {Visinelli}}]{Carr:2020mqm}%
  \BibitemOpen
  \bibfield  {author} {\bibinfo {author} {\bibfnamefont {B.}~\bibnamefont
  {Carr}}, \bibinfo {author} {\bibfnamefont {F.}~\bibnamefont {Kuhnel}}, \ and\
  \bibinfo {author} {\bibfnamefont {L.}~\bibnamefont {Visinelli}},\ }\href
  {\doibase 10.1093/mnras/stab1930} {\bibfield  {journal} {\bibinfo  {journal}
  {Mon. Not. Roy. Astron. Soc.}\ }\textbf {\bibinfo {volume} {506}},\ \bibinfo
  {pages} {3648} (\bibinfo {year} {2021}{\natexlab{c}})},\ \Eprint
  {http://arxiv.org/abs/2011.01930} {arXiv:2011.01930 [astro-ph.CO]}
  \BibitemShut {NoStop}%
\bibitem [{\citenamefont {{Boudaud}}\ \emph {et~al.}(2021)\citenamefont
  {{Boudaud}}, \citenamefont {{Lacroix}}, \citenamefont {{Stref}},
  \citenamefont {{Lavalle}},\ and\ \citenamefont {{Salati}}}]{Boudaud:2021irr}%
  \BibitemOpen
  \bibfield  {author} {\bibinfo {author} {\bibfnamefont {M.}~\bibnamefont
  {{Boudaud}}}, \bibinfo {author} {\bibfnamefont {T.}~\bibnamefont
  {{Lacroix}}}, \bibinfo {author} {\bibfnamefont {M.}~\bibnamefont {{Stref}}},
  \bibinfo {author} {\bibfnamefont {J.}~\bibnamefont {{Lavalle}}}, \ and\
  \bibinfo {author} {\bibfnamefont {P.}~\bibnamefont {{Salati}}},\ }\href@noop
  {} {\bibfield  {journal} {\bibinfo  {journal} {arXiv e-prints}\ ,\ \bibinfo
  {eid} {arXiv:2106.07480}} (\bibinfo {year} {2021})},\ \Eprint
  {http://arxiv.org/abs/2106.07480} {arXiv:2106.07480 [astro-ph.CO]}
  \BibitemShut {NoStop}%
\bibitem [{\citenamefont {Bertone}\ \emph {et~al.}(2019)\citenamefont
  {Bertone}, \citenamefont {Coogan}, \citenamefont {Gaggero}, \citenamefont
  {Kavanagh},\ and\ \citenamefont {Weniger}}]{Bertone:2019vsk}%
  \BibitemOpen
  \bibfield  {author} {\bibinfo {author} {\bibfnamefont {G.}~\bibnamefont
  {Bertone}}, \bibinfo {author} {\bibfnamefont {A.~M.}\ \bibnamefont {Coogan}},
  \bibinfo {author} {\bibfnamefont {D.}~\bibnamefont {Gaggero}}, \bibinfo
  {author} {\bibfnamefont {B.~J.}\ \bibnamefont {Kavanagh}}, \ and\ \bibinfo
  {author} {\bibfnamefont {C.}~\bibnamefont {Weniger}},\ }\href {\doibase
  10.1103/PhysRevD.100.123013} {\bibfield  {journal} {\bibinfo  {journal}
  {Phys. Rev. D}\ }\textbf {\bibinfo {volume} {100}},\ \bibinfo {pages}
  {123013} (\bibinfo {year} {2019})},\ \Eprint
  {http://arxiv.org/abs/1905.01238} {arXiv:1905.01238 [hep-ph]} \BibitemShut
  {NoStop}%
\bibitem [{\citenamefont {Abbott}\ \emph {et~al.}(2016)\citenamefont {Abbott}
  \emph {et~al.}}]{LIGOScientific:2016aoc}%
  \BibitemOpen
  \bibfield  {author} {\bibinfo {author} {\bibfnamefont {B.~P.}\ \bibnamefont
  {Abbott}} \emph {et~al.} (\bibinfo {collaboration} {LIGO Scientific,
  Virgo}),\ }\href {\doibase 10.1103/PhysRevLett.116.061102} {\bibfield
  {journal} {\bibinfo  {journal} {Phys. Rev. Lett.}\ }\textbf {\bibinfo
  {volume} {116}},\ \bibinfo {pages} {061102} (\bibinfo {year} {2016})},\
  \Eprint {http://arxiv.org/abs/1602.03837} {arXiv:1602.03837 [gr-qc]}
  \BibitemShut {NoStop}%
\bibitem [{\citenamefont {Bird}\ \emph {et~al.}(2016)\citenamefont {Bird},
  \citenamefont {Cholis}, \citenamefont {Mu\~noz}, \citenamefont
  {Ali-Ha\"\i{}moud}, \citenamefont {Kamionkowski}, \citenamefont {Kovetz},
  \citenamefont {Raccanelli},\ and\ \citenamefont {Riess}}]{Bird:2016dcv}%
  \BibitemOpen
  \bibfield  {author} {\bibinfo {author} {\bibfnamefont {S.}~\bibnamefont
  {Bird}}, \bibinfo {author} {\bibfnamefont {I.}~\bibnamefont {Cholis}},
  \bibinfo {author} {\bibfnamefont {J.~B.}\ \bibnamefont {Mu\~noz}}, \bibinfo
  {author} {\bibfnamefont {Y.}~\bibnamefont {Ali-Ha\"\i{}moud}}, \bibinfo
  {author} {\bibfnamefont {M.}~\bibnamefont {Kamionkowski}}, \bibinfo {author}
  {\bibfnamefont {E.~D.}\ \bibnamefont {Kovetz}}, \bibinfo {author}
  {\bibfnamefont {A.}~\bibnamefont {Raccanelli}}, \ and\ \bibinfo {author}
  {\bibfnamefont {A.~G.}\ \bibnamefont {Riess}},\ }\href {\doibase
  10.1103/PhysRevLett.116.201301} {\bibfield  {journal} {\bibinfo  {journal}
  {Phys. Rev. Lett.}\ }\textbf {\bibinfo {volume} {116}},\ \bibinfo {pages}
  {201301} (\bibinfo {year} {2016})},\ \Eprint
  {http://arxiv.org/abs/1603.00464} {arXiv:1603.00464 [astro-ph.CO]}
  \BibitemShut {NoStop}%
\bibitem [{\citenamefont {Clesse}\ and\ \citenamefont
  {Garc\'\i{}a-Bellido}(2017)}]{Clesse:2016vqa}%
  \BibitemOpen
  \bibfield  {author} {\bibinfo {author} {\bibfnamefont {S.}~\bibnamefont
  {Clesse}}\ and\ \bibinfo {author} {\bibfnamefont {J.}~\bibnamefont
  {Garc\'\i{}a-Bellido}},\ }\href {\doibase 10.1016/j.dark.2016.10.002}
  {\bibfield  {journal} {\bibinfo  {journal} {Phys. Dark Univ.}\ }\textbf
  {\bibinfo {volume} {15}},\ \bibinfo {pages} {142} (\bibinfo {year} {2017})},\
  \Eprint {http://arxiv.org/abs/1603.05234} {arXiv:1603.05234 [astro-ph.CO]}
  \BibitemShut {NoStop}%
\bibitem [{\citenamefont {Sasaki}\ \emph {et~al.}(2016)\citenamefont {Sasaki},
  \citenamefont {Suyama}, \citenamefont {Tanaka},\ and\ \citenamefont
  {Yokoyama}}]{Sasaki:2016jop}%
  \BibitemOpen
  \bibfield  {author} {\bibinfo {author} {\bibfnamefont {M.}~\bibnamefont
  {Sasaki}}, \bibinfo {author} {\bibfnamefont {T.}~\bibnamefont {Suyama}},
  \bibinfo {author} {\bibfnamefont {T.}~\bibnamefont {Tanaka}}, \ and\ \bibinfo
  {author} {\bibfnamefont {S.}~\bibnamefont {Yokoyama}},\ }\href {\doibase
  10.1103/PhysRevLett.117.061101} {\bibfield  {journal} {\bibinfo  {journal}
  {Phys. Rev. Lett.}\ }\textbf {\bibinfo {volume} {117}},\ \bibinfo {pages}
  {061101} (\bibinfo {year} {2016})},\ \bibinfo {note} {[Erratum:
  Phys.Rev.Lett. 121, 059901 (2018)]},\ \Eprint
  {http://arxiv.org/abs/1603.08338} {arXiv:1603.08338 [astro-ph.CO]}
  \BibitemShut {NoStop}%
\bibitem [{\citenamefont {Garc{\'\i}a-Bellido}(2017)}]{Garcia-Bellido:2017fdg}%
  \BibitemOpen
  \bibfield  {author} {\bibinfo {author} {\bibfnamefont {J.}~\bibnamefont
  {Garc{\'\i}a-Bellido}},\ }\bibfield  {booktitle} {\emph {\bibinfo {booktitle}
  {{Proceedings, 11th Int.~LISA Symposium: Zurich, Switzerland, September 5-9,
  2016}}},\ }\href {\doibase 10.1088/1742-6596/840/1/012032} {\bibfield
  {journal} {\bibinfo  {journal} {J.~Phys.~Conf.~Ser.}\ }\textbf {\bibinfo
  {volume} {840}},\ \bibinfo {pages} {012032} (\bibinfo {year} {2017})},\
  \Eprint {http://arxiv.org/abs/1702.08275} {arXiv:1702.08275 [astro-ph.CO]}
  \BibitemShut {NoStop}%
\bibitem [{\citenamefont {{Sasaki}}\ \emph {et~al.}(2018)\citenamefont
  {{Sasaki}}, \citenamefont {{Suyama}}, \citenamefont {{Tanaka}},\ and\
  \citenamefont {{Yokoyama}}}]{2018CQGra..35f3001S}%
  \BibitemOpen
  \bibfield  {author} {\bibinfo {author} {\bibfnamefont {M.}~\bibnamefont
  {{Sasaki}}}, \bibinfo {author} {\bibfnamefont {T.}~\bibnamefont {{Suyama}}},
  \bibinfo {author} {\bibfnamefont {T.}~\bibnamefont {{Tanaka}}}, \ and\
  \bibinfo {author} {\bibfnamefont {S.}~\bibnamefont {{Yokoyama}}},\ }\href
  {\doibase 10.1088/1361-6382/aaa7b4} {\bibfield  {journal} {\bibinfo
  {journal} {Classical and Quantum Gravity}\ }\textbf {\bibinfo {volume}
  {35}},\ \bibinfo {eid} {063001} (\bibinfo {year} {2018})},\ \Eprint
  {http://arxiv.org/abs/1801.05235} {arXiv:1801.05235 [astro-ph.CO]}
  \BibitemShut {NoStop}%
\bibitem [{\citenamefont {Hall}\ \emph {et~al.}(2020)\citenamefont {Hall},
  \citenamefont {Gow},\ and\ \citenamefont {Byrnes}}]{Hall:2020daa}%
  \BibitemOpen
  \bibfield  {author} {\bibinfo {author} {\bibfnamefont {A.}~\bibnamefont
  {Hall}}, \bibinfo {author} {\bibfnamefont {A.~D.}\ \bibnamefont {Gow}}, \
  and\ \bibinfo {author} {\bibfnamefont {C.~T.}\ \bibnamefont {Byrnes}},\
  }\href {\doibase 10.1103/PhysRevD.102.123524} {\bibfield  {journal} {\bibinfo
   {journal} {Phys. Rev. D}\ }\textbf {\bibinfo {volume} {102}},\ \bibinfo
  {pages} {123524} (\bibinfo {year} {2020})},\ \Eprint
  {http://arxiv.org/abs/2008.13704} {arXiv:2008.13704 [astro-ph.CO]}
  \BibitemShut {NoStop}%
\bibitem [{\citenamefont {De~Luca}\ \emph {et~al.}(2020)\citenamefont
  {De~Luca}, \citenamefont {Franciolini}, \citenamefont {Pani},\ and\
  \citenamefont {Riotto}}]{DeLuca:2020qqa}%
  \BibitemOpen
  \bibfield  {author} {\bibinfo {author} {\bibfnamefont {V.}~\bibnamefont
  {De~Luca}}, \bibinfo {author} {\bibfnamefont {G.}~\bibnamefont
  {Franciolini}}, \bibinfo {author} {\bibfnamefont {P.}~\bibnamefont {Pani}}, \
  and\ \bibinfo {author} {\bibfnamefont {A.}~\bibnamefont {Riotto}},\ }\href
  {\doibase 10.1088/1475-7516/2020/06/044} {\bibfield  {journal} {\bibinfo
  {journal} {JCAP}\ }\textbf {\bibinfo {volume} {06}},\ \bibinfo {pages} {044}
  (\bibinfo {year} {2020})},\ \Eprint {http://arxiv.org/abs/2005.05641}
  {arXiv:2005.05641 [astro-ph.CO]} \BibitemShut {NoStop}%
\bibitem [{\citenamefont {H\"utsi}\ \emph {et~al.}(2021)\citenamefont
  {H\"utsi}, \citenamefont {Raidal}, \citenamefont {Vaskonen},\ and\
  \citenamefont {Veerm\"ae}}]{Hutsi:2020sol}%
  \BibitemOpen
  \bibfield  {author} {\bibinfo {author} {\bibfnamefont {G.}~\bibnamefont
  {H\"utsi}}, \bibinfo {author} {\bibfnamefont {M.}~\bibnamefont {Raidal}},
  \bibinfo {author} {\bibfnamefont {V.}~\bibnamefont {Vaskonen}}, \ and\
  \bibinfo {author} {\bibfnamefont {H.}~\bibnamefont {Veerm\"ae}},\ }\href
  {\doibase 10.1088/1475-7516/2021/03/068} {\bibfield  {journal} {\bibinfo
  {journal} {JCAP}\ }\textbf {\bibinfo {volume} {03}},\ \bibinfo {pages} {068}
  (\bibinfo {year} {2021})},\ \Eprint {http://arxiv.org/abs/2012.02786}
  {arXiv:2012.02786 [astro-ph.CO]} \BibitemShut {NoStop}%
\bibitem [{\citenamefont {Franciolini}\ \emph {et~al.}(2022)\citenamefont
  {Franciolini}, \citenamefont {Baibhav}, \citenamefont {De~Luca},
  \citenamefont {Ng}, \citenamefont {Wong}, \citenamefont {Berti},
  \citenamefont {Pani}, \citenamefont {Riotto},\ and\ \citenamefont
  {Vitale}}]{Franciolini:2021tla}%
  \BibitemOpen
  \bibfield  {author} {\bibinfo {author} {\bibfnamefont {G.}~\bibnamefont
  {Franciolini}}, \bibinfo {author} {\bibfnamefont {V.}~\bibnamefont
  {Baibhav}}, \bibinfo {author} {\bibfnamefont {V.}~\bibnamefont {De~Luca}},
  \bibinfo {author} {\bibfnamefont {K.~K.~Y.}\ \bibnamefont {Ng}}, \bibinfo
  {author} {\bibfnamefont {K.~W.~K.}\ \bibnamefont {Wong}}, \bibinfo {author}
  {\bibfnamefont {E.}~\bibnamefont {Berti}}, \bibinfo {author} {\bibfnamefont
  {P.}~\bibnamefont {Pani}}, \bibinfo {author} {\bibfnamefont {A.}~\bibnamefont
  {Riotto}}, \ and\ \bibinfo {author} {\bibfnamefont {S.}~\bibnamefont
  {Vitale}},\ }\href {\doibase 10.1103/PhysRevD.105.083526} {\bibfield
  {journal} {\bibinfo  {journal} {Phys. Rev. D}\ }\textbf {\bibinfo {volume}
  {105}},\ \bibinfo {pages} {083526} (\bibinfo {year} {2022})},\ \Eprint
  {http://arxiv.org/abs/2105.03349} {arXiv:2105.03349 [gr-qc]} \BibitemShut
  {NoStop}%
\bibitem [{\citenamefont {Kovetz}(2017)}]{Kovetz:2017rvv}%
  \BibitemOpen
  \bibfield  {author} {\bibinfo {author} {\bibfnamefont {E.}~\bibnamefont
  {Kovetz}},\ }\href {\doibase 10.1103/PhysRevLett.119.131301} {\bibfield
  {journal} {\bibinfo  {journal} {Phys. Rev. Lett.}\ }\textbf {\bibinfo
  {volume} {119}},\ \bibinfo {pages} {131301} (\bibinfo {year} {2017})},\
  \Eprint {http://arxiv.org/abs/1705.09182} {arXiv:1705.09182 [astro-ph.CO]}
  \BibitemShut {NoStop}%
\bibitem [{\citenamefont {Niikura}\ \emph {et~al.}(2019)\citenamefont {Niikura}
  \emph {et~al.}}]{Niikura:2017zjd}%
  \BibitemOpen
  \bibfield  {author} {\bibinfo {author} {\bibfnamefont {H.}~\bibnamefont
  {Niikura}} \emph {et~al.},\ }\href {\doibase 10.1038/s41550-019-0723-1}
  {\bibfield  {journal} {\bibinfo  {journal} {Nature Astron.}\ }\textbf
  {\bibinfo {volume} {3}},\ \bibinfo {pages} {524} (\bibinfo {year} {2019})},\
  \Eprint {http://arxiv.org/abs/1701.02151} {arXiv:1701.02151 [astro-ph.CO]}
  \BibitemShut {NoStop}%
\bibitem [{\citenamefont {Smyth}\ \emph {et~al.}(2020)\citenamefont {Smyth},
  \citenamefont {Profumo}, \citenamefont {English}, \citenamefont {Jeltema},
  \citenamefont {McKinnon},\ and\ \citenamefont
  {Guhathakurta}}]{Smyth:2019whb}%
  \BibitemOpen
  \bibfield  {author} {\bibinfo {author} {\bibfnamefont {N.}~\bibnamefont
  {Smyth}}, \bibinfo {author} {\bibfnamefont {S.}~\bibnamefont {Profumo}},
  \bibinfo {author} {\bibfnamefont {S.}~\bibnamefont {English}}, \bibinfo
  {author} {\bibfnamefont {T.}~\bibnamefont {Jeltema}}, \bibinfo {author}
  {\bibfnamefont {K.}~\bibnamefont {McKinnon}}, \ and\ \bibinfo {author}
  {\bibfnamefont {P.}~\bibnamefont {Guhathakurta}},\ }\href {\doibase
  10.1103/PhysRevD.101.063005} {\bibfield  {journal} {\bibinfo  {journal}
  {Phys. Rev. D}\ }\textbf {\bibinfo {volume} {101}},\ \bibinfo {pages}
  {063005} (\bibinfo {year} {2020})},\ \Eprint
  {http://arxiv.org/abs/1910.01285} {arXiv:1910.01285 [astro-ph.CO]}
  \BibitemShut {NoStop}%
\bibitem [{\citenamefont {Croon}\ \emph {et~al.}(2020)\citenamefont {Croon},
  \citenamefont {McKeen}, \citenamefont {Raj},\ and\ \citenamefont
  {Wang}}]{Croon:2020ouk}%
  \BibitemOpen
  \bibfield  {author} {\bibinfo {author} {\bibfnamefont {D.}~\bibnamefont
  {Croon}}, \bibinfo {author} {\bibfnamefont {D.}~\bibnamefont {McKeen}},
  \bibinfo {author} {\bibfnamefont {N.}~\bibnamefont {Raj}}, \ and\ \bibinfo
  {author} {\bibfnamefont {Z.}~\bibnamefont {Wang}},\ }\href {\doibase
  10.1103/PhysRevD.102.083021} {\bibfield  {journal} {\bibinfo  {journal}
  {Phys. Rev. D}\ }\textbf {\bibinfo {volume} {102}},\ \bibinfo {pages}
  {083021} (\bibinfo {year} {2020})},\ \Eprint
  {http://arxiv.org/abs/2007.12697} {arXiv:2007.12697 [astro-ph.CO]}
  \BibitemShut {NoStop}%
\bibitem [{\citenamefont {Sugiyama}\ \emph {et~al.}(2020)\citenamefont
  {Sugiyama}, \citenamefont {Kurita},\ and\ \citenamefont
  {Takada}}]{Sugiyama:2019dgt}%
  \BibitemOpen
  \bibfield  {author} {\bibinfo {author} {\bibfnamefont {S.}~\bibnamefont
  {Sugiyama}}, \bibinfo {author} {\bibfnamefont {T.}~\bibnamefont {Kurita}}, \
  and\ \bibinfo {author} {\bibfnamefont {M.}~\bibnamefont {Takada}},\ }\href
  {\doibase 10.1093/mnras/staa407} {\bibfield  {journal} {\bibinfo  {journal}
  {Mon. Not. Roy. Astron. Soc.}\ }\textbf {\bibinfo {volume} {493}},\ \bibinfo
  {pages} {3632} (\bibinfo {year} {2020})},\ \Eprint
  {http://arxiv.org/abs/1905.06066} {arXiv:1905.06066 [astro-ph.CO]}
  \BibitemShut {NoStop}%
\bibitem [{\citenamefont {Blaineau}\ \emph {et~al.}(2022)\citenamefont
  {Blaineau} \emph {et~al.}}]{Blaineau:2022nhy}%
  \BibitemOpen
  \bibfield  {author} {\bibinfo {author} {\bibfnamefont {T.}~\bibnamefont
  {Blaineau}} \emph {et~al.},\ }\href {\doibase 10.1051/0004-6361/202243430}
  {\bibfield  {journal} {\bibinfo  {journal} {Astron. Astrophys.}\ }\textbf
  {\bibinfo {volume} {664}},\ \bibinfo {pages} {A106} (\bibinfo {year}
  {2022})},\ \Eprint {http://arxiv.org/abs/2202.13819} {arXiv:2202.13819
  [astro-ph.GA]} \BibitemShut {NoStop}%
\bibitem [{\citenamefont {Mroz}\ \emph {et~al.}(2024)\citenamefont {Mroz} \emph
  {et~al.}}]{Mroz:2024mse}%
  \BibitemOpen
  \bibfield  {author} {\bibinfo {author} {\bibfnamefont {P.}~\bibnamefont
  {Mroz}} \emph {et~al.},\ }\href@noop {} {\  (\bibinfo {year} {2024})},\
  \Eprint {http://arxiv.org/abs/2403.02386} {arXiv:2403.02386 [astro-ph.GA]}
  \BibitemShut {NoStop}%
\bibitem [{\citenamefont {Garcia-Bellido}\ and\ \citenamefont
  {Hawkins}(2024)}]{Garcia-Bellido:2024yaz}%
  \BibitemOpen
  \bibfield  {author} {\bibinfo {author} {\bibfnamefont {J.}~\bibnamefont
  {Garcia-Bellido}}\ and\ \bibinfo {author} {\bibfnamefont {M.}~\bibnamefont
  {Hawkins}},\ }\href@noop {} {\  (\bibinfo {year} {2024})},\ \Eprint
  {http://arxiv.org/abs/2402.00212} {arXiv:2402.00212 [astro-ph.GA]}
  \BibitemShut {NoStop}%
\bibitem [{\citenamefont {Zumalacarregui}\ and\ \citenamefont
  {Seljak}(2018)}]{Zumalacarregui:2017qqd}%
  \BibitemOpen
  \bibfield  {author} {\bibinfo {author} {\bibfnamefont {M.}~\bibnamefont
  {Zumalacarregui}}\ and\ \bibinfo {author} {\bibfnamefont {U.}~\bibnamefont
  {Seljak}},\ }\href {\doibase 10.1103/PhysRevLett.121.141101} {\bibfield
  {journal} {\bibinfo  {journal} {Phys. Rev. Lett.}\ }\textbf {\bibinfo
  {volume} {121}},\ \bibinfo {pages} {141101} (\bibinfo {year} {2018})},\
  \Eprint {http://arxiv.org/abs/1712.02240} {arXiv:1712.02240 [astro-ph.CO]}
  \BibitemShut {NoStop}%
\bibitem [{\citenamefont {Oguri}\ \emph {et~al.}(2018)\citenamefont {Oguri},
  \citenamefont {Diego}, \citenamefont {Kaiser}, \citenamefont {Kelly},\ and\
  \citenamefont {Broadhurst}}]{Oguri:2017ock}%
  \BibitemOpen
  \bibfield  {author} {\bibinfo {author} {\bibfnamefont {M.}~\bibnamefont
  {Oguri}}, \bibinfo {author} {\bibfnamefont {J.~M.}\ \bibnamefont {Diego}},
  \bibinfo {author} {\bibfnamefont {N.}~\bibnamefont {Kaiser}}, \bibinfo
  {author} {\bibfnamefont {P.~L.}\ \bibnamefont {Kelly}}, \ and\ \bibinfo
  {author} {\bibfnamefont {T.}~\bibnamefont {Broadhurst}},\ }\href {\doibase
  10.1103/PhysRevD.97.023518} {\bibfield  {journal} {\bibinfo  {journal} {Phys.
  Rev. D}\ }\textbf {\bibinfo {volume} {97}},\ \bibinfo {pages} {023518}
  (\bibinfo {year} {2018})},\ \Eprint {http://arxiv.org/abs/1710.00148}
  {arXiv:1710.00148 [astro-ph.CO]} \BibitemShut {NoStop}%
\bibitem [{\citenamefont {M\"uller}\ and\ \citenamefont
  {Miralda-Escud\'e}(2024)}]{Muller:2024pwn}%
  \BibitemOpen
  \bibfield  {author} {\bibinfo {author} {\bibfnamefont {C.~V.}\ \bibnamefont
  {M\"uller}}\ and\ \bibinfo {author} {\bibfnamefont {J.}~\bibnamefont
  {Miralda-Escud\'e}},\ }\href@noop {} {\  (\bibinfo {year} {2024})},\ \Eprint
  {http://arxiv.org/abs/2403.16989} {arXiv:2403.16989 [astro-ph.CO]}
  \BibitemShut {NoStop}%
\bibitem [{\citenamefont {Venumadhav}\ \emph {et~al.}(2017)\citenamefont
  {Venumadhav}, \citenamefont {Dai},\ and\ \citenamefont
  {Miralda-Escud\'e}}]{Venumadhav:2017pps}%
  \BibitemOpen
  \bibfield  {author} {\bibinfo {author} {\bibfnamefont {T.}~\bibnamefont
  {Venumadhav}}, \bibinfo {author} {\bibfnamefont {L.}~\bibnamefont {Dai}}, \
  and\ \bibinfo {author} {\bibfnamefont {J.}~\bibnamefont {Miralda-Escud\'e}},\
  }\href {\doibase 10.3847/1538-4357/aa9575} {\bibfield  {journal} {\bibinfo
  {journal} {Astrophys. J.}\ }\textbf {\bibinfo {volume} {850}},\ \bibinfo
  {pages} {49} (\bibinfo {year} {2017})},\ \Eprint
  {http://arxiv.org/abs/1707.00003} {arXiv:1707.00003 [astro-ph.CO]}
  \BibitemShut {NoStop}%
\bibitem [{\citenamefont {Esteban-Guti\'errez}\ \emph
  {et~al.}(2023)\citenamefont {Esteban-Guti\'errez}, \citenamefont
  {Mediavilla}, \citenamefont {Jim\'enez-Vicente},\ and\ \citenamefont
  {Mu\~noz}}]{Esteban-Gutierrez:2023qcz}%
  \BibitemOpen
  \bibfield  {author} {\bibinfo {author} {\bibfnamefont {A.}~\bibnamefont
  {Esteban-Guti\'errez}}, \bibinfo {author} {\bibfnamefont {E.}~\bibnamefont
  {Mediavilla}}, \bibinfo {author} {\bibfnamefont {J.}~\bibnamefont
  {Jim\'enez-Vicente}}, \ and\ \bibinfo {author} {\bibfnamefont {J.~A.}\
  \bibnamefont {Mu\~noz}},\ }\href {\doibase 10.3847/1538-4357/ace62f}
  {\bibfield  {journal} {\bibinfo  {journal} {Astrophys. J.}\ }\textbf
  {\bibinfo {volume} {954}},\ \bibinfo {pages} {172} (\bibinfo {year}
  {2023})},\ \Eprint {http://arxiv.org/abs/2307.07473} {arXiv:2307.07473
  [astro-ph.CO]} \BibitemShut {NoStop}%
\bibitem [{\citenamefont {Arbey}\ and\ \citenamefont
  {Auffinger}(2019)}]{Arbey:2019mbc}%
  \BibitemOpen
  \bibfield  {author} {\bibinfo {author} {\bibfnamefont {A.}~\bibnamefont
  {Arbey}}\ and\ \bibinfo {author} {\bibfnamefont {J.}~\bibnamefont
  {Auffinger}},\ }\href {\doibase 10.1140/epjc/s10052-019-7161-1} {\bibfield
  {journal} {\bibinfo  {journal} {Eur. Phys. J. C}\ }\textbf {\bibinfo {volume}
  {79}},\ \bibinfo {pages} {693} (\bibinfo {year} {2019})},\ \Eprint
  {http://arxiv.org/abs/1905.04268} {arXiv:1905.04268 [gr-qc]} \BibitemShut
  {NoStop}%
\bibitem [{\citenamefont {Arbey}\ and\ \citenamefont
  {Auffinger}(2021)}]{Arbey:2021mbl}%
  \BibitemOpen
  \bibfield  {author} {\bibinfo {author} {\bibfnamefont {A.}~\bibnamefont
  {Arbey}}\ and\ \bibinfo {author} {\bibfnamefont {J.}~\bibnamefont
  {Auffinger}},\ }\href {\doibase 10.1140/epjc/s10052-021-09702-8} {\bibfield
  {journal} {\bibinfo  {journal} {Eur. Phys. J. C}\ }\textbf {\bibinfo {volume}
  {81}},\ \bibinfo {pages} {910} (\bibinfo {year} {2021})},\ \Eprint
  {http://arxiv.org/abs/2108.02737} {arXiv:2108.02737 [gr-qc]} \BibitemShut
  {NoStop}%
\bibitem [{\citenamefont {Auffinger}(2023)}]{Auffinger:2022khh}%
  \BibitemOpen
  \bibfield  {author} {\bibinfo {author} {\bibfnamefont {J.}~\bibnamefont
  {Auffinger}},\ }\href {\doibase 10.1016/j.ppnp.2023.104040} {\bibfield
  {journal} {\bibinfo  {journal} {Prog. Part. Nucl. Phys.}\ }\textbf {\bibinfo
  {volume} {131}},\ \bibinfo {pages} {104040} (\bibinfo {year} {2023})},\
  \Eprint {http://arxiv.org/abs/2206.02672} {arXiv:2206.02672 [astro-ph.CO]}
  \BibitemShut {NoStop}%
\bibitem [{\citenamefont {Tian}\ \emph {et~al.}(2019)\citenamefont {Tian},
  \citenamefont {El-Badry}, \citenamefont {Rix},\ and\ \citenamefont
  {Gould}}]{Tian_2019}%
  \BibitemOpen
  \bibfield  {author} {\bibinfo {author} {\bibfnamefont {H.-J.}\ \bibnamefont
  {Tian}}, \bibinfo {author} {\bibfnamefont {K.}~\bibnamefont {El-Badry}},
  \bibinfo {author} {\bibfnamefont {H.-W.}\ \bibnamefont {Rix}}, \ and\
  \bibinfo {author} {\bibfnamefont {A.}~\bibnamefont {Gould}},\ }\href
  {\doibase 10.3847/1538-4365/ab54c4} {\bibfield  {journal} {\bibinfo
  {journal} {The Astrophysical Journal Supplement Series}\ }\textbf {\bibinfo
  {volume} {246}},\ \bibinfo {pages} {4} (\bibinfo {year} {2019})}\BibitemShut
  {NoStop}%
\bibitem [{\citenamefont {Brandt}(2016)}]{Brandt:2016aco}%
  \BibitemOpen
  \bibfield  {author} {\bibinfo {author} {\bibfnamefont {T.~D.}\ \bibnamefont
  {Brandt}},\ }\href {\doibase 10.3847/2041-8205/824/2/L31} {\bibfield
  {journal} {\bibinfo  {journal} {Astrophys. J. Lett.}\ }\textbf {\bibinfo
  {volume} {824}},\ \bibinfo {pages} {L31} (\bibinfo {year} {2016})},\ \Eprint
  {http://arxiv.org/abs/1605.03665} {arXiv:1605.03665 [astro-ph.GA]}
  \BibitemShut {NoStop}%
\bibitem [{\citenamefont {Zhu}\ \emph {et~al.}(2018)\citenamefont {Zhu},
  \citenamefont {Vasiliev}, \citenamefont {Li},\ and\ \citenamefont
  {Jing}}]{Zhu:2017plg}%
  \BibitemOpen
  \bibfield  {author} {\bibinfo {author} {\bibfnamefont {Q.}~\bibnamefont
  {Zhu}}, \bibinfo {author} {\bibfnamefont {E.}~\bibnamefont {Vasiliev}},
  \bibinfo {author} {\bibfnamefont {Y.}~\bibnamefont {Li}}, \ and\ \bibinfo
  {author} {\bibfnamefont {Y.}~\bibnamefont {Jing}},\ }\href {\doibase
  10.1093/mnras/sty079} {\bibfield  {journal} {\bibinfo  {journal} {Mon. Not.
  Roy. Astron. Soc.}\ }\textbf {\bibinfo {volume} {476}},\ \bibinfo {pages} {2}
  (\bibinfo {year} {2018})},\ \Eprint {http://arxiv.org/abs/1710.05032}
  {arXiv:1710.05032 [astro-ph.CO]} \BibitemShut {NoStop}%
\bibitem [{\citenamefont {Stegmann}\ \emph {et~al.}(2020)\citenamefont
  {Stegmann}, \citenamefont {Capelo}, \citenamefont {Bortolas},\ and\
  \citenamefont {Mayer}}]{Stegmann:2019wyz}%
  \BibitemOpen
  \bibfield  {author} {\bibinfo {author} {\bibfnamefont {J.}~\bibnamefont
  {Stegmann}}, \bibinfo {author} {\bibfnamefont {P.~R.}\ \bibnamefont
  {Capelo}}, \bibinfo {author} {\bibfnamefont {E.}~\bibnamefont {Bortolas}}, \
  and\ \bibinfo {author} {\bibfnamefont {L.}~\bibnamefont {Mayer}},\ }\href
  {\doibase 10.1093/mnras/staa170} {\bibfield  {journal} {\bibinfo  {journal}
  {Mon. Not. Roy. Astron. Soc.}\ }\textbf {\bibinfo {volume} {492}},\ \bibinfo
  {pages} {5247} (\bibinfo {year} {2020})},\ \Eprint
  {http://arxiv.org/abs/1910.04793} {arXiv:1910.04793 [astro-ph.GA]}
  \BibitemShut {NoStop}%
\bibitem [{\citenamefont {Poulin}\ \emph {et~al.}(2017)\citenamefont {Poulin},
  \citenamefont {Serpico}, \citenamefont {Calore}, \citenamefont {Clesse},\
  and\ \citenamefont {Kohri}}]{Poulin:2017bwe}%
  \BibitemOpen
  \bibfield  {author} {\bibinfo {author} {\bibfnamefont {V.}~\bibnamefont
  {Poulin}}, \bibinfo {author} {\bibfnamefont {P.~D.}\ \bibnamefont {Serpico}},
  \bibinfo {author} {\bibfnamefont {F.}~\bibnamefont {Calore}}, \bibinfo
  {author} {\bibfnamefont {S.}~\bibnamefont {Clesse}}, \ and\ \bibinfo {author}
  {\bibfnamefont {K.}~\bibnamefont {Kohri}},\ }\href {\doibase
  10.1103/PhysRevD.96.083524} {\bibfield  {journal} {\bibinfo  {journal} {Phys.
  Rev. D}\ }\textbf {\bibinfo {volume} {96}},\ \bibinfo {pages} {083524}
  (\bibinfo {year} {2017})},\ \Eprint {http://arxiv.org/abs/1707.04206}
  {arXiv:1707.04206 [astro-ph.CO]} \BibitemShut {NoStop}%
\bibitem [{\citenamefont {Gaggero}\ \emph {et~al.}(2017)\citenamefont
  {Gaggero}, \citenamefont {Bertone}, \citenamefont {Calore}, \citenamefont
  {Connors}, \citenamefont {Lovell}, \citenamefont {Markoff},\ and\
  \citenamefont {Storm}}]{Gaggero:2016dpq}%
  \BibitemOpen
  \bibfield  {author} {\bibinfo {author} {\bibfnamefont {D.}~\bibnamefont
  {Gaggero}}, \bibinfo {author} {\bibfnamefont {G.}~\bibnamefont {Bertone}},
  \bibinfo {author} {\bibfnamefont {F.}~\bibnamefont {Calore}}, \bibinfo
  {author} {\bibfnamefont {R.~M.~T.}\ \bibnamefont {Connors}}, \bibinfo
  {author} {\bibfnamefont {M.}~\bibnamefont {Lovell}}, \bibinfo {author}
  {\bibfnamefont {S.}~\bibnamefont {Markoff}}, \ and\ \bibinfo {author}
  {\bibfnamefont {E.}~\bibnamefont {Storm}},\ }\href {\doibase
  10.1103/PhysRevLett.118.241101} {\bibfield  {journal} {\bibinfo  {journal}
  {Phys. Rev. Lett.}\ }\textbf {\bibinfo {volume} {118}},\ \bibinfo {pages}
  {241101} (\bibinfo {year} {2017})},\ \Eprint
  {http://arxiv.org/abs/1612.00457} {arXiv:1612.00457 [astro-ph.HE]}
  \BibitemShut {NoStop}%
\bibitem [{\citenamefont {Manshanden}\ \emph {et~al.}(2019)\citenamefont
  {Manshanden}, \citenamefont {Gaggero}, \citenamefont {Bertone}, \citenamefont
  {Connors},\ and\ \citenamefont {Ricotti}}]{Manshanden:2018tze}%
  \BibitemOpen
  \bibfield  {author} {\bibinfo {author} {\bibfnamefont {J.}~\bibnamefont
  {Manshanden}}, \bibinfo {author} {\bibfnamefont {D.}~\bibnamefont {Gaggero}},
  \bibinfo {author} {\bibfnamefont {G.}~\bibnamefont {Bertone}}, \bibinfo
  {author} {\bibfnamefont {R.~M.~T.}\ \bibnamefont {Connors}}, \ and\ \bibinfo
  {author} {\bibfnamefont {M.}~\bibnamefont {Ricotti}},\ }\href {\doibase
  10.1088/1475-7516/2019/06/026} {\bibfield  {journal} {\bibinfo  {journal}
  {JCAP}\ }\textbf {\bibinfo {volume} {06}},\ \bibinfo {pages} {026} (\bibinfo
  {year} {2019})},\ \Eprint {http://arxiv.org/abs/1812.07967} {arXiv:1812.07967
  [astro-ph.HE]} \BibitemShut {NoStop}%
\bibitem [{\citenamefont {Inoue}\ and\ \citenamefont
  {Kusenko}(2017)}]{Inoue:2017csr}%
  \BibitemOpen
  \bibfield  {author} {\bibinfo {author} {\bibfnamefont {Y.}~\bibnamefont
  {Inoue}}\ and\ \bibinfo {author} {\bibfnamefont {A.}~\bibnamefont
  {Kusenko}},\ }\href {\doibase 10.1088/1475-7516/2017/10/034} {\bibfield
  {journal} {\bibinfo  {journal} {JCAP}\ }\textbf {\bibinfo {volume} {10}},\
  \bibinfo {pages} {034} (\bibinfo {year} {2017})},\ \Eprint
  {http://arxiv.org/abs/1705.00791} {arXiv:1705.00791 [astro-ph.CO]}
  \BibitemShut {NoStop}%
\bibitem [{\citenamefont {Lu}\ \emph {et~al.}(2021)\citenamefont {Lu},
  \citenamefont {Takhistov}, \citenamefont {Gelmini}, \citenamefont {Hayashi},
  \citenamefont {Inoue},\ and\ \citenamefont {Kusenko}}]{Lu:2020bmd}%
  \BibitemOpen
  \bibfield  {author} {\bibinfo {author} {\bibfnamefont {P.}~\bibnamefont
  {Lu}}, \bibinfo {author} {\bibfnamefont {V.}~\bibnamefont {Takhistov}},
  \bibinfo {author} {\bibfnamefont {G.~B.}\ \bibnamefont {Gelmini}}, \bibinfo
  {author} {\bibfnamefont {K.}~\bibnamefont {Hayashi}}, \bibinfo {author}
  {\bibfnamefont {Y.}~\bibnamefont {Inoue}}, \ and\ \bibinfo {author}
  {\bibfnamefont {A.}~\bibnamefont {Kusenko}},\ }\href {\doibase
  10.3847/2041-8213/abdcb6} {\bibfield  {journal} {\bibinfo  {journal}
  {Astrophys. J. Lett.}\ }\textbf {\bibinfo {volume} {908}},\ \bibinfo {pages}
  {L23} (\bibinfo {year} {2021})},\ \Eprint {http://arxiv.org/abs/2007.02213}
  {arXiv:2007.02213 [astro-ph.CO]} \BibitemShut {NoStop}%
\bibitem [{\citenamefont {Fuller}\ \emph {et~al.}(2017)\citenamefont {Fuller},
  \citenamefont {Kusenko},\ and\ \citenamefont {Takhistov}}]{Fuller:2017uyd}%
  \BibitemOpen
  \bibfield  {author} {\bibinfo {author} {\bibfnamefont {G.}~\bibnamefont
  {Fuller}}, \bibinfo {author} {\bibfnamefont {A.}~\bibnamefont {Kusenko}}, \
  and\ \bibinfo {author} {\bibfnamefont {V.}~\bibnamefont {Takhistov}},\ }\href
  {\doibase 10.1103/PhysRevLett.119.061101} {\bibfield  {journal} {\bibinfo
  {journal} {Phys. Rev. Lett.}\ }\textbf {\bibinfo {volume} {119}},\ \bibinfo
  {pages} {061101} (\bibinfo {year} {2017})},\ \Eprint
  {http://arxiv.org/abs/1704.01129} {arXiv:1704.01129 [astro-ph.HE]}
  \BibitemShut {NoStop}%
\bibitem [{\citenamefont {Abramowicz}\ \emph {et~al.}(2018)\citenamefont
  {Abramowicz}, \citenamefont {Bejger},\ and\ \citenamefont
  {Wielgus}}]{Abramowicz:2017zbp}%
  \BibitemOpen
  \bibfield  {author} {\bibinfo {author} {\bibfnamefont {M.}~\bibnamefont
  {Abramowicz}}, \bibinfo {author} {\bibfnamefont {M.}~\bibnamefont {Bejger}},
  \ and\ \bibinfo {author} {\bibfnamefont {M.}~\bibnamefont {Wielgus}},\ }\href
  {\doibase 10.3847/1538-4357/aae64a} {\bibfield  {journal} {\bibinfo
  {journal} {Astrophys. J.}\ }\textbf {\bibinfo {volume} {868}},\ \bibinfo
  {pages} {17} (\bibinfo {year} {2018})},\ \Eprint
  {http://arxiv.org/abs/1704.05931} {arXiv:1704.05931 [astro-ph.HE]}
  \BibitemShut {NoStop}%
\bibitem [{\citenamefont {Carr}\ and\ \citenamefont
  {Silk}(2018)}]{Carr:2018rid}%
  \BibitemOpen
  \bibfield  {author} {\bibinfo {author} {\bibfnamefont {B.}~\bibnamefont
  {Carr}}\ and\ \bibinfo {author} {\bibfnamefont {J.}~\bibnamefont {Silk}},\
  }\href {\doibase 10.1093/mnras/sty1204} {\bibfield  {journal} {\bibinfo
  {journal} {Mon. Not. Roy. Astron. Soc.}\ }\textbf {\bibinfo {volume} {478}},\
  \bibinfo {pages} {3756} (\bibinfo {year} {2018})}\BibitemShut {NoStop}%
\bibitem [{\citenamefont {{Kashlinsky}}\ \emph {et~al.}(2005)\citenamefont
  {{Kashlinsky}}, \citenamefont {{Arendt}}, \citenamefont {{Mather}},\ and\
  \citenamefont {{Moseley}}}]{2005Natur.438...45K}%
  \BibitemOpen
  \bibfield  {author} {\bibinfo {author} {\bibfnamefont {A.}~\bibnamefont
  {{Kashlinsky}}}, \bibinfo {author} {\bibfnamefont {R.~G.}\ \bibnamefont
  {{Arendt}}}, \bibinfo {author} {\bibfnamefont {J.}~\bibnamefont {{Mather}}},
  \ and\ \bibinfo {author} {\bibfnamefont {S.~H.}\ \bibnamefont {{Moseley}}},\
  }\href {\doibase 10.1038/nature04143} {\bibfield  {journal} {\bibinfo
  {journal} {\nat}\ }\textbf {\bibinfo {volume} {438}},\ \bibinfo {pages} {45}
  (\bibinfo {year} {2005})},\ \Eprint {http://arxiv.org/abs/astro-ph/0511105}
  {arXiv:astro-ph/0511105 [astro-ph]} \BibitemShut {NoStop}%
\bibitem [{\citenamefont {{Nico Cappelluti {\it et
  al.}}}(2013)}]{2013ApJ...769...68C}%
  \BibitemOpen
  \bibfield  {author} {\bibinfo {author} {\bibnamefont {{Nico Cappelluti {\it
  et al.}}}},\ }\href {\doibase 10.1088/0004-637X/769/1/68} {\bibfield
  {journal} {\bibinfo  {journal} {Astrophys. J.}\ }\textbf {\bibinfo {volume}
  {769}},\ \bibinfo {eid} {68} (\bibinfo {year} {2013})},\ \Eprint
  {http://arxiv.org/abs/1210.5302} {arXiv:1210.5302 [astro-ph.CO]} \BibitemShut
  {NoStop}%
\bibitem [{\citenamefont {{Kashlinsky}}\ \emph {et~al.}(2018)\citenamefont
  {{Kashlinsky}}, \citenamefont {{Arendt}}, \citenamefont {{Atrio-Barandela}},
  \citenamefont {{Cappelluti}}, \citenamefont {{Ferrara}},\ and\ \citenamefont
  {{Hasinger}}}]{2018RvMP...90b5006K}%
  \BibitemOpen
  \bibfield  {author} {\bibinfo {author} {\bibfnamefont {A.}~\bibnamefont
  {{Kashlinsky}}}, \bibinfo {author} {\bibfnamefont {R.}~\bibnamefont
  {{Arendt}}}, \bibinfo {author} {\bibfnamefont {F.}~\bibnamefont
  {{Atrio-Barandela}}}, \bibinfo {author} {\bibfnamefont {N.}~\bibnamefont
  {{Cappelluti}}}, \bibinfo {author} {\bibfnamefont {A.}~\bibnamefont
  {{Ferrara}}}, \ and\ \bibinfo {author} {\bibfnamefont {G.}~\bibnamefont
  {{Hasinger}}},\ }\href {\doibase 10.1103/RevModPhys.90.025006} {\bibfield
  {journal} {\bibinfo  {journal} {Reviews of Modern Physics}\ }\textbf
  {\bibinfo {volume} {90}},\ \bibinfo {eid} {025006} (\bibinfo {year}
  {2018})},\ \Eprint {http://arxiv.org/abs/1802.07774} {arXiv:1802.07774
  [astro-ph.CO]} \BibitemShut {NoStop}%
\bibitem [{\citenamefont {{Kashlinsky}}\ \emph {et~al.}(2019)\citenamefont
  {{Kashlinsky}}, \citenamefont {{Arendt}}, \citenamefont {{Cappelluti}},
  \citenamefont {{Finoguenov}}, \citenamefont {{Hasinger}}, \citenamefont
  {{Helgason}},\ and\ \citenamefont {Kari}}]{2019ApJ...871L...6K}%
  \BibitemOpen
  \bibfield  {author} {\bibinfo {author} {\bibfnamefont {A.}~\bibnamefont
  {{Kashlinsky}}}, \bibinfo {author} {\bibfnamefont {R.}~\bibnamefont
  {{Arendt}}}, \bibinfo {author} {\bibfnamefont {N.}~\bibnamefont
  {{Cappelluti}}}, \bibinfo {author} {\bibfnamefont {A.}~\bibnamefont
  {{Finoguenov}}}, \bibinfo {author} {\bibfnamefont {G.}~\bibnamefont
  {{Hasinger}}}, \bibinfo {author} {\bibnamefont {{Helgason}}}, \ and\ \bibinfo
  {author} {\bibfnamefont {A.}~\bibnamefont {Kari}, \bibfnamefont
  {{Merloni}}},\ }\href {\doibase 10.3847/2041-8213/aafaf6} {\bibfield
  {journal} {\bibinfo  {journal} {Astrophys. J. Lett.}\ }\textbf {\bibinfo
  {volume} {871}},\ \bibinfo {eid} {L6} (\bibinfo {year} {2019})},\ \Eprint
  {http://arxiv.org/abs/1812.01535} {arXiv:1812.01535 [astro-ph.CO]}
  \BibitemShut {NoStop}%
\bibitem [{\citenamefont {Hasinger}(2020)}]{Hasinger:2020ptw}%
  \BibitemOpen
  \bibfield  {author} {\bibinfo {author} {\bibfnamefont {G.}~\bibnamefont
  {Hasinger}},\ }\href {\doibase 10.1088/1475-7516/2020/07/022} {\bibfield
  {journal} {\bibinfo  {journal} {JCAP}\ }\textbf {\bibinfo {volume} {07}},\
  \bibinfo {pages} {022} (\bibinfo {year} {2020})},\ \Eprint
  {http://arxiv.org/abs/2003.05150} {arXiv:2003.05150 [astro-ph.CO]}
  \BibitemShut {NoStop}%
\bibitem [{\citenamefont {Cappelluti}\ \emph {et~al.}(2022)\citenamefont
  {Cappelluti}, \citenamefont {Hasinger},\ and\ \citenamefont
  {Natarajan}}]{Cappelluti:2021usg}%
  \BibitemOpen
  \bibfield  {author} {\bibinfo {author} {\bibfnamefont {N.}~\bibnamefont
  {Cappelluti}}, \bibinfo {author} {\bibfnamefont {G.}~\bibnamefont
  {Hasinger}}, \ and\ \bibinfo {author} {\bibfnamefont {P.}~\bibnamefont
  {Natarajan}},\ }\href {\doibase 10.3847/1538-4357/ac332d} {\bibfield
  {journal} {\bibinfo  {journal} {Astrophys. J.}\ }\textbf {\bibinfo {volume}
  {926}},\ \bibinfo {pages} {205} (\bibinfo {year} {2022})},\ \Eprint
  {http://arxiv.org/abs/2109.08701} {arXiv:2109.08701 [astro-ph.CO]}
  \BibitemShut {NoStop}%
\bibitem [{\citenamefont {Silk}(2017)}]{Silk:2017yai}%
  \BibitemOpen
  \bibfield  {author} {\bibinfo {author} {\bibfnamefont {J.}~\bibnamefont
  {Silk}},\ }\href {\doibase 10.3847/2041-8213/aa67da} {\bibfield  {journal}
  {\bibinfo  {journal} {Astrophys. J. Lett.}\ }\textbf {\bibinfo {volume}
  {839}},\ \bibinfo {pages} {L13} (\bibinfo {year} {2017})},\ \Eprint
  {http://arxiv.org/abs/1703.08553} {arXiv:1703.08553 [astro-ph.GA]}
  \BibitemShut {NoStop}%
\bibitem [{\citenamefont {{Kormendy}}\ and\ \citenamefont
  {{Ho}}(2013)}]{2013ARAA..51..511K}%
  \BibitemOpen
  \bibfield  {author} {\bibinfo {author} {\bibfnamefont {J.}~\bibnamefont
  {{Kormendy}}}\ and\ \bibinfo {author} {\bibfnamefont {L.~C.}\ \bibnamefont
  {{Ho}}},\ }\href {\doibase 10.1146/annurev-astro-082708-101811} {\bibfield
  {journal} {\bibinfo  {journal} {Annual Review Astronomy and Astrophysics}\
  }\textbf {\bibinfo {volume} {51}},\ \bibinfo {pages} {511} (\bibinfo {year}
  {2013})},\ \Eprint {http://arxiv.org/abs/1304.7762} {arXiv:1304.7762
  [astro-ph.CO]} \BibitemShut {NoStop}%
\bibitem [{\citenamefont {{Pardo}}\ \emph {et~al.}(2016)\citenamefont {{Pardo}}
  \emph {et~al.}}]{2016ApJ...831..203P}%
  \BibitemOpen
  \bibfield  {author} {\bibinfo {author} {\bibfnamefont {K.}~\bibnamefont
  {{Pardo}}} \emph {et~al.},\ }\href {\doibase 10.3847/0004-637X/831/2/203}
  {\bibfield  {journal} {\bibinfo  {journal} {Astrophys. J.}\ }\textbf
  {\bibinfo {volume} {831}},\ \bibinfo {eid} {203} (\bibinfo {year} {2016})},\
  \Eprint {http://arxiv.org/abs/1603.01622} {arXiv:1603.01622} \BibitemShut
  {NoStop}%
\bibitem [{\citenamefont {Baldassare}\ \emph {et~al.}(2017)\citenamefont
  {Baldassare}, \citenamefont {Reines}, \citenamefont {Gallo},\ and\
  \citenamefont {Greene}}]{Baldassare:2016cox}%
  \BibitemOpen
  \bibfield  {author} {\bibinfo {author} {\bibfnamefont {V.}~\bibnamefont
  {Baldassare}}, \bibinfo {author} {\bibfnamefont {A.}~\bibnamefont {Reines}},
  \bibinfo {author} {\bibfnamefont {E.}~\bibnamefont {Gallo}}, \ and\ \bibinfo
  {author} {\bibfnamefont {J.}~\bibnamefont {Greene}},\ }\href {\doibase
  10.3847/1538-4357/836/1/20} {\bibfield  {journal} {\bibinfo  {journal}
  {Astrophys. J.}\ }\textbf {\bibinfo {volume} {836}},\ \bibinfo {pages} {20}
  (\bibinfo {year} {2017})},\ \Eprint {http://arxiv.org/abs/1609.07148}
  {arXiv:1609.07148 [astro-ph.HE]} \BibitemShut {NoStop}%
\bibitem [{\citenamefont {Labb{\'{e}}}\ \emph {et~al.}(2023)\citenamefont
  {Labb{\'{e}}} \emph {et~al.}}]{Labb__2023}%
  \BibitemOpen
  \bibfield  {author} {\bibinfo {author} {\bibfnamefont {I.}~\bibnamefont
  {Labb{\'{e}}}} \emph {et~al.},\ }\href {\doibase 10.1038/s41586-023-05786-2}
  {\bibfield  {journal} {\bibinfo  {journal} {Nature}\ }\textbf {\bibinfo
  {volume} {616}},\ \bibinfo {pages} {266} (\bibinfo {year}
  {2023})}\BibitemShut {NoStop}%
\bibitem [{\citenamefont {Endsley}\ \emph {et~al.}(2023)\citenamefont
  {Endsley}, \citenamefont {Stark}, \citenamefont {Lyu}, \citenamefont {Wang},
  \citenamefont {Yang}, \citenamefont {Fan}, \citenamefont {Smit},
  \citenamefont {Bouwens}, \citenamefont {Hainline},\ and\ \citenamefont
  {Schouws}}]{Endsley_2023}%
  \BibitemOpen
  \bibfield  {author} {\bibinfo {author} {\bibfnamefont {R.}~\bibnamefont
  {Endsley}}, \bibinfo {author} {\bibfnamefont {D.}~\bibnamefont {Stark}},
  \bibinfo {author} {\bibfnamefont {J.}~\bibnamefont {Lyu}}, \bibinfo {author}
  {\bibfnamefont {F.}~\bibnamefont {Wang}}, \bibinfo {author} {\bibfnamefont
  {J.}~\bibnamefont {Yang}}, \bibinfo {author} {\bibfnamefont {X.}~\bibnamefont
  {Fan}}, \bibinfo {author} {\bibfnamefont {R.}~\bibnamefont {Smit}}, \bibinfo
  {author} {\bibfnamefont {R.}~\bibnamefont {Bouwens}}, \bibinfo {author}
  {\bibfnamefont {K.}~\bibnamefont {Hainline}}, \ and\ \bibinfo {author}
  {\bibfnamefont {S.}~\bibnamefont {Schouws}},\ }\href {\doibase
  10.1093/mnras/stad266} {\ \textbf {\bibinfo {volume} {520}},\ \bibinfo
  {pages} {4609} (\bibinfo {year} {2023})}\BibitemShut {NoStop}%
\bibitem [{\citenamefont {Dolgov}(2023)}]{Dolgov:2023ijt}%
  \BibitemOpen
  \bibfield  {author} {\bibinfo {author} {\bibfnamefont {A.}~\bibnamefont
  {Dolgov}},\ }in\ \href@noop {} {\emph {\bibinfo {booktitle} {{14th Frascati
  workshop on Multifrequency Behaviour of High Energy Cosmic Sources}}}}\
  (\bibinfo {year} {2023})\ \Eprint {http://arxiv.org/abs/2310.00671}
  {arXiv:2310.00671 [astro-ph.CO]} \BibitemShut {NoStop}%
\bibitem [{\citenamefont {Liu}\ and\ \citenamefont
  {Bromm}(2022)}]{Liu:2022bvr}%
  \BibitemOpen
  \bibfield  {author} {\bibinfo {author} {\bibfnamefont {B.}~\bibnamefont
  {Liu}}\ and\ \bibinfo {author} {\bibfnamefont {V.}~\bibnamefont {Bromm}},\
  }\href {\doibase 10.3847/2041-8213/ac927f} {\bibfield  {journal} {\bibinfo
  {journal} {Astrophys. J. Lett.}\ }\textbf {\bibinfo {volume} {937}},\
  \bibinfo {pages} {L30} (\bibinfo {year} {2022})},\ \Eprint
  {http://arxiv.org/abs/2208.13178} {arXiv:2208.13178 [astro-ph.CO]}
  \BibitemShut {NoStop}%
\bibitem [{\citenamefont {Byrnes}\ \emph {et~al.}(2019)\citenamefont {Byrnes},
  \citenamefont {Cole},\ and\ \citenamefont {Patil}}]{Byrnes:2018txb}%
  \BibitemOpen
  \bibfield  {author} {\bibinfo {author} {\bibfnamefont {C.}~\bibnamefont
  {Byrnes}}, \bibinfo {author} {\bibfnamefont {P.}~\bibnamefont {Cole}}, \ and\
  \bibinfo {author} {\bibfnamefont {S.}~\bibnamefont {Patil}},\ }\href
  {\doibase 10.1088/1475-7516/2019/06/028} {\bibfield  {journal} {\bibinfo
  {journal} {JCAP}\ }\textbf {\bibinfo {volume} {1906}},\ \bibinfo {pages}
  {028} (\bibinfo {year} {2019})},\ \Eprint {http://arxiv.org/abs/1811.11158}
  {arXiv:1811.11158 [astro-ph.CO]} \BibitemShut {NoStop}%
\bibitem [{\citenamefont {Chluba}\ \emph {et~al.}(2012)\citenamefont {Chluba},
  \citenamefont {Erickcek},\ and\ \citenamefont {Ben-Dayan}}]{Chluba:2012we}%
  \BibitemOpen
  \bibfield  {author} {\bibinfo {author} {\bibfnamefont {J.}~\bibnamefont
  {Chluba}}, \bibinfo {author} {\bibfnamefont {A.~L.}\ \bibnamefont
  {Erickcek}}, \ and\ \bibinfo {author} {\bibfnamefont {I.}~\bibnamefont
  {Ben-Dayan}},\ }\href {\doibase 10.1088/0004-637X/758/2/76} {\bibfield
  {journal} {\bibinfo  {journal} {Astrophys. J.}\ }\textbf {\bibinfo {volume}
  {758}},\ \bibinfo {pages} {76} (\bibinfo {year} {2012})},\ \Eprint
  {http://arxiv.org/abs/1203.2681} {arXiv:1203.2681 [astro-ph.CO]} \BibitemShut
  {NoStop}%
\end{thebibliography}%
\end{document}